\documentclass{aa}  

\usepackage{graphicx}
\usepackage{txfonts}
\usepackage{lipsum}
\usepackage{subcaption}         
\usepackage{lscape}             
\usepackage{placeins}           
\usepackage{afterpage}
                                
\usepackage{hyperref}
\usepackage{amsmath}	
\usepackage{bm}
\usepackage{xcolor}     
\renewcommand{\linenumbers}{}

\begin{document}

   \title{New perspectives on MASCARA-1b: A combined analysis of pre- and post-eclipse emission data using CRIRES+}
   \titlerunning{Combined Pre- and Post-eclipse Observations of MASCARA-1b with CRIRES+}


   \author{Swaetha Ramkumar\inst{1}\thanks{E-mail: ramkumas@tcd.ie},
        Neale P. Gibson\inst{1},
        Stevanus K. Nugroho\inst{2}\fnmsep\inst{3},
        Mark Fortune\inst{1},
        \and Cathal Maguire\inst{1}
        }
    \authorrunning{S. Ramkumar et al.}

   \institute{School of Physics, Trinity College Dublin, University of Dublin, Dublin-2, Ireland
            \and Astrobiology Center, NINS, 2-21-1 Osawa, Mitaka, Tokyo 181-8588, Japan
            \and National Astronomical Observatory of Japan, 2-21-1 Osawa, Mitaka, Tokyo 181-8588, Japan\\}

   \date{Received XXXX; accepted XXXX}

 
  \abstract
  {} 
   {We present high-resolution emission spectroscopy observations of the ultra-hot Jupiter MASCARA-1b with CRIRES+ in the K-band, covering the post-eclipse phases of the planet's orbit. These observations complement previously published pre-eclipse data.} 
   {The stellar and telluric features were removed from the data using the \textsc{SysRem} algorithm, and the planetary signal was analysed with the cross-correlation technique. After confirming the presence of chemical species in our atmospheric model, we combined the pre- and post-eclipse datasets for a joint analysis. By employing a Bayesian retrieval framework with various parametrisations, this joint retrieval enabled us to constrain the spatially varying temperature-pressure ($T$-$P$) profile and the atmospheric carbon-to-oxygen ($\rm C/O$) ratio.}
   {We detected strong emission signatures of CO and H$_2$O in the post-eclipse and the combined datasets. A well-mixed retrieval model results in a super-solar $\rm C/O$, whereas allowing for vertically varying chemistry by incorporating a chemical model results in $\rm C/O$ values consistent with solar within ${\approx}1.1\sigma$. The retrieved parameters are not only consistent across the datasets but also across different chemical regimes. We did not identify any significant velocity shifts between the detected species or across the datasets, which could otherwise serve as proxies for possible atmospheric dynamics. We also explored potential phase dependence through the model scaling factor and found no substantial changes in the atmospheric properties throughout the observed phases.}
   {Due to strong degeneracies between the temperature gradient and chemical abundances, our retrieved temperatures are broadly consistent with either a full redistribution of heat or strong day-night contrasts. While this complicates direct comparisons with recent Spitzer phase curve analyses suggesting inefficient recirculation, we find no clear evidence of spatial variation in the chemical or temperature structure of MASCARA-1b from pre- to post-eclipse, nor temporal variation over a period of $\approx$2 years.}


   \maketitle

\section{Introduction}\label{sect:1}
One of the most profound discoveries in the last decades in astronomy is that exoplanets are extremely common in the Universe. This discovery has not only introduced us to a range of strange and exciting worlds fundamentally different from the planets in our Solar System but has also allowed us to gain a deeper insight into their formation and evolution. Transit observations of exoplanets are crucial for comparative studies, and when combined with radial velocity data, these measurements allow the inference of the planet's mass, radius, and composition. However, little is known about the processes that determine the atmospheric structure of these planets, which highlights the need for spectroscopic observations using space-based and ground-based instruments.

In particular, high-resolution cross-correlation spectroscopy (HRCCS) has emerged as a powerful technique for characterising exoplanetary atmospheres. With significant advancements in ground-based high-spectral resolution platforms \citep[e.g. CRIRES+, ESPRESSO, IGRINS, IRD;][]{2014Msngr.156....7D,2023A&A...671A..24D,2021A&A...645A..96P,2014SPIE.9147E..1DP,2018SPIE10702E..0QM,Tamura2012,Kotani2018}, HRCCS has become instrumental in constraining basic atmospheric properties. First demonstrated by the detection of CO in the hot Jupiter HD 209458b \citep{2010Natur.465.1049S} using the CRyogenic high-resolution InfraRed Echelle Spectrograph \citep[CRIRES;][]{2004SPIE.5492.1218K}, HRCCS observations leverage the planetary Doppler shift of the dense forest of individually resolved spectral lines to detect both atomic and molecular species robustly. Since then, it has been used to detect several chemical species in the atmospheres of exoplanets, particularly those of close-in giant planets, such as hot- and ultra-hot Jupiters (UHJs). Due to their proximity to their host stars, these tidally locked gas giants are a fascinating population without similar counterparts in our Solar System. Their high equilibrium temperatures ($T_\mathrm{eq}$\,$\gtrsim$\,$2{,}200$ K) not only create conditions where major C- and O-bearing species (such as CO, OH and atomic O) are expected to exist in gaseous form but they also have a significant impact on their atmospheric structure and composition, such as the occurrence of inversion layers due to the presence of molecular compounds like TiO and VO \citep[][]{2022NatAs...6..449P,2023Natur.619..491P,2024A&A...687A..49M} high up in the atmospheres as well as due to absorption by atomic metals and metal hydrides \citep{2018ApJ...866...27L,2019MNRAS.485.5817G,2023ApJ...953L..19F}. This makes UHJs ideal targets for atmospheric characterisation wherein they serve as a treasure trove for numerous chemical species (e.g. $\rm Fe$\hspace{0.3mm}\textsc{i}, $\rm Fe$\hspace{0.3mm}\textsc{ii}, $\rm Ca$\hspace{0.3mm}\textsc{i}, $\rm Ca$\hspace{0.3mm}\textsc{ii}, $\rm OH$, $\rm CO$, $\rm H_2O$, etc.) which have been detected using high-resolution transmission and emission spectroscopy \citep[e.g.][]{2019A&A...627A.165H, 2020MNRAS.493.2215G, 2020ApJ...894L..27P, 2023MNRAS.519.1030M, 2021ApJ...910L...9N, 2023AJ....165...91B, 2023MNRAS.525.2985R}.

The HRCCS technique is not limited to transmission spectroscopy; it is also powerful when applied to emission spectroscopy and phase curve studies of close-in gas giant planets. By analysing the thermal emission from the planet's day-side and the variations in brightness throughout its orbit, we can obtain measurements of the temperature structure and chemical composition, which provides crucial information about day-night temperature differences, atmospheric dynamics, and heat redistribution. In addition, resolved phase-curve observations also enable us to obtain spectra as the planet rotates, allowing us to effectively spatially resolve the surface \citep[e.g.][]{2019ApJ...872....1R,2021MNRAS.501...78P}. When combined with high-resolution Bayesian methods \citep[e.g.][]{2019AJ....157..114B, 2020MNRAS.493.2215G}, these techniques allow us to recover vital constraints on the atmosphere, such as the abundance of chemical species, as well as the atmospheric $\rm C/O$ ratios and metallicities \citep[e.g.][]{2021Natur.598..580L, 2023AJ....165...91B, 2023MNRAS.525.2985R, 2024AJ....167..110S}; quantities that can provide potential insights into the formation and subsequent evolution history of these extreme planets \citep[e.g.][]{2011ApJ...743L..16O, 2011Natur.469...64M}.

Recently, CRIRES has been upgraded to a cross-dispersed echelle spectrograph \citep{2014Msngr.156....7D, 2014SPIE.9147E..19F} and the improved spectrograph, now known as CRIRES+, offers a ten-fold improvement in spectral coverage, making it ideal for characterising the atmospheres of exoplanets. MASCARA-1b, an ultra-hot Jupiter orbiting a fast-rotating bright A8 star ($V$\,=\,$8.3$, $K$\,=\,7.7) with an orbital period of $\approx$\,2.15 days \citep{2017A&A...606A..73T,2022A&A...658A..75H}, is a prime target for such studies due to its high equilibrium temperature \citep[2594 K;][]{2022A&A...658A..75H}. Analyses using high-resolution emission spectroscopy (e.g. CRIRES+, PEPSI, CARMENES) have detected both molecular \citep[CO, H$_2$O;][]{2022AJ....164...79H,2023MNRAS.525.2985R} and atomic species \citep[$\rm Fe$\hspace{0.3mm}\textsc{i}, $\rm Ti$\hspace{0.3mm}\textsc{i}, and $\rm Cr$\hspace{0.3mm}\textsc{i};][]{2023A&A...674A..58S,2023MNRAS.525.2985R,2024A&A...687A.103G} in the day-side atmosphere of this gas giant, underscoring the planet’s suitability for atmospheric characterisation at high resolution.

We have recently analysed the CRIRES+ phase-curve observations of MASCARA-1b, taken during the SV run in September 2021 \citep[program 107.22TQ.001;][]{2023MNRAS.525.2985R}. The observations were taken in the K-band at the peak of the thermal emission from the planet just before the secondary eclipse (covering orbital phase\,$\sim$\,0.34-0.43) and have enabled high-significance detections of CO, H$_2$O, and Fe. The species are found as emission signatures, implying a temperature inversion in the atmosphere, and by using a Bayesian retrieval framework, we were able to place constraints on the $\rm C/O$ ratio and temperature-pressure ($T$-$P$) profile, confirming the consistency of $\rm C/O$ with the solar value. Our analysis also revealed intriguing evidence that we are sensitive to temporal and spatial variations in the planet's atmosphere. For instance, our cross-correlation maps, as well as the retrievals, hint that the Fe feature originates in different regions of the atmosphere than the CO/H$_2$O, with the rotation ($v_{\rm eq}$$\sim$\,3.6 km s$^{-1}$ assuming tidal locking) and winds \citep[>5 km s$^{-1}$, e.g.][]{2016ApJ...821....9K} separating the features in velocity space. To explore potential variations and understand the atmosphere's chemical composition and temperature patterns, we observed MASCARA-1b using the same instrumental setup as the previous observations but after the secondary eclipse, thus extending the phase coverage. Analysing phase curves separately, for example, before and after the eclipse, can facilitate an immediate comparison of the $T$-$P$ profiles, $\rm C/O$ ratios, and metallicities. Furthermore, incorporating velocity constraints will provide valuable information about how atmospheric properties vary as the planet rotates and will help explore the importance of 3D effects. When combined with an atmospheric retrieval, constraints on the parametric phase curve, as well as spatially varying $T$-$P$ profile and abundances, can be placed, enabling us to pinpoint the chemical and temperature composition of the atmosphere.

Here, we present a complementary analysis of high-resolution secondary eclipse observations taken two years apart with the CRIRES+ spectrograph, covering the post-eclipse phases of MASCARA-1b's orbit. The paper is organised as follows: in Section~\ref{sect:2}, we present our observations and data reduction; in Section~\ref{sect:3}, we describe our forward model atmosphere for emission and detail our methodology, including the cross-correlation technique and present our detection results. We then outline our atmospheric retrieval frameworks and present our retrieval results for the new data and the joint analysis in Section~\ref{sect:4}. Finally, we discuss our iron detection, examine signatures of atmospheric dynamics, and contextualise MASCARA-1b's atmosphere in Section~\ref{sect:5} before concluding the study in Section~\ref{sect:6}.

\section{Observations and data reduction}\label{sect:2}
\begin{figure*}
\centering
\begin{minipage}[htbp]{0.333\textwidth}
  \includegraphics[width=\linewidth]{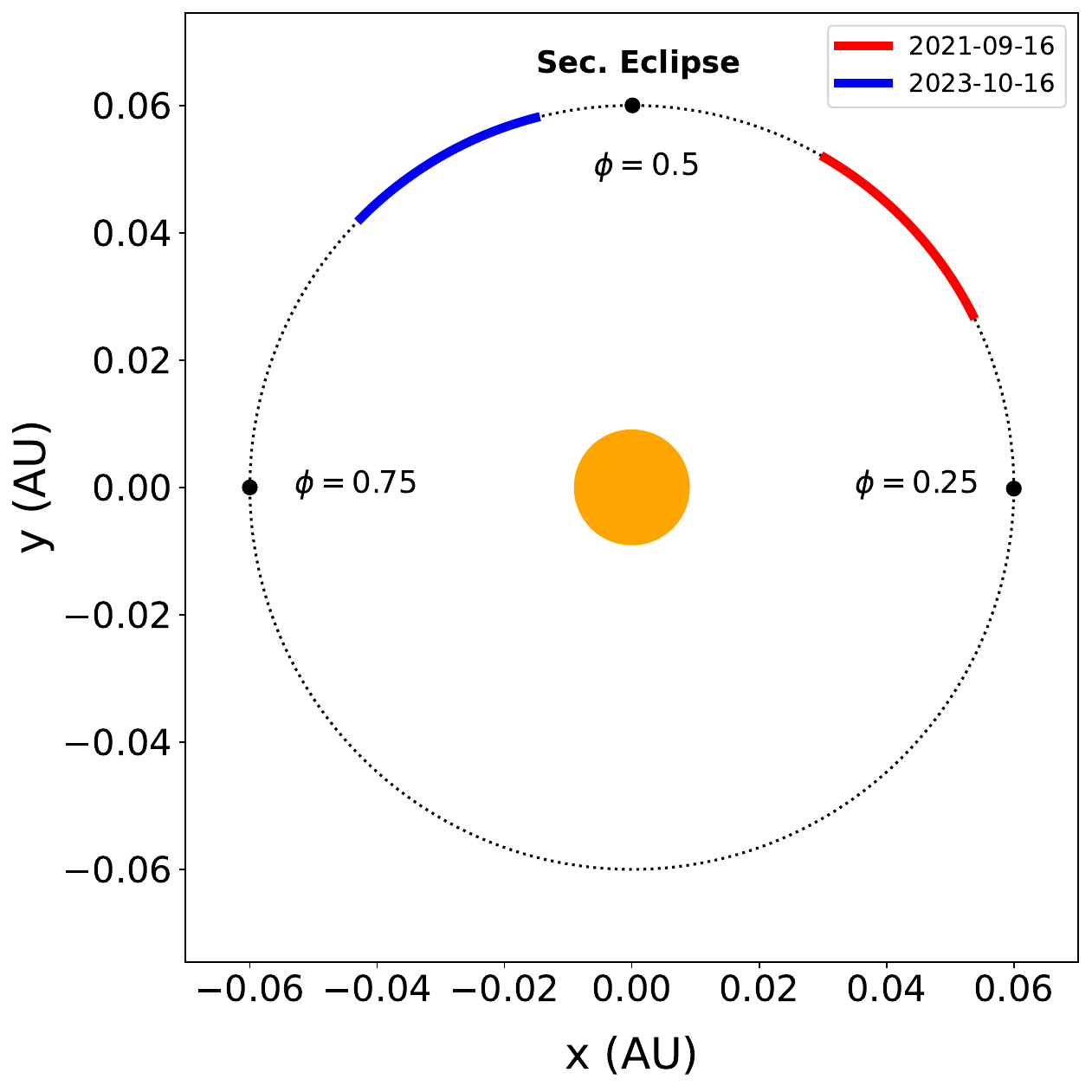}
\end{minipage}%
\hfill 
\begin{minipage}[htbp]{0.333\textwidth}
  \includegraphics[width=\linewidth]{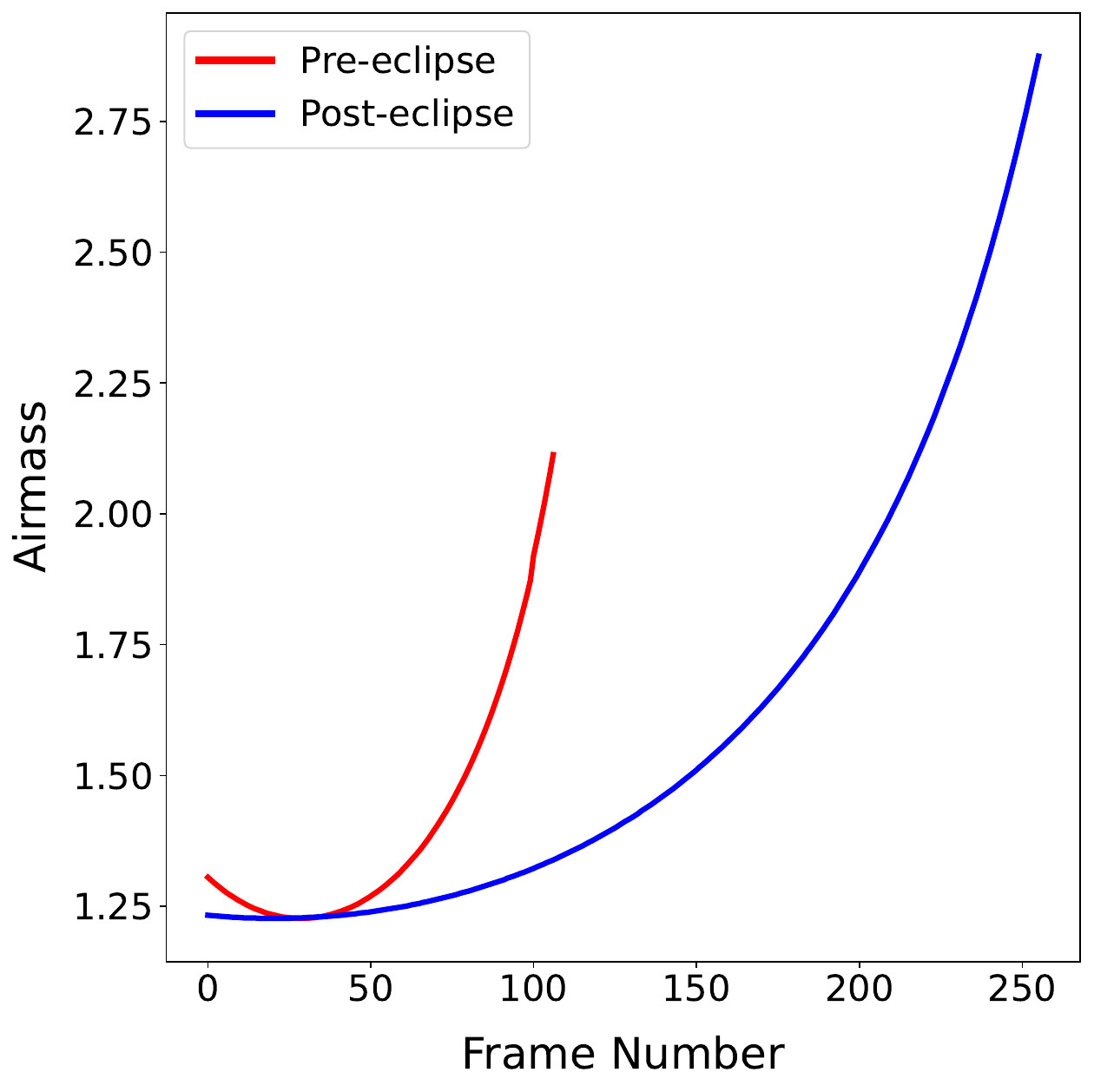}
\end{minipage}%
\hfill
\begin{minipage}[htbp]{0.333\textwidth}
  \includegraphics[width=\linewidth]{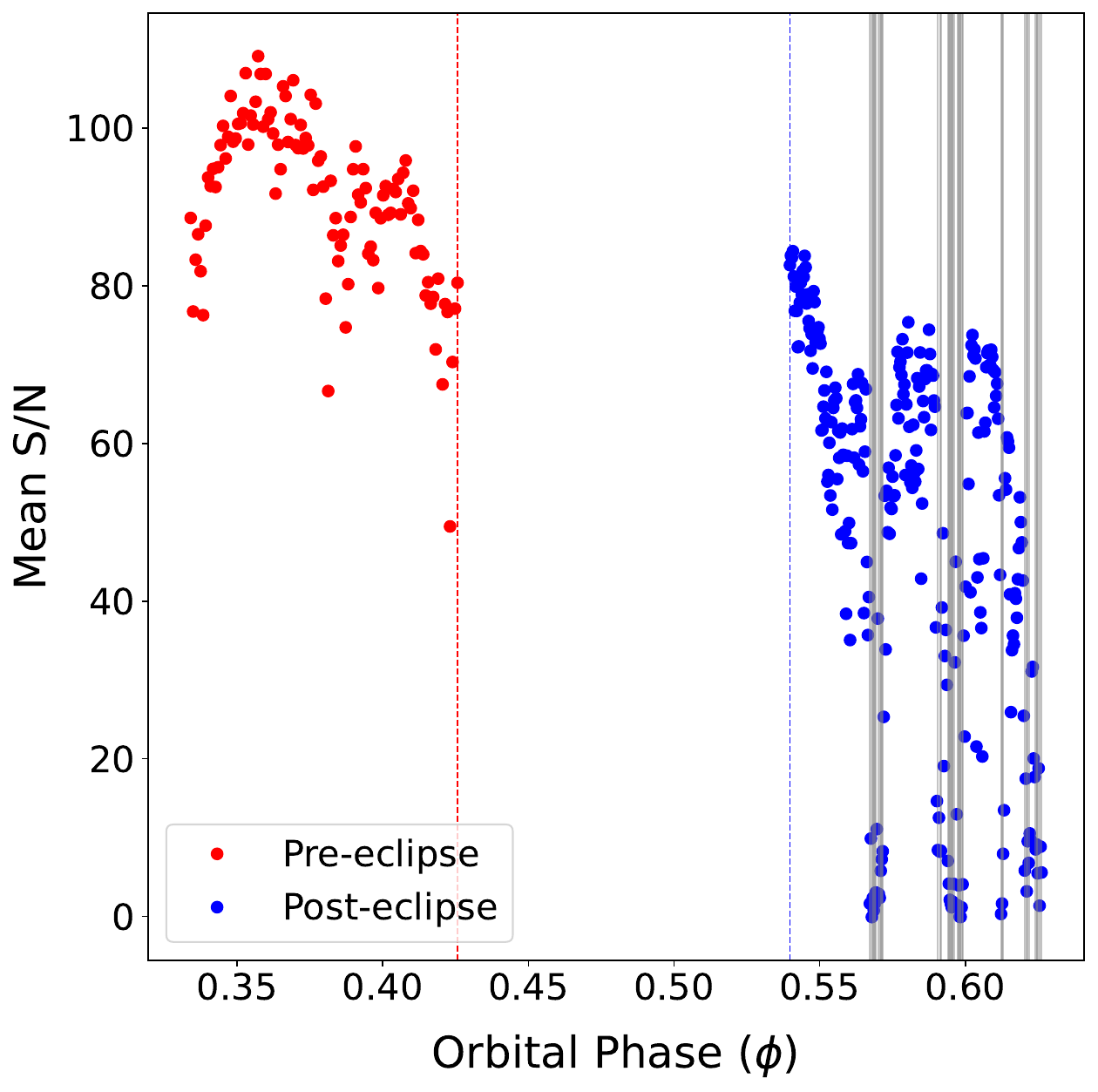}
\end{minipage}%
\caption{\emph{Left:} Phase coverage of our CRIRES+ pre- and post-eclipse observations. \emph{Middle and Right:} Observing conditions showing the variation of airmass and mean signal-to-noise ratio (SNR) with orbital phase. Vertical grey bands denote cloud-affected regions.}
\label{fig1}
\end{figure*}

Observations of MASCARA-1b were taken two years apart for two half nights in September 2021 and October 2023 as part of programs 107.22TQ.001 and 112.260X.001 (PI: Gibson, dPI: Ramkumar) using the upgraded CRIRES+ spectrograph \citep[$R\,{\sim}$\,100,000; $\lambda{\sim}1921$-$2472$ nm (K-band);][]{2014SPIE.9147E..19F} installed on UT3 of the VLT. The data taken on 2021 September 16, covering the pre-eclipse phases, were previously presented and analysed in \citet{2023MNRAS.525.2985R}, and we refer the reader there for further details. The data taken on 2023 October 16, covering the post-eclipse phases, are presented here for the first time. This observing night was interrupted by thick clouds at intervals, resulting in the loss of the target for $\sim$30 minutes over the entire observing night (visible as horizontal bands in the cross-correlation map, e.g. Fig.~\ref{fig6}). We obtained 256 exposures covering the orbital phase of MASCARA-1b from $\phi$$\sim$0.54 to 0.63 (where $\phi$ = 0.5 corresponds to the secondary eclipse) and observed in the K-band using the K2166 wavelength setting with (interrupted) coverage from $\lambda$$\sim$1921 to 2472 nm. We calculated the orbital phase using the transit epoch from \citet{2022A&A...658A..75H} and computed the barycentric velocity correction for each exposure using the online tool from \citet{2014PASP..126..838W}. An overview of the observations is shown in Table~\ref{table:1}, and the observing conditions are presented in Fig.~\ref{fig1}.
\begin{table*}
\caption{A summary of the CRIRES+ observations of MASCARA-1b.}     
\label{table:1}                     
\centering                          
\renewcommand{\arraystretch}{1.4}
\begin{tabular}{c c c c c c}        
\hline\hline                        
& Date & Phase coverage & Exposure time [s] & N$_{\rm spectra}$ & Airmass \\    
\hline                        
Pre-eclipse & 2021-09-16 & $\approx$0.334-0.425 & 150 & 107 & 1.305-2.113 \\   
Post-eclipse & 2023-10-16 & $\approx$0.539-0.626 & 60 & 256 & 1.233-2.874\\
\hline                                   
\end{tabular}
\end{table*}

The basic calibration and extraction of the time-series spectra for each spectral order were performed with the ESO CR2RES data pipeline (version 1.1.5) and executed using the command-line interface {\sc Esorex}\footnote{Documentation available at the ESO website \url{http://www.eso.org/sci/software/pipelines/}} (version 3.13.5). First, we combined each set of raw dark frames into master dark frames (which also produced a bad pixel map). Next, we compute master flat frames and perform trace detection. Following this, we perform wavelength calibration of the extracted spectra using the Fabry-P\'erot Etalon (FPET) frame and also check for additional wavelength shifts during the observation by cross-correlating the data with Doppler-shifted telluric templates produced using ESO Sky Model Calculator \citep[SkyCalc,][]{2012A&A...543A..92N,2013A&A...560A..91J} over a velocity range of $-$300 to 300 km s$^{-1}$ in steps of 0.01 km s$^{-1}$. However, we did not find any significant shifts (<1 km s$^{-1}$); and therefore did not apply any additional refinement to the wavelength calibration.
Finally, we calibrate and extract the science spectra (1D spectra as a function of order). The extraction is done in pieces by dividing the image into several swaths with widths set to 400 pixels.
In total, we obtain 7 echelle orders for the K2166 wavelength setting. Additionally, with CRIRES+, each spectral order is observed across 3 separate detectors (CHIP1, CHIP2, and CHIP3), so we treat each separate chip independently and hereafter refer to them as orders. We produce a 3D data cube (order\,$\times$\,time/phase\,$ \times$\,wavelength) with 21 spectral orders ranging from $\lambda$\,$\approx$\,1920 to 2470 $\mathrm{nm}$.

Following the reduction and spectral extraction, we perform similar pre-processing steps as \citet{2023MNRAS.525.2985R} to clean and blaze correct the extracted spectra. The noise estimates for each order were then extracted by assuming Poisson-dominated noise following \citet{2020MNRAS.493.2215G, 2022MNRAS.512.4618G}, to which we refer the reader. However, while this approach is well-suited for optical data \citep[e.g.][]{2020MNRAS.493.2215G}, it may not fully account for the final noise amplitude in near-infrared data, particularly for poorly corrected telluric, stellar, or systematic effects. Therefore, as an alternative to the optimisation presented above, we also estimated the uncertainties of each order post-\textsc{SysRem} by taking the outer product of the standard deviation of each wavelength channel and spectrum, normalised by the overall mean. Re-running \textsc{SysRem} with these new uncertainties, we found that the choice of error estimation method did not have a discernible impact on our detections or retrieved values.
Since the post-eclipse observations were affected by clouds, we note that applying the same treatment as the pre-eclipse data is not ideal, as the blaze correction does not perform optimally for these frames and affects the shape of the posteriors, though in small ways. Therefore, to ensure that the pre-processing steps did not significantly distort the underlying exoplanetary signal, we performed injection/recovery tests detailed in Sect.~\ref{sect:3.2}.

\begin{figure}[htbp]
   \centering
   \includegraphics[width=\hsize]{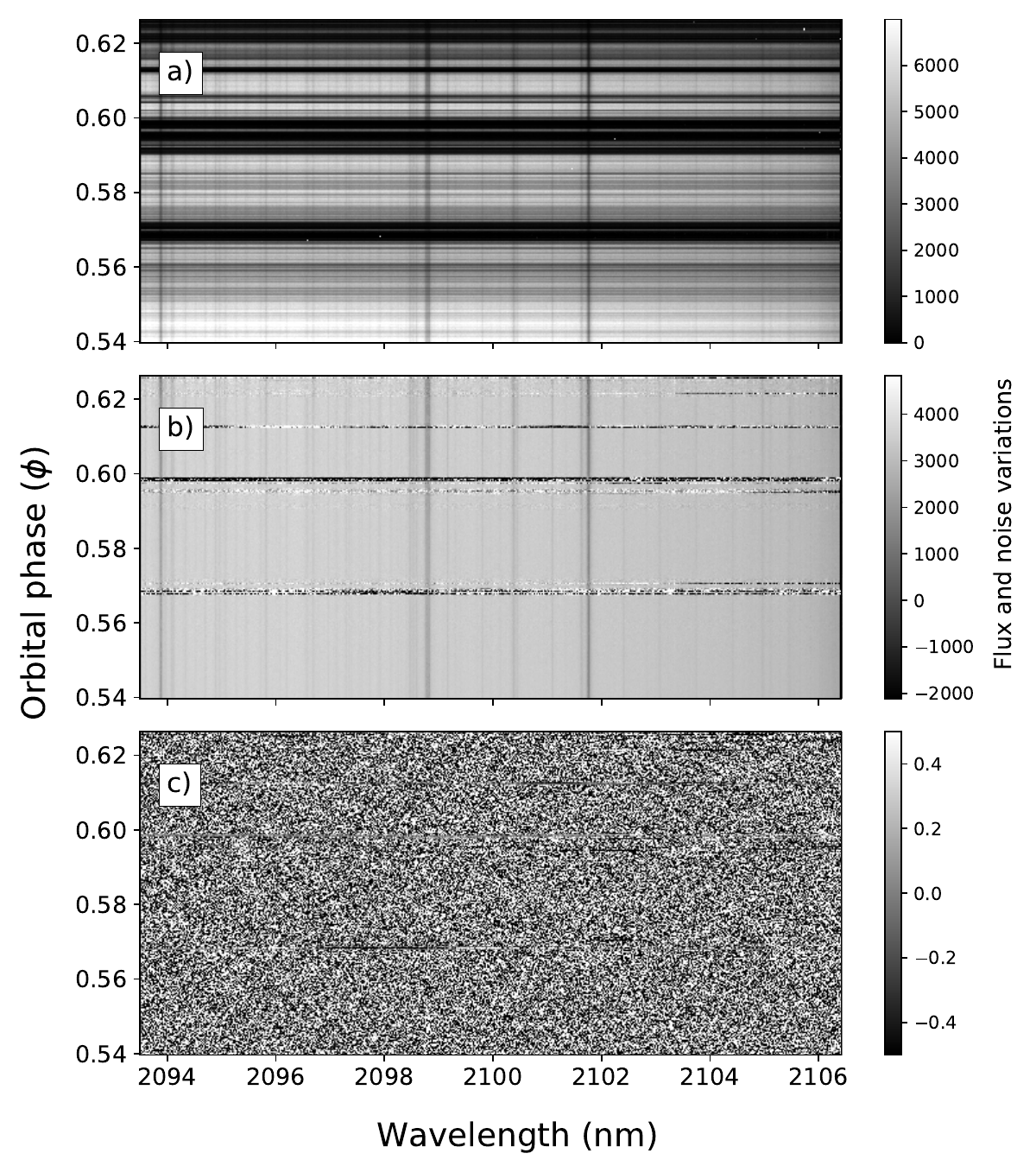}
      \caption{The data pre-processing steps, applied to a single order: (a) raw spectra, (b) cleaned and blaze corrected spectra, and (c) residual spectra after division by the median spectrum and subtraction of the \textsc{SysRem} model (shown weighted by the uncertainties). We note that the blaze correction does not work well for cloud-affected frames, resulting in some horizontal structure/residuals in these regions (the last two panels). However, they only have a minimal impact on our retrievals.}
         \label{fig2}
   \end{figure}

Before employing the cross-correlation technique to search for any exoplanetary signal, we removed the stellar and telluric lines using a de-trending algorithm, \textsc{SysRem} \citep{2005MNRAS.356.1466T}, which was initially adapted for high-resolution spectroscopy by \citet{2013MNRAS.436L..35B} and has since been successfully utilised for transmission and emission spectra \citep[e.g.][]{2017ApJ...839L...2B, 2017AJ....154..221N, 2020MNRAS.496..504N,  2020ApJ...898L..31N, 2020MNRAS.493.2215G, 2022MNRAS.512.4618G, 2023MNRAS.525.2985R}. Following the procedure outlined in \citet{2022MNRAS.512.4618G}, we first normalise the data by dividing through the median spectrum in each order. Then, we perform \textsc{SysRem} on the normalised data and corresponding uncertainties for every iteration, generate a resultant model, subtract it from the input data to get the processed (residual) matrix, and repeat the procedure for the subsequent iteration. Thus, after $N$ iterations, the \textsc{SysRem} model for a single order is:
\begin{equation}
    D =  \sum_{i=1}^{N}\boldsymbol{u_i}\boldsymbol{w_i}^{\rm T} = {\mathbf{U}\mathbf{W}^{\rm T}}
    \label{eqn1}
\end{equation}
where $\mathbf{U}$ and $\mathbf{W}$ are matrices containing column vectors $\boldsymbol{u_i}$ and $\boldsymbol{w_i}$. $\mathbf{U}$ is then stored for processing the forward model (see Sect.~\ref{sect:3}). We apply 15 passes of \textsc{SysRem}; while that is somewhat arbitrary, we rerun the analysis in Sect.~\ref{sect:4} with different values of \textsc{SysRem} ($N$ = 5, 10 and 20). A step-by-step overview of our pre-processing steps is shown in Fig.~\ref{fig2} for a single order.

\section{A search for atmospheric species}\label{sect:3}
\subsection{Model spectra}\label{sect:3.1}
To search for chemical species in the atmospheres of exoplanets and place quantitative constraints on atmospheric properties, both the cross-correlation and likelihood approaches require spectral templates. Therefore, for our analysis, we computed the 1D atmospheric models using \textsc{irradiator} \citep{2022MNRAS.512.4618G}, where we defined 80 atmospheric layers equally spaced in log pressure from 10$^{-8}$ bar to 100 bar and computed the $T$-$P$ profile at each layer using the parametric model from \citet{2010A&A...520A..27G} with four input terms $T_{\rm irr}$ (irradiation temperature), $\kappa_{\rm IR}$ (mean infrared opacity), $\gamma$ (ratio of visible-to-infrared opacity) and $T_{\rm int}$ (internal temperature). Next, we set the volume mixing ratios (VMRs), $\chi_{\rm species}$, for each species of interest at each layer. These are either given as free parameters, that is, the species are assumed to be well-mixed, or computed using \textsc{FastChem}\footnote{Documentation:\hspace{0.5mm}\url{https://newstrangeworlds.github.io/FastChem/}} \citep[version 3.1.1,][]{2018MNRAS.479..865S, 2022MNRAS.517.4070S, 2024MNRAS.527.7263K} using an additional free parameter to specify the metallicity (relative to solar) and $\rm C/O$ ratio, where we adjust both C and O abundances to have a desired ratio but preserve their sum. We compute the emission by integrating through the atmosphere for every point in the wavelength grid, assuming black-body emission from the deepest layer. The spectra are calculated across a wavelength range of 1800-2850 nm at a constant resolution of $R$\,=\,$200{,}000$.
In our atmospheric models, we account for the opacity contributions from chemical species prominent in the atmospheres of UHJs at near-infrared wavelengths \citep[e.g.][]{2010Natur.465.1049S,2013MNRAS.436L..35B,2020ApJ...898L..31N,2023AJ....165...91B}. For atomic and molecular absorption, we use the high-resolution pre-computed opacity grids provided by \texttt{petitRADTRANS}\footnote{Documentation:\hspace{0.5mm}\url{https://petitradtrans.readthedocs.io/en/latest/}} \citep[version 2.7.7, ][]{2019A&A...627A..67M, 2020A&A...640A.131M} and also include collision-induced opacity due to H$_2$-H$_2$ and H$_2$-He and bound-free and free-free absorption from H$^{-}$, using the cross-sections from \citet{2005oasp.book.....G} and \citet{doi:10.1021/jp109441f}. Additionally, we incorporate a parameterised cloud deck pressure ($P_{\rm cloud}$), assuming infinite opacity below the atmosphere. Lastly, since data pre-processing methods such as \textsc{SysRem} can alter the underlying planet signal, applying these same processing steps to our forward models as we do to the data is necessary. Thus, we implement the model filtering technique introduced by \citet{2022MNRAS.512.4618G}, which makes use of the output matrices $\mathbf{U}$ and $\mathbf{W}$ (Sect.~\ref{sect:2}), containing the column and row vectors $\mathbf{u}$ and $\mathbf{w}$ for each \textsc{SysRem} iteration. Before applying this technique, we convolve our model emission spectra with a broadening kernel with a standard deviation of $W_{\rm conv}$, followed by linear interpolation to match the 2D wavelength grid of our data. After this, we Doppler shift this model to a planetary velocity, $v_{\rm p}$, for each order:
\begin{equation}
    v_\mathrm{p} = K_\mathrm{p} \sin(2\pi\phi) + v_\mathrm{sys} + v_\mathrm{bary} \label{eqn2}
\end{equation}
where $ K_\mathrm{p} $, $\phi$, $v_{\rm sys}$ and $v_{\rm bary}$ are the radial velocity semi-amplitude of the planet's orbit, orbital phase, the systemic velocity offset and barycentric velocity, respectively. This produces a 3D shifted forward model, which is then filtered using the \textsc{SysRem} basis vectors via the procedure outlined in \citet{2022MNRAS.512.4618G}.

\begin{figure*}[htbp]
\centering
    \includegraphics[width=18.5cm]{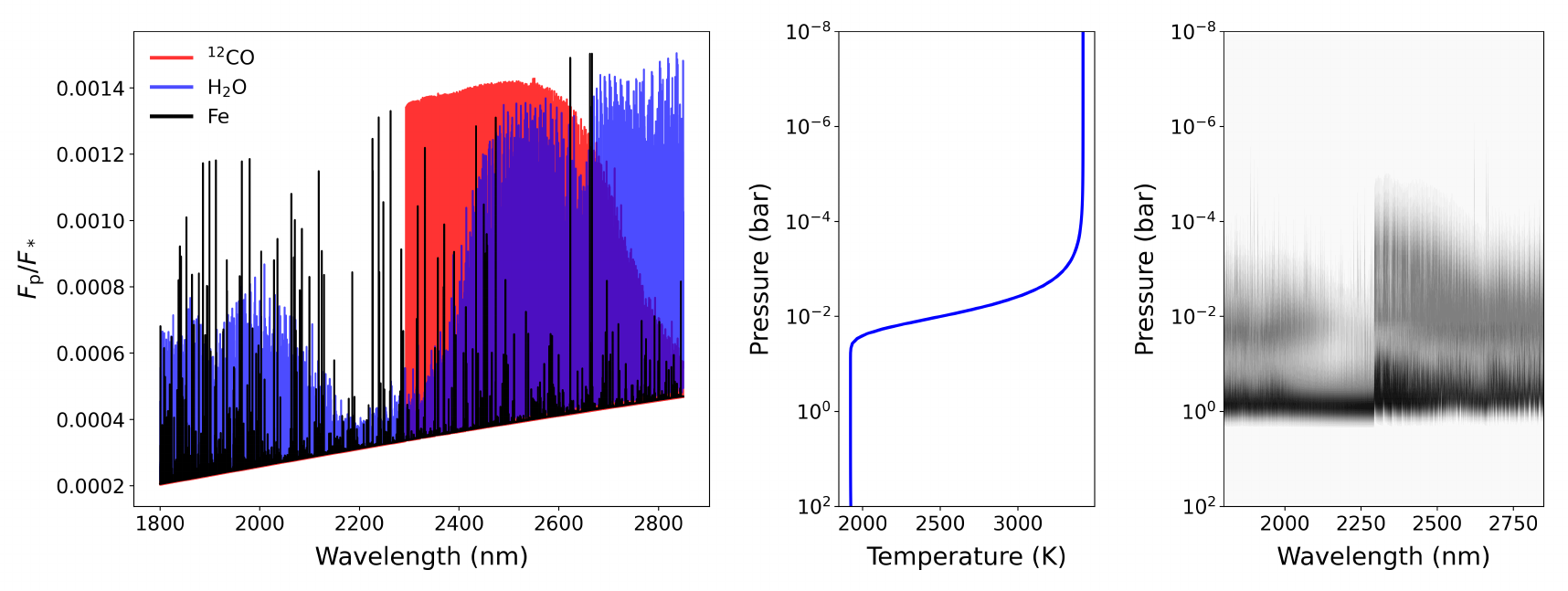}
    \caption{\emph{Left panel:} the best-fit model spectra (constant VMRs) of MASCARA-1b with contributions from the chemical species shown individually. \emph{Middle and Right panels:} the parametric $T$-$P$ profile from \citet{2010A&A...520A..27G} and combined contribution function.}
    \label{fig3}
\end{figure*}

In summary, we specify our 1D atmosphere using the parametric model from \citet{2010A&A...520A..27G} and two different chemical regimes, assuming either that the species of interest are well-mixed (constant VMRs with altitude) or by using \textsc{FastChem} to compute the chemical profiles after specifying a $\rm C/O$ ratio and metallicity. We apply various combinations of these parametrisations to our cross-correlation and retrieval analyses, which we discuss in more detail in Sect.~\ref{sect:4}. The model spectrum (constant VMRs) and corresponding $T$-$P$ profile are shown in Fig.~\ref{fig3}.

\subsection{Injection/recovery tests}\label{sect:3.2}
Before performing a retrieval analysis (outlined in Sect.~\ref{sect:4}) using our forward model, we conducted injection and recovery tests to ensure the pre-processing and filtering steps did not alter the underlying planet signal. We injected an atmospheric model containing CO, H$_2$O, and Fe with a negative value of the expected $K{\rm p}$ and performed our retrieval analysis on the injected data. We found the retrieved model parameters and $T$-$P$ profiles to be in agreement with the injected values (see Appendix~\ref{appendix:A}). Additionally, although the cloud-affected frames effectively contain weaker signal due to low S/N and could be filtered out, given the already limited phase coverage and data, we chose not to exclude these frames. To ensure their inclusion did not bias the results, we ran retrievals (Sect.~\ref{sect:4}) with and without the cloud-affected frames. We found the retrieved parameters and $T$-$P$ profiles to be consistent within ${\approx}1\sigma$ (see Appendix~\ref{appendix:B}). Although including these frames introduced small changes to the temperature gradient and irradiation temperature ($T_{\rm irr}$), they did not impact the overall results.

\subsection{Cross-correlation} \label{sect:3.3}
After removing the stellar and telluric lines from our data using \textsc{SysRem}, the remaining data is dominated by noise (panel (c) of Fig.~\ref{fig2}). To extract the planetary signal, we employ the traditional cross-correlation technique \citep[e.g.][]{2010Natur.465.1049S} and generate the cross-correlation function (CCF) by summing the product of the residual data ($f_i$) and the shifted (+filtered) model ($m_i$), weighted on the variance of the data ($\sigma_i$) over both spectral order and wavelength:
\begin{equation}\label{eqn3}
    {\rm CCF}(\Delta v_{\rm sys}) = \sum^{N}_{i=1} \frac{f_im_i(\Delta v_{\rm sys})}{\sigma_i^2}
\end{equation}
This produces a cross-correlation array of phase vs $\Delta v_{\rm sys}$, referred to as the cross-correlation map, where $\Delta v_{\rm sys}$ represents the residual velocity shift relative to the stellar rest frame after correcting for systemic and barycentric velocities\footnote{The expected $v_{\rm sys}$ is zero, as the spectra have already been shifted to the stellar rest frame.}. For strong planetary signals, the CCF will trace out the predicted radial velocity with phase in the CC-map, allowing us to confirm the presence of atmospheric species in exoplanets. However, the strength of the cross-correlation (CC) signal is often much weaker, and we need to integrate the cross-correlation map over a range of predicted planetary velocities ($v_{\rm p}$). Since the exact value for the radial velocity semi-amplitude ($K_{\rm p}$) is unknown and to evaluate noise, we select a range of $K_{\rm p}$ values close to the predicted value from radial-velocity measurements (Table.~\ref{table:2}) and shift the CCF to a new planetary velocity (according to Eqn.~\ref{eqn2}) and then integrate over time to produce a $K_{\rm p}$-$\Delta v_{\rm sys}$ map. Integration helps isolate the source of the signal in velocity space, and we can be confident of the detection of species in the atmosphere.
\begin{table}[htbp]
\caption{Parameters for MASCARA-1 system.}
\label{table:2}                     
\renewcommand{\arraystretch}{1.12} 
\begin{tabular}{lcc}
\hline \hline
Parameter [unit] & Symbol & Value \\
\hline
\emph{stellar parameters} & & \\
Effective temperature [K] & $T_{\rm eff}$ & 7554 $\pm$ 150$^{(a)}$ \\
Stellar radius [$ \textup{R}_\odot\ $] & $R_{\star}$ & $2.082\pm0.02$$^{(b)}$ \\
Stellar mass [$ \textup{M}_\odot\ $] & $M_{\star}$ & 1.72 $\pm$ 0.07$^{(a)}$ \\
Metallicity [dex] & $\rm [Fe/H]$ & 0$^{(a)}$ \\
& & 0.15 $\pm$ 0.15$^{(b)}$\\
\hline
\emph{planet parameters} & & \\
Planet radius [$R_{\rm J}$] & $R_{\rm p}$ & 1.5 $\pm$ 0.3$^{(a)}$\\
Planet mass [$M_{\rm J}$] & $M_{\rm p}$ & 3.7 $\pm$ 0.9$^{(a)}$\\
Equilibrium temperature [K] & $T_{\rm eq}$ & 2570$^{+50}_{-30}$$^{(a)}$ \\
& & 2594.3$^{+1.6}_{-1.5}$$^{(b)}$\\
\hline
\emph{system parameters} & & \\
Epoch [$\rm BJD_{TDB}$] & $T_{\rm 0}$ & 2458833.488151 \\
& & $\pm$0.00009$^{(b)}$ \\
Orbital period [day] & $P$ & 2.14877381\\
& & $\pm$0.0000009$^{(b)}$ \\
Semi-major axis [AU] & $a$ & 0.040352 $\pm$ 0.00005$^{(b)}$\\
Eccentricity & $e$ & 0 (fixed) \\
Inclination [degree] & $i$ & 87$^{+2}_{-3}$$^{(a)}$ \\
RV semi-amplitude [$ \rm km $ $\rm s^{-1}$] & $K_{\rm p}$ & 204.2 $\pm$ 0.2$^{(b)}$\\
& & 217 $\pm$ 25$^{(a)}$ \\
Systemic velocity [$ \rm km $ $\rm s^{-1}$] & $\gamma$, $v_{\rm sys}$ & 11.20 $\pm$ 0.08\\
& & 8.52 $\pm$ 0.02$^{(c)}$\\
\hline
\end{tabular}
\tablebib{(a)~\citet{2017A&A...606A..73T};
(b) \citet{2022A&A...658A..75H}; (c) \citet{2017A&A...606A..73T} found an offset in the systemic velocities derived from two different data sets.}
\end{table}

The final step in securing the spectroscopic detection of the exoplanet is to determine the significance level of the signal, which measures the signal amplitude. There are several ways to do this \citep[see, e.g.][]{2010Natur.465.1049S, 2013MNRAS.436L..35B}. We subtract the $K_{\rm p}$-$\Delta v_{\rm sys}$ map by the mean in regions away from the peak before dividing through by the standard deviation \citep{2012Natur.486..502B, 2018A&A...615A..16B}. Since it is difficult to interpret the significance of these values statistically due to the arbitrary selection of this region, we utilise the likelihood distribution (see Sect.~\ref{sect:4.1}) of our model scale factor, $\alpha$, to compute the detection significance, similar to the approach of \citet{2020MNRAS.493.2215G}, which we discuss in the following sections.
\begin{figure*}
\begin{minipage}[htbp]{0.333\textwidth}
  \includegraphics[width=\linewidth]{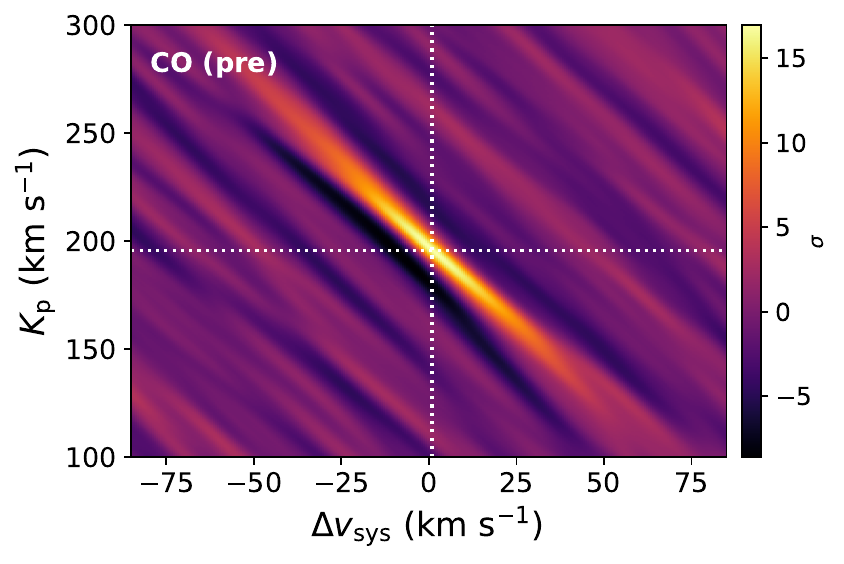}
\end{minipage}%
\hfill 
\begin{minipage}[htbp]{0.333\textwidth}
  \includegraphics[width=\linewidth]{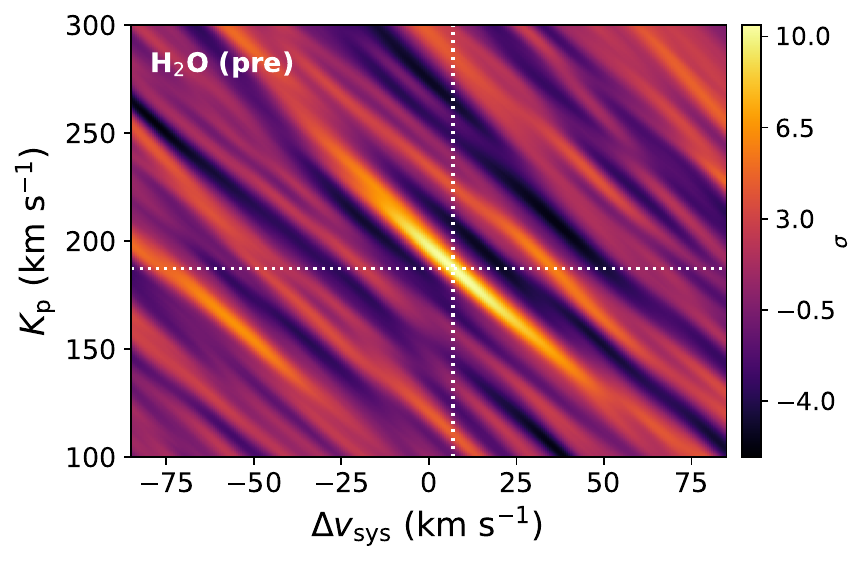}
\end{minipage}%
\hfill
\begin{minipage}[htbp]{0.333\textwidth}
  \includegraphics[width=\linewidth]{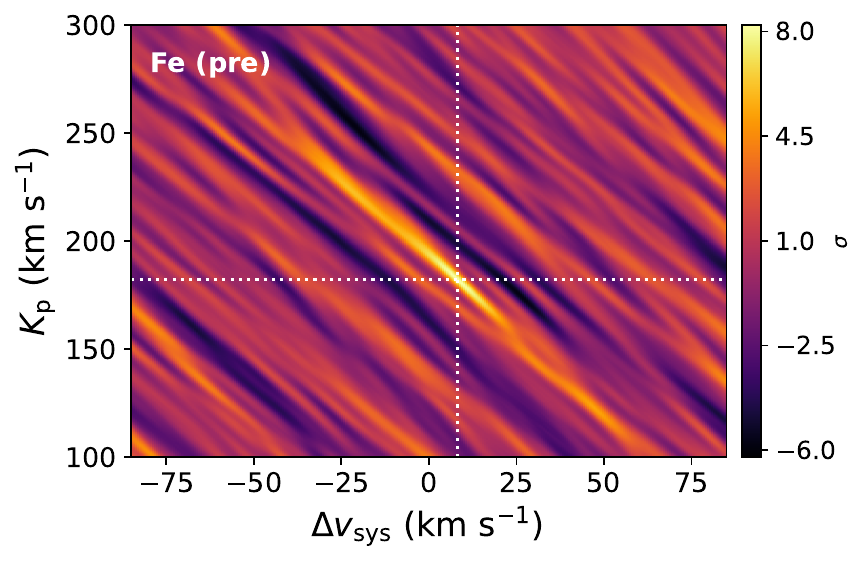}
\end{minipage}%
\end{figure*}

\begin{figure*}
\begin{minipage}[htbp]{0.333\textwidth}
  \includegraphics[width=\linewidth]{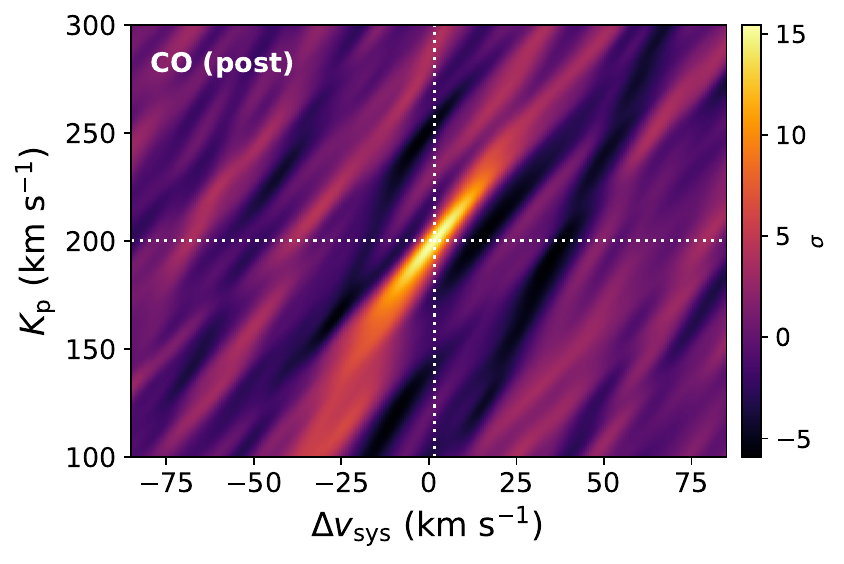}
\end{minipage}%
\hfill 
\begin{minipage}[htbp]{0.333\textwidth}
  \includegraphics[width=\linewidth]{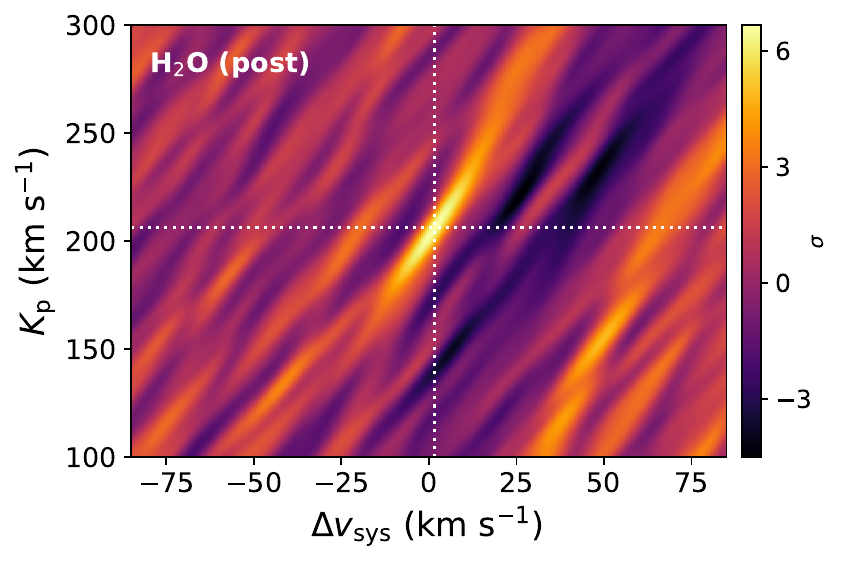}
\end{minipage}%
\hfill
\begin{minipage}[htbp]{0.333\textwidth}
  \includegraphics[width=\linewidth]{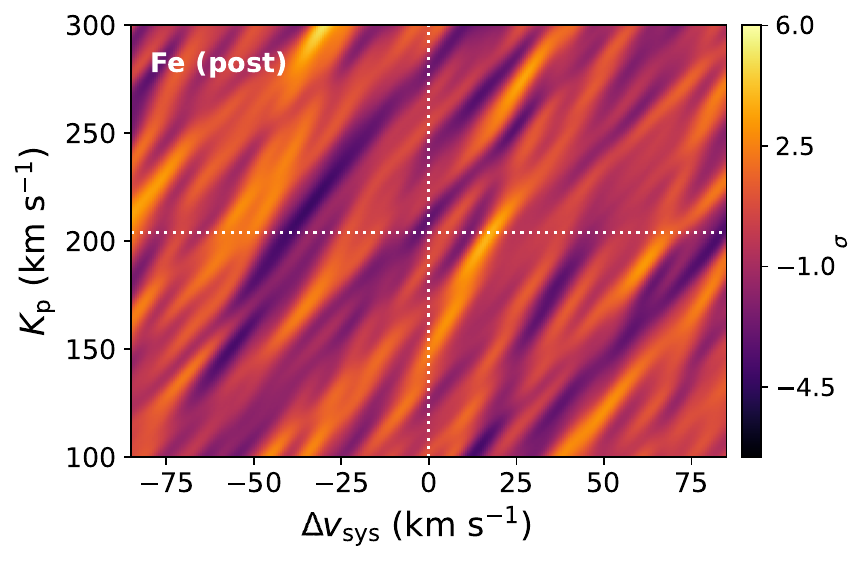}
\end{minipage}%
\end{figure*}

\begin{figure*}
\begin{minipage}[htbp]{0.333\textwidth}
  \includegraphics[width=\linewidth]{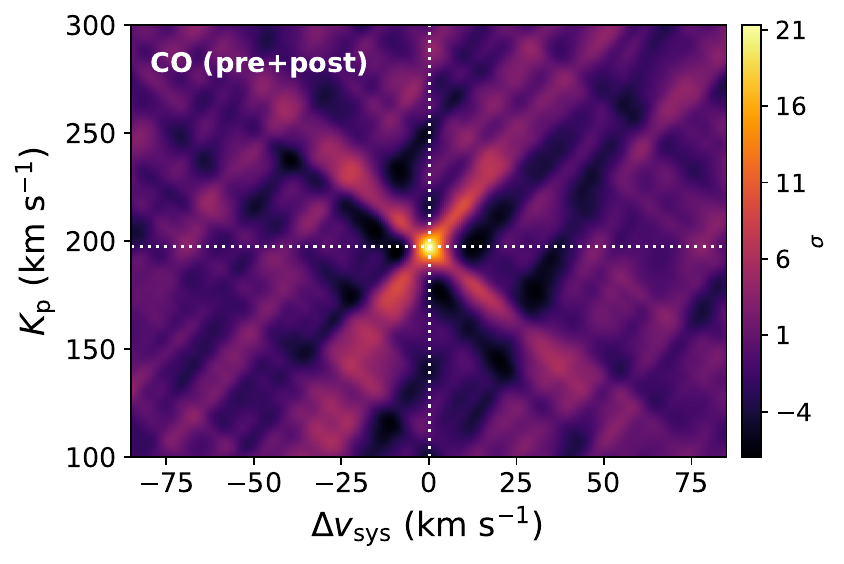}
\end{minipage}%
\hfill 
\begin{minipage}[htbp]{0.333\textwidth}
  \includegraphics[width=\linewidth]{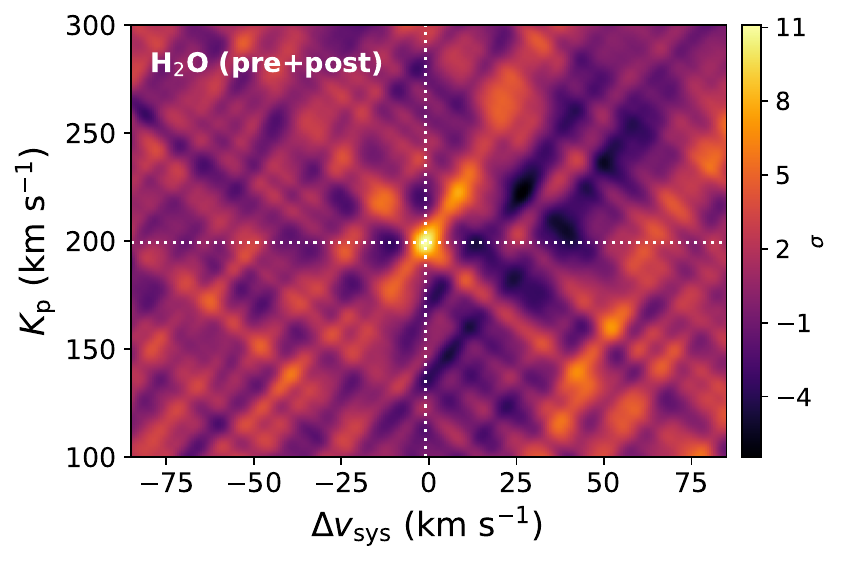}
\end{minipage}%
\hfill
\begin{minipage}[htbp]{0.333\textwidth}
  \includegraphics[width=\linewidth]{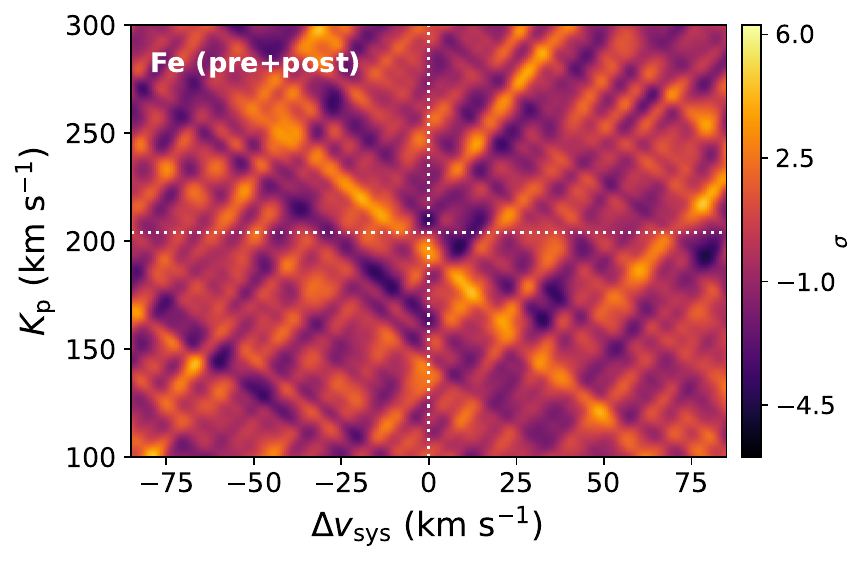}
\end{minipage}%
\end{figure*}

\begin{figure*}
\begin{minipage}[htbp]{0.333\textwidth}
  \includegraphics[width=\linewidth]{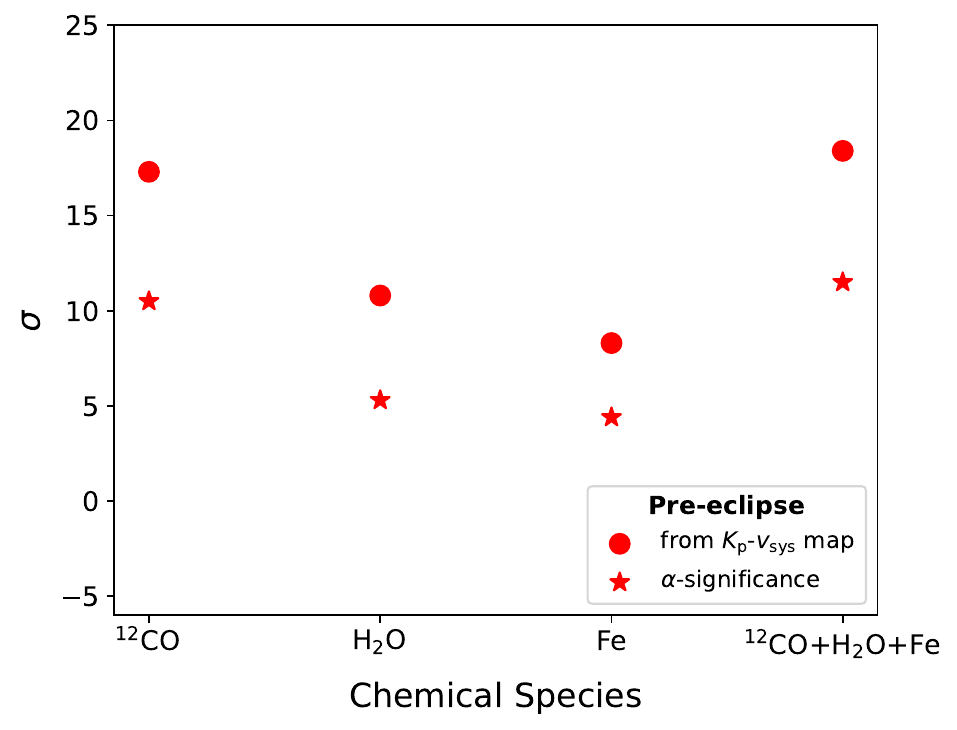}
\end{minipage}%
\hfill 
\begin{minipage}[htbp]{0.333\textwidth}
  \includegraphics[width=\linewidth]{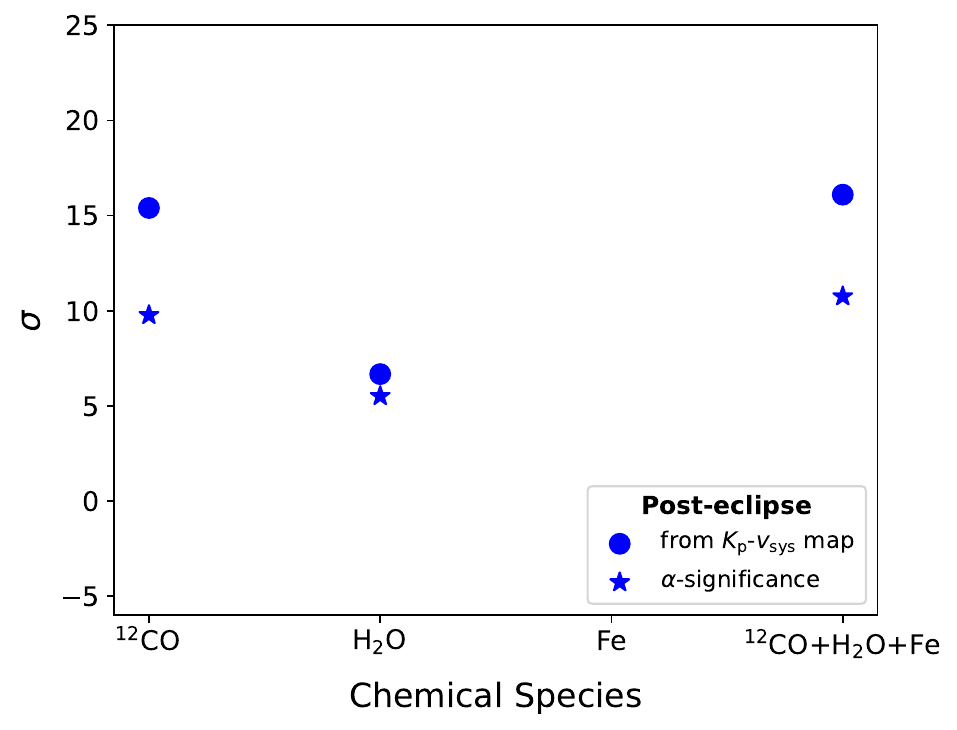}
\end{minipage}%
\hfill
\begin{minipage}[htbp]{0.333\textwidth}
  \includegraphics[width=\linewidth]{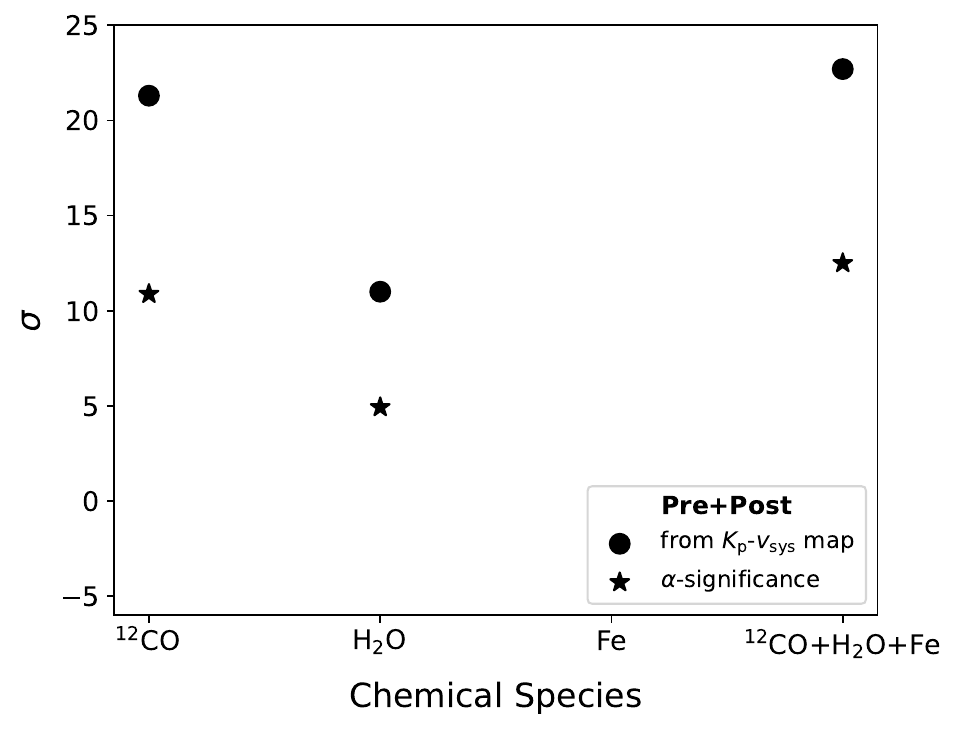}
\end{minipage}%
\end{figure*}

\begin{figure*}
\begin{minipage}[htbp]{0.333\textwidth}
  \includegraphics[width=\linewidth]{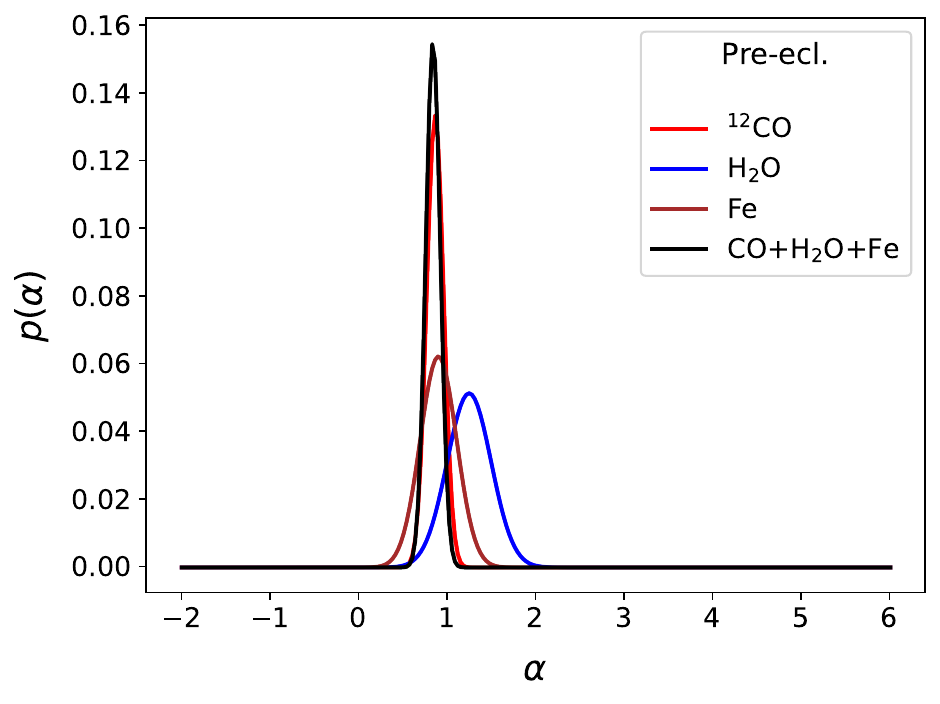}
\end{minipage}%
\hfill 
\begin{minipage}[htbp]{0.333\textwidth}
  \includegraphics[width=\linewidth]{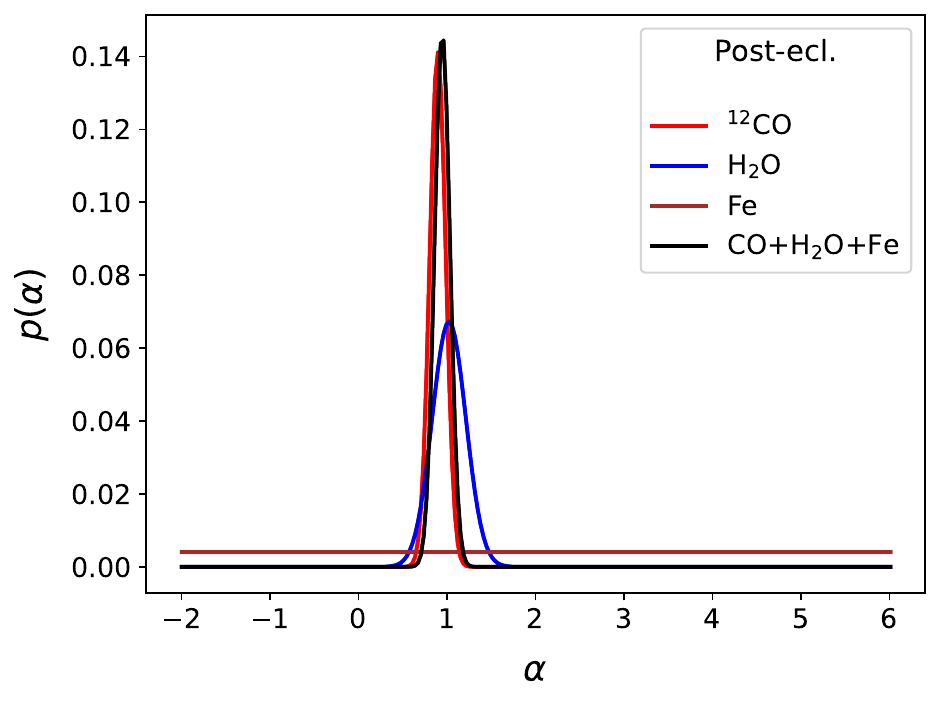}
\end{minipage}%
\hfill
\begin{minipage}[htbp]{0.333\textwidth}
  \includegraphics[width=\linewidth]{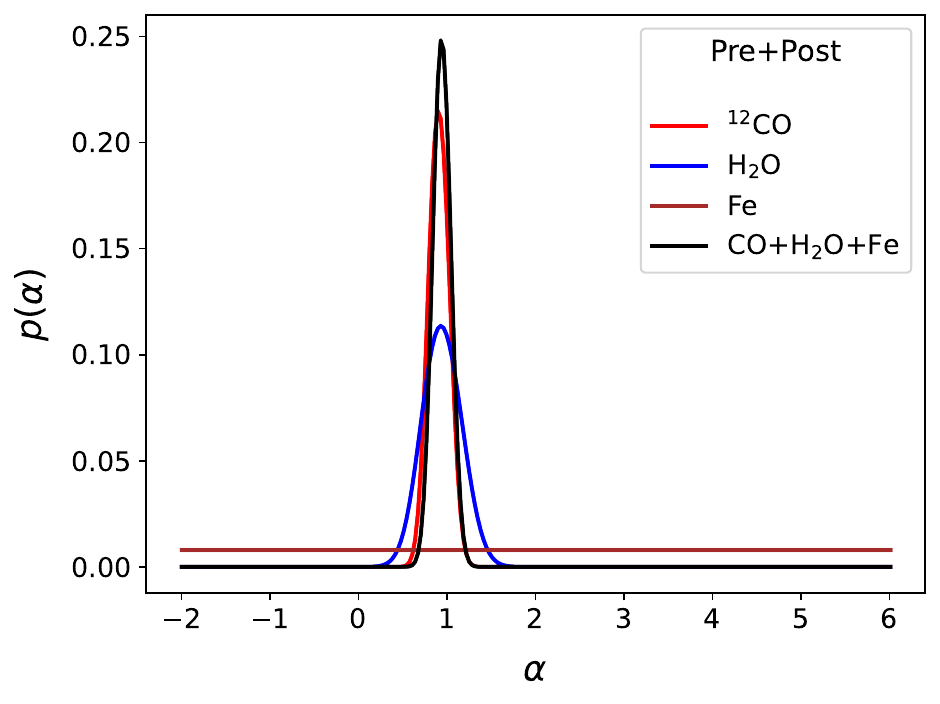}
\end{minipage}%
\caption{Detection of individual species in the atmosphere of MASCARA-1b. \emph{Top, second and middle panels:} $K_{\rm p}$-$\Delta v_{\rm sys}$ maps for the pre-eclipse, post-eclipse and pre+post datasets. \emph{Penultimate panel:} detection significance computed using two ways: $\alpha$-method is less sensitive to deep side lobes and is more objective (see text and Sect.~\ref{sect:4}). \emph{Bottom panel:} conditional likelihood distribution of $\alpha$.}
\label{fig4}
\end{figure*}
\begin{figure*}
\begin{minipage}[htbp]{0.333\textwidth}
  \includegraphics[width=\linewidth]{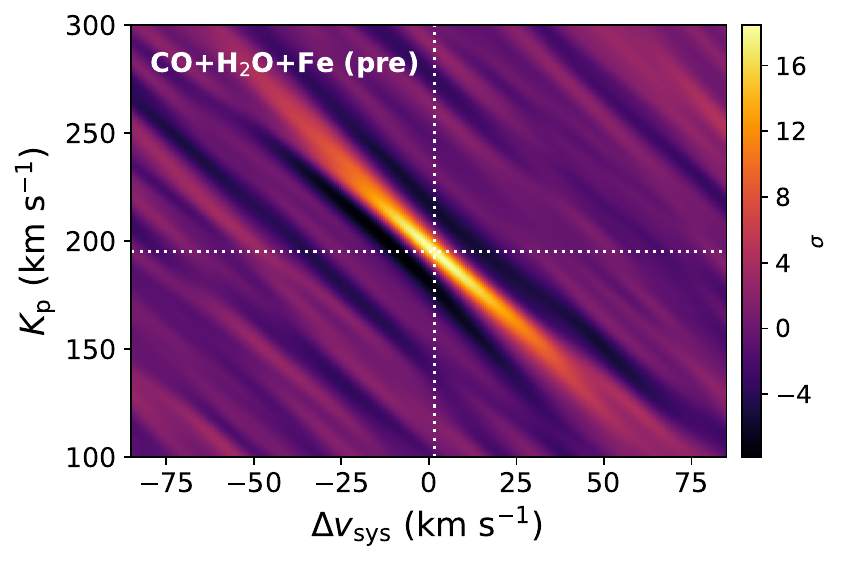}
\end{minipage}%
\hfill 
\begin{minipage}[htbp]{0.333\textwidth}
  \includegraphics[width=\linewidth]{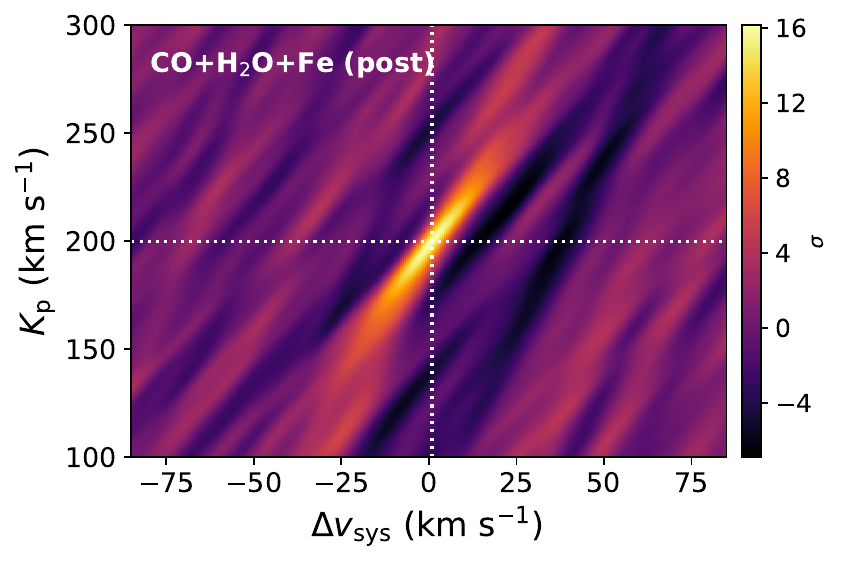}
\end{minipage}%
\hfill
\begin{minipage}[htbp]{0.333\textwidth}
  \includegraphics[width=\linewidth]{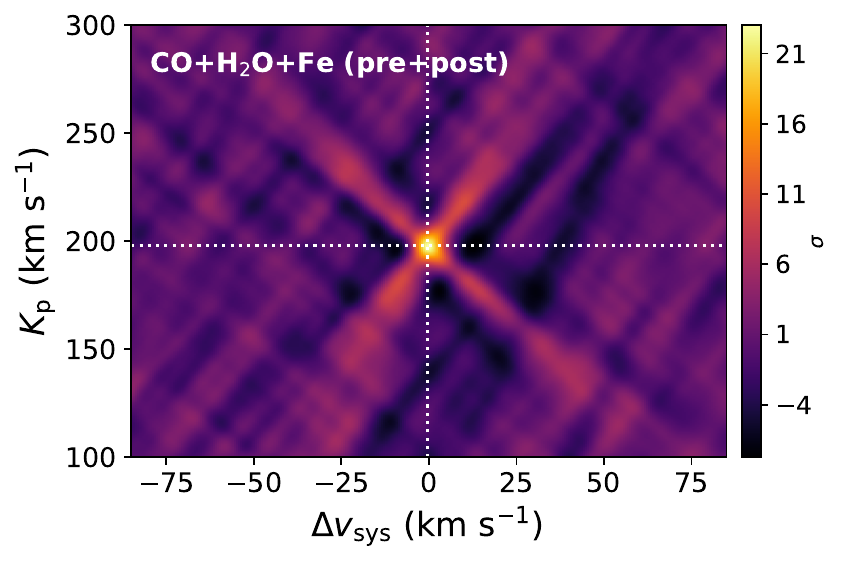}
\end{minipage}%
\caption{$K_{\rm p}$-$\Delta v_{\rm sys}$ maps of the combined CO, H$_2$O and Fe detection in the atmosphere of MASCARA-1b for the pre-, post- and pre+post eclipse datasets. The atmospheric model used for the cross-correlation has combined contributions from the species and a thermal inversion as discussed in Sect.~\ref{sect:3.1}. The white dotted lines mark the detection peak, and the colour bar shows the detection significance computed from maps. The corresponding ``alpha'' detection significance and the conditional likelihood distribution of $\alpha$ are plotted in Fig.~\ref{fig4}.}
\label{fig5}
\end{figure*}

\subsection{Cross-correlation results}\label{sect:3.4}
Following the procedure outlined in the previous section, we searched for individual species of interest by cross-correlating our data with a filtered model emission spectrum using the best-fitting model parameters from our atmospheric retrieval (see Sect.~\ref{sect:4}). We calculated the $K_{\rm p}$-$\Delta v_{\rm sys}$ maps by shifting the CCFs to the rest frame of the planet over a range of $K_{\rm p}$ and $\Delta v_{\rm sys}$ from $-$300 to $+$300 km s$^{-1}$, in steps of 0.3 and 0.2 km s$^{-1}$, respectively, and summed over time. To estimate the detection significance \citep[sometimes referred to as signal-to-noise, e.g.][]{2018A&A...615A..16B}, we divided the $K_{\rm p}$-$\Delta v_{\rm sys}$ maps by their standard deviation calculated by avoiding the area $\pm$50 km s$^{-1}$ from the expected planet signal.

\begin{figure}[h!]
   \centering
   \includegraphics[width=\hsize]{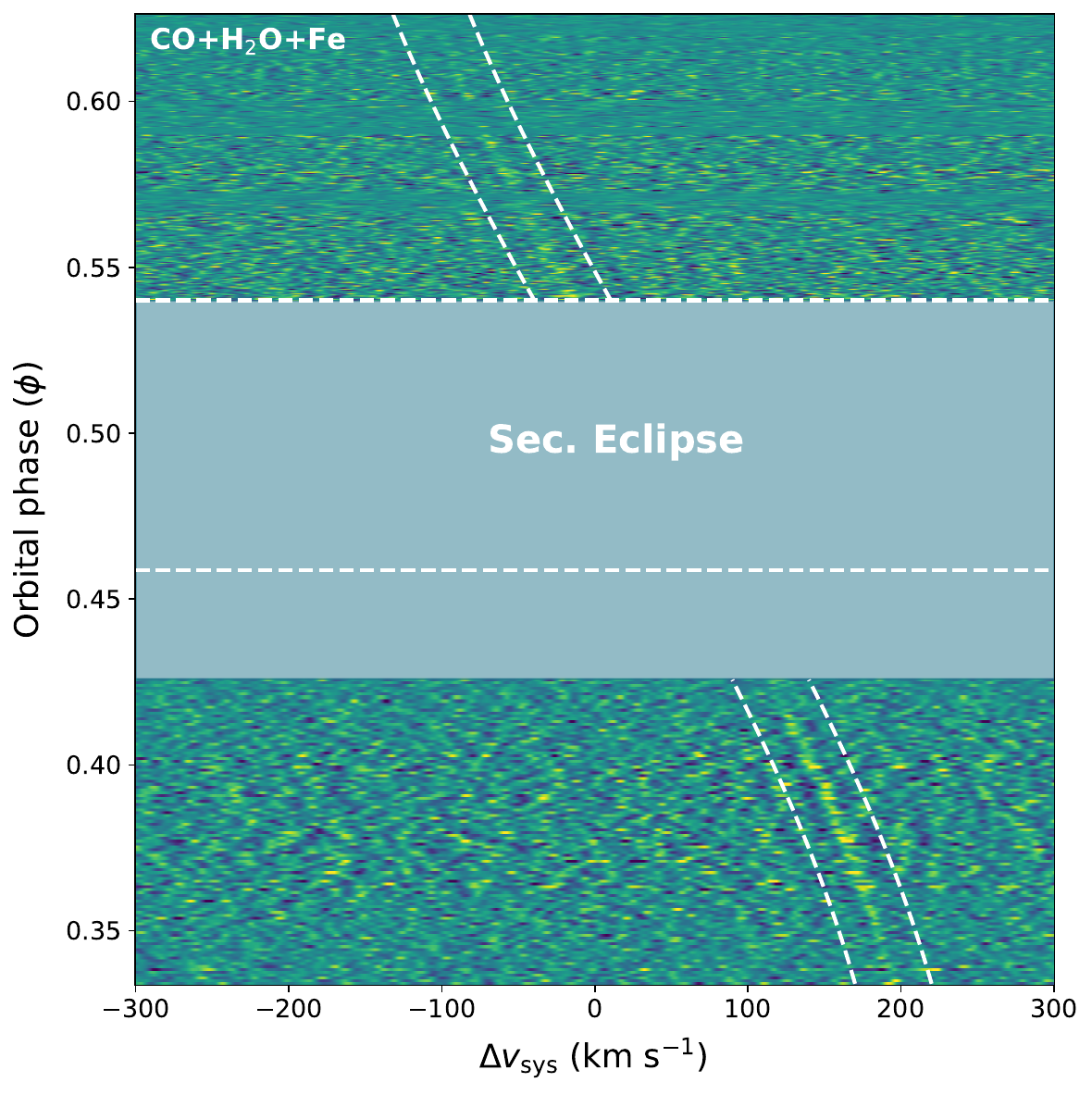}
      \caption{Cross-correlation map for CO+H$_2$O+Fe model, tracing out the planet signal throughout the observed phases. The dashed lines on either side of the CCF trail show the expected planet radial velocity (offset by $\pm$25 $\rm km$ $\rm s^{-1}$), and the horizontal dashed lines indicate the secondary eclipse. The grey space between the end of our pre-eclipse observations and the start of the eclipse denotes phases at which no data were taken.}
         \label{fig6}
   \end{figure}

In our search for chemical species in the day-side atmosphere of MASCARA-1b, we detected strong emission signatures of CO, H$_2$O and Fe in the pre-eclipse data and CO and H$_2$O in the post-eclipse data (second panel of Fig.~\ref{fig4}), respectively. These emission features are also detected at high significance when combining the two nights (middle panel of Fig.~\ref{fig4}). Despite the lower SNR of the post-eclipse data, CO and H$_2$O are detected at similar strengths as in the pre-eclipse data. We do not detect iron in the post-eclipse and the combined dataset. The species are detected roughly at the expected $K_{\rm p}$ and $\Delta v_{\rm sys}$, except for the pre-eclipse Fe signal, which has a slight offset in both $K_{\rm p}$ and $\Delta v_{\rm sys}$. We discuss our iron detection in more detail in Sect.~\ref{sect:5.2}. The $K_{\rm p}$-$\Delta v_{\rm sys}$ maps of the individual species and the full atmosphere for the pre-, post- and pre+post datasets are shown in Figures~\ref{fig4} and~\ref{fig5}, respectively. The colour bar shows the detection significance\footnote{Sometimes referred to as signal-to-noise in the literature \citep[e.g.][]{2018A&A...615A..16B}} values computed from the maps and the strong negative values are deep side lobes near the signal. We note that the $\alpha$ method is less sensitive to these choices, offering a more objective and robust assessment of the detection significance.
For both datasets, the emission signal for the entire atmosphere is strong and directly observable on the CCF map (see Fig.~\ref{fig6}). In addition, we also applied the cross-correlation method to search for $^{13}$CO, Mg, CO$_2$, OH, HCN, Si and Ca but no significant detections are found in the $K_{\rm p}$-$\Delta v_{\rm sys}$ maps of the species (see Appendix~\ref{app:c}).

We emphasise that the detection significance values computed from the maps can change when cross-correlating with the same models due to the arbitrary selection of the noise region. Therefore, we followed the method outlined in \citet{2020MNRAS.493.2215G} to evaluate the detection significance. We first computed the log-likelihood distribution (Sect.~\ref{sect:4.1}) based on the model scale factor, $\alpha$, conditioned on the optimal values of all atmospheric parameters. Next, we calculated the mean of this distribution and divided it by the standard deviation to determine the ``alpha'' detection significance. This value indicates how many standard deviations the maximum signal deviates from an alpha of 0, corresponding to a null signal. Because the alpha detection significance is based on likelihood, the results are more objective, less arbitrary (i.e. require fewer assumptions) and provide a more conservative representation of the data. The penultimate panel of Fig.~\ref{fig4} shows a plot of the detection significance computed from the $K_{\rm p}$-$\Delta v_{\rm sys}$ maps and the ``alpha'' detection significance of the detected species for the datasets.

\section{Atmospheric retrievals}\label{sect:4}
While cross-correlation allows us to obtain detections of chemical species in the atmospheres of exoplanets, it does not provide quantitative information about the atmospheric properties. Additionally, this technique does not allow direct comparisons between observations and atmospheric models. However, recent advances in high-resolution Bayesian methods \citep[e.g.][]{2019AJ....157..114B, 2020MNRAS.493.2215G} have allowed us to ``map'' cross-correlation values of a given atmospheric model to a likelihood value. This will enable us to perform atmospheric `retrievals', which help constrain properties such as the temperature profile, chemical/elemental abundances, and subsequent products like the atmospheric $\rm C/O$ ratio and metallicity. We apply the same retrieval framework described in \citet{2023MNRAS.525.2985R}, which follows the approach of \citet{2020MNRAS.493.2215G,2022MNRAS.512.4618G}, to estimate the chemical composition and thermal structure of MASCARA-1b. We briefly outline our retrieval setup and present our results in the following sections.

\subsection{Retrieval framework}\label{sect:4.1}
To perform our retrieval analysis on the datasets, we computed our forward model, $m_i$, by taking into account the atmospheric model parameters, $\bm{\theta} =$ \{$\kappa_{\rm IR}$, $\gamma$, $T_{\rm irr}$, $T_{\rm int}$, $P_{\rm cloud}$, $\chi_{\rm species}$$\times$$N_{\rm species}$\}, the broadening kernel width \{$W_{\rm conv}$\}, planetary velocities \{$K_{\rm p}$, $\Delta v_{\rm sys}$\}, and the scaling factors for the model (\{$\alpha$\}) and noise (\{$\beta$\}). The forward model was then filtered using the procedure outlined in \citet{2022MNRAS.512.4618G}. Finally, the log-likelihood for a given set of model parameters is computed as follows:
\begin{equation}
    \ln\mathcal{L} = -N\ln \beta - \frac{1}{2}\chi^2 \label{eqn4}
\end{equation}
with
\begin{equation}
    \chi^2 = \sum_{i=1}^{N}\frac{(f_i-\alpha m_i)^2}{(\beta \sigma_i)^2} \label{eqn5}
\end{equation}
where, $f_{\rm i}$ and $\sigma_{\rm i}$ are the data and the standard deviation, respectively. Assuming uniform priors (defined over linear or logarithmic parameter spacing, as listed in Tables~\ref{table:3} and~\ref{table:4}), we compute the log-posterior by computing the log-prior and then adding the log-likelihood (i.e. Eq.~\ref{eqn4}), which was folded into a Markov chain Monte Carlo (MCMC) framework to retrieve posterior distributions of the model parameters for each dataset. We use a Differential-Evolution Markov Chain (DEMC) \citep[][]{2006S&C....16..239T, 2013PASP..125...83E}, running an MCMC chain with 192 walkers, with a burn-in length of 200 and a chain length of 300. This setup results in 96,000 samples, of which 38,400 are discarded. To assess convergence, we apply the Gelman-Rubin statistic \citep[][]{1992StaSc...7..457G} after splitting the chains into four separate groups.

For our retrieval analysis, we focus on the chemical species detected in the day-side atmosphere of MASCARA-1b. Therefore, the number of atmospheric species under consideration, $N_\mathrm{species}$, is three (i.e. $\rm CO$, $\rm H_2O$, and $\rm Fe$). However, we note that this parameter may vary in different MCMC runs as we also include other $\rm C$- and $\rm O$-bearing species for the sake of completeness, in which case, the value of $N_{\rm species}$ is mentioned within parentheses (e.g. Sect.~\ref{sect:5.3}). We fixed $T_{\rm int}$ to be 100 K and $ \alpha $ to be 1 in the model fits, and although $\alpha$ is fixed in the retrievals, it can be easily computed within the likelihood mapping, post-MCMC. That is, defining alpha as a free parameter in the likelihood rather than incorporating it into the model means we can quickly compute the conditional likelihood as a function of alpha from a fixed cross-correlation map. The last panel of Fig.~\ref{fig4} shows the conditional likelihood distribution of $\alpha$ of the detected species for the pre, post and pre+post datasets.

\subsection{Retrieval results}\label{sect:4.2}
\begin{figure*}
\begin{minipage}[htbp]{0.333\textwidth}
  \includegraphics[width=\linewidth]{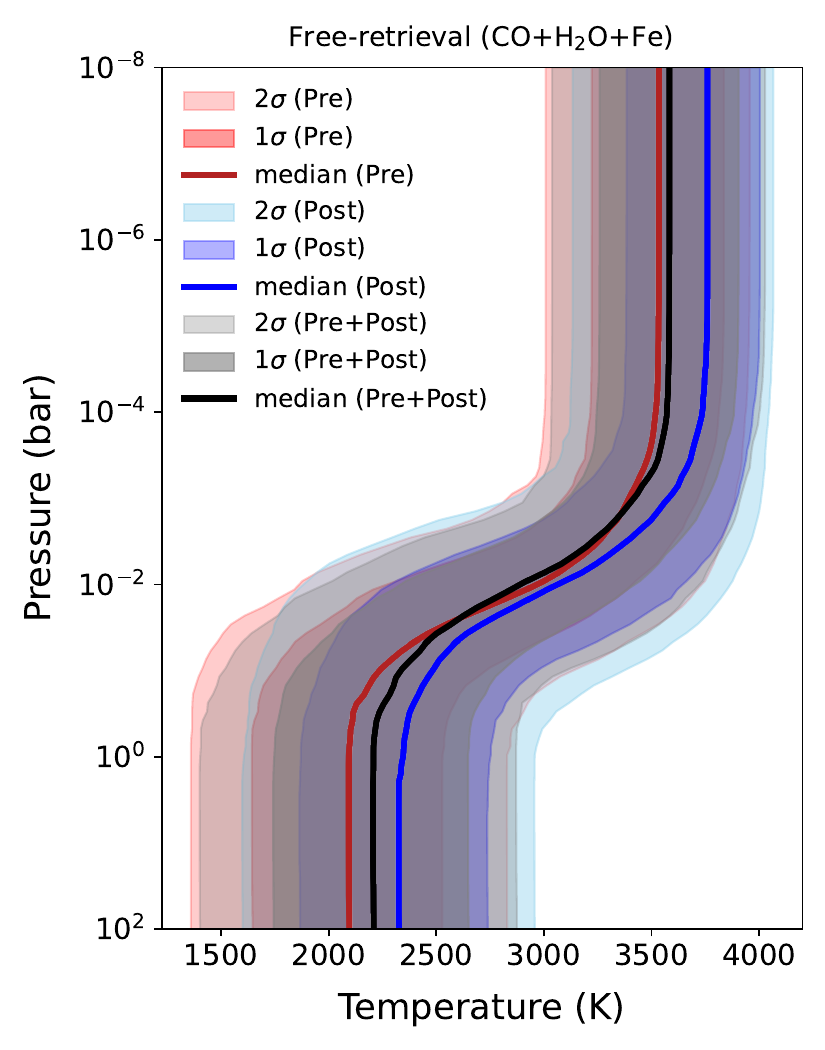}
\end{minipage}%
\hfill 
\begin{minipage}[htbp]{0.333\textwidth}
  \includegraphics[width=\linewidth]{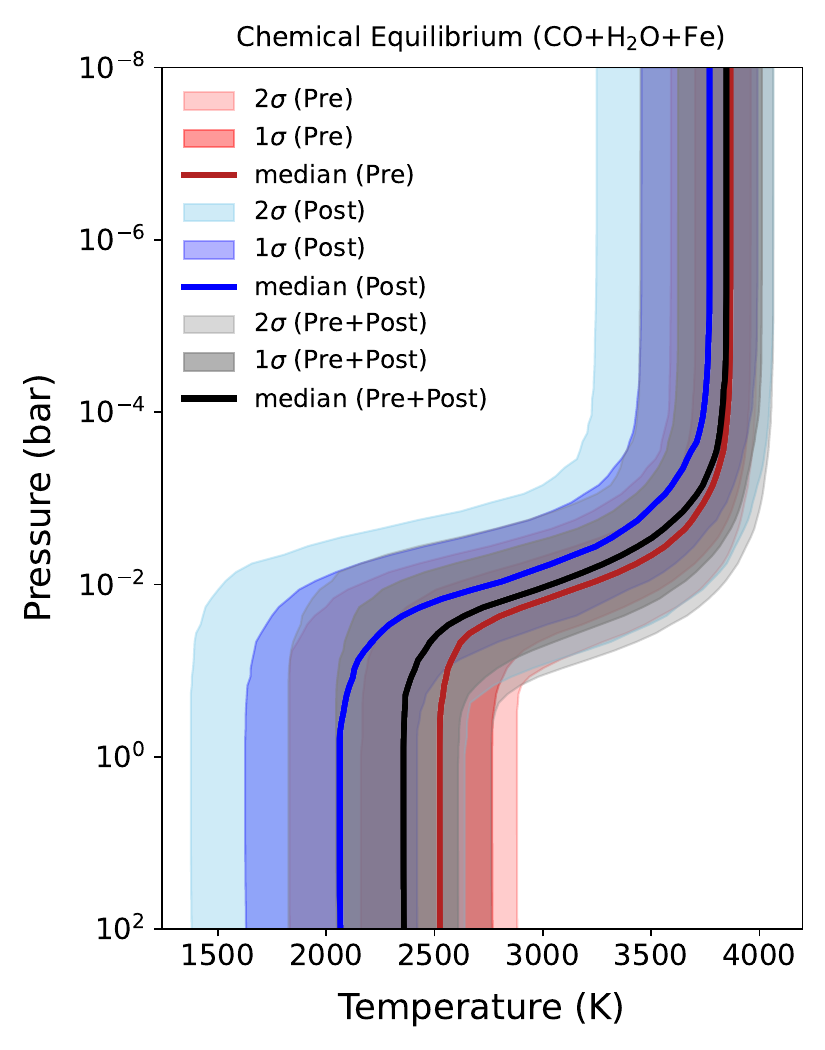}
\end{minipage}%
\hfill
\begin{minipage}[htbp]{0.333\textwidth}
  \includegraphics[width=\linewidth]{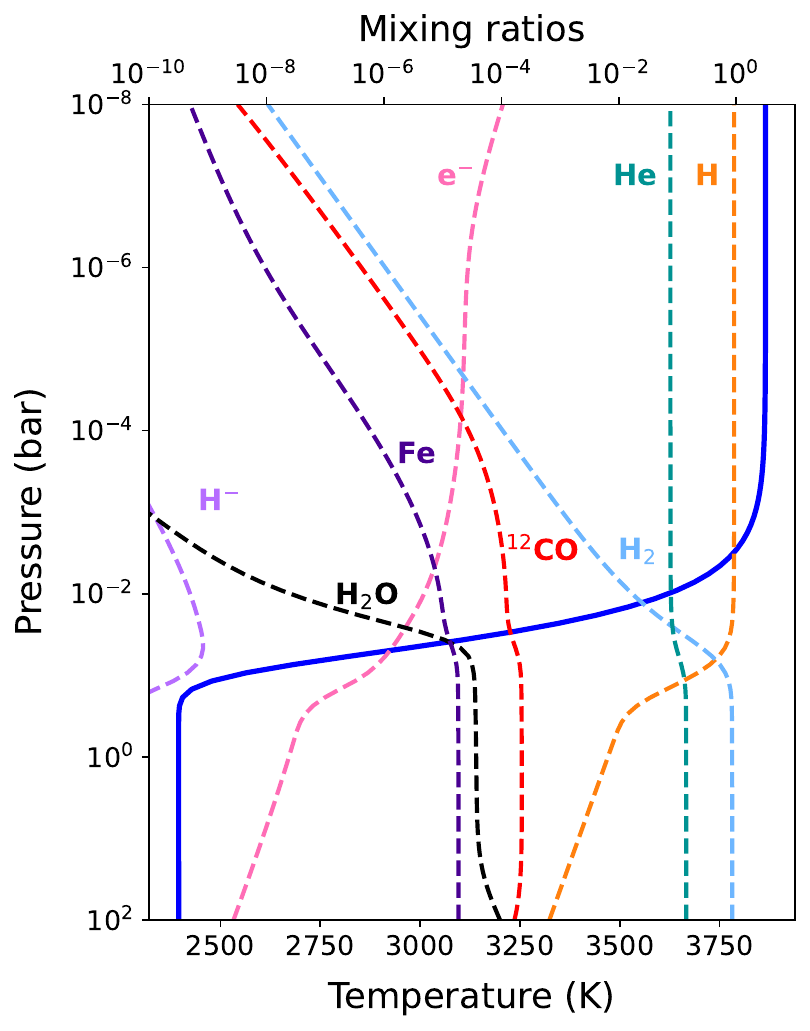}
\end{minipage}%
\caption{\emph{Left and middle panel:} retrieved $T$-$P$ profiles \citep{2010A&A...520A..27G} for the pre-eclipse (red), post-eclipse (blue) and pre+post (black) datasets under the ``free-retrieval'' and chemical equilibrium setups. The red, blue and grey shading marks the 1$\sigma$ and 2$\sigma$ recovered distribution computed from 10,000 samples from the MCMC for both regimes. \emph{Right panel:} the atmospheric structure for the post-eclipse data from the best-fitting chemical model. The volume mixing ratio profiles for the continuum and detected species are shown as dashed lines (calculated using \textsc{FastChem}), with the parametric $T$-$P$ profile shown as a solid blue line.}
\label{fig7}
\end{figure*}

We employ two approximations to describe the atmospheric chemical composition of MASCARA-1b in our retrieval setup. The first approach involves a free retrieval of the mixing ratios for the species, while the second approach uses \textsc{FastChem} \citep[][]{2018MNRAS.479..865S,2022MNRAS.517.4070S,2024MNRAS.527.7263K} to calculate the abundances self-consistently. The pre-eclipse data are analysed in \citet{2023MNRAS.525.2985R} and we refer the reader there for a detailed discussion. The retrieved $T$-$P$ profiles for the two approximations are over-plotted in Fig.~\ref{fig7} and the corresponding posterior distributions are plotted in Figures~\ref{figD1} and~\ref{figD2}, respectively.

\subsubsection{Free retrieval} \label{sect:4.2.1}
We begin the retrieval process with a ``free-retrieval'' paradigm, which assumes constant mixing ratios with altitude and uses the parametric model from \citet{2010A&A...520A..27G}. The retrieval parameters and their prior ranges for the datasets are listed in Table~\ref{table:3}.
For the pre-eclipse, post-eclipse, and joint analysis (pre+post), we place bounded constraints on the abundance of H$_2$O while obtaining lower limits on CO and Fe as their retrieved volume mixing ratios approach the upper bounds of the prior distribution\footnote{Our corner plots display the marginalised posterior medians and $1\sigma$ confidence interval, but we report $1\sigma$ lower limits for posterior distributions against the upper prior.}. We did not detect Fe in the post-eclipse and combined datasets but obtained its 1$\sigma$ lower limit as $\log_{10}(\chi_{\rm Fe}) > -4.97$ for the latter and did not constrain it in the post-eclipse data. Despite the difference in signal-to-noise between the two datasets, the retrieved parameters for the post-eclipse (shown in blue in Fig.~\ref{figD1}) are consistent with the pre-eclipse values, though with slightly lower precision. When we combine the two nights of CRIRES+ data (shown in black in Fig.~\ref{figD1}), the abundance constraints are consistent with the pre- and post-eclipse sequences. However, they are not more precise than those provided by the pre-eclipse data alone, suggesting that the post-eclipse data contributed little new compositional information. We conclude that the precision is dominated by some degeneracies in the model that are not constrained at those wavelengths. Additionally, it is possible that the planet's radial velocity trail is not perfectly sinusoidal for all species, which could introduce limitations when combining datasets if a suboptimal model is used. Nevertheless, combining the datasets enabled precise constraints on the planetary velocities ($K_{\rm p}$, $\Delta v_{\rm sys}$). Overall, the abundances derived from our free-retrieval setup are consistent across the two datasets (pre-eclipse and post-eclipse) as well as the joint analysis (pre+post), with CO dominating relative to H$_2$O on the day-side of MASCARA-1b. However, this does not necessarily imply a carbon-rich atmosphere, as other oxygen-bearing species may influence the interpretation of the $\rm C/O$ ratio (see Sect.~\ref{sect:5.1}).

Lastly, as noted in Sect.~\ref{sect:2}, we use 15 \textsc{SysRem} iterations for our cross-correlation and retrieval analysis presented in this work. While this number is arbitrary, we re-run the analysis with different values of \textsc{SysRem} for the post-eclipse and combined dataset. The results of this analysis are shown in Figs.~\ref{figE1} and~\ref{figE2}. We find that using $N=5,10$ and $20$ iterations give us consistent results for the post-eclipse dataset, whereas $N=20$ filtered out some of our exoplanet signal for the pre-eclipse \citep[][{Fig.}10]{2023MNRAS.525.2985R} and combined datasets.

\subsubsection{Chemical equilibrium}\label{sect:4.2.2}
For our retrievals assuming chemical equilibrium, we find that the H$_2$O, H$_2$ and H abundances change quite drastically with altitude in this temperature regime (e.g. right panel of Fig.~\ref{fig7}). Therefore, instead of retrieving the individual gas volume mixing ratios, we used \textsc{FastChem} \citep[version 3.1.1,][]{2018MNRAS.479..865S,2022MNRAS.517.4070S,2024MNRAS.527.7263K} to calculate the abundances of chemical species. This chemical regime is preferred over a well-mixed model, as the latter is known to break down for ultra-hot Jupiters, where strong altitude-dependent variations in composition occur. The results of this analysis (i.e. assuming chemical equilibrium) are shown in Fig.~\ref{figD2}, and the marginalised distributions for each parameter and dataset are summarised in Table~\ref{table:4}. We placed bounded constraints on the planetary velocities and the $T$-$P$ profile parameters for the two datasets and the joint analysis. The retrieved temperature profiles under this setup are over-plotted in Fig.~\ref{fig7} (middle panel). Similar to the free-retrieval setup, constraints on the retrieved parameters for the post-eclipse data remain consistent with the pre-eclipse values, although less precise. In addition, combining the pre- and post-eclipse data provides consistent constraints, albeit weighted more in favour of the pre-eclipse data, which appears to be driving these inferences. The retrieved values for the $T$-$P$ profile parameters between the two retrieval frameworks remain consistent for the individual datasets and the joint analysis (pre+post) within $\approx$1.1$\sigma$.

Lastly, in our retrieval framework, H$^{-}$ bound-free and free-free opacities are included in the chemical equilibrium retrievals but not the free-chemistry retrievals. While the former is less relevant in the K-band, omitting free-free opacity in the free-chemistry retrieval does not significantly affect our results. To verify this, we re-computed our best-fit model emission spectra (assuming chemical equilibrium) with and without the H$^{-}$ free-free opacity (and CIA) and found that the relative flux difference is ${\sim}10^{-5}$. This indicates that although H$^{-}$ free-free opacity is present, its impact on the overall emission spectra in the observed K-band wavelength range (1921–2472 nm) is minimal, suggesting that the omission of H$^{-}$ opacities in the free retrievals does not significantly affect the model outcomes for this wavelength range.

\FloatBarrier
\begin{table*}[htbp]
\caption{Parameters recovered for the combined fits (CO+H$_2$O+Fe) of MASCARA-1b under the free-retrieval (FR) setup.}
\label{table:3}                     
\centering                                      
\renewcommand{\arraystretch}{1.49} 
\begin{tabular}{lcccc}
\hline \hline
Parameter [unit] & Prior & \multicolumn{3}{c}{Retrieved value (FR)} \\
\cline{3-5}
                 &       & Pre-eclipse & Post-eclipse & Pre+Post\\
\hline
$\alpha$ & - & fixed & fixed & fixed\\
$\beta$ & $\mathcal{U}$(0.1, 2) & 0.76$\pm$0.0003 & 0.82$\pm$0.0003 & -\\
$K_{\rm p}$ [$\rm km$ $\rm s^{-1}$] & $\mathcal{U}$(185, 215) & 194.7$^{+2.8}_{-2.7}$ & 197.4$\pm$2.4 & 197.8$^{+0.3}_{-0.4}$\\
$\Delta v_{\rm sys}$ [$\rm km$ $\rm s^{-1}$] & $\mathcal{U}$($-15$, 15) & 1.77$\pm$1.90 & $-0.59^{+1.3}_{-1.2}$ & $-0.34\pm$0.20\\
$W_{\rm conv}$ & $\mathcal{U}$(1, 50) & 1.65$^{+0.33}_{-0.31}$ & 1.48$\pm$0.30 & 1.48$\pm$0.20\\
$\log_{10}(\kappa_{\rm IR})$ [$\rm m^2$ $\rm kg^{-1}$] & $\mathcal{U}$($-5$, 0) & $-2.80^{+0.67}_{-0.65}$ & $-2.61^{+0.71}_{-0.82}$ & $-2.58^{+0.66}_{-0.71}$\\
$\log_{10}(\gamma)$ & $\mathcal{U}$($-2$, 2) & 0.98$^{+0.40}_{-0.28}$ & 0.94$^{+0.36}_{-0.25}$ & 0.86$^{+0.34}_{-0.20}$\\
$T_{\rm irr}$ [K] & $\mathcal{U}$(1000, 4100) & 2367$^{+502}_{-549}$ & 2536$^{+491}_{-574}$ & 2583$^{+474}_{-565}$\\
$T_{\rm int}$ [K] & - & fixed & fixed & fixed\\
$\log_{10}(P_{\rm cloud})$ [bar] & $\mathcal{U}$($-4$, 2) & 0.26$^{+1.20}_{-1.28}$ & 0.21$^{+1.23}_{-1.30}$ & 0.04$^{+1.29}_{-1.21}$\\
$\log_{10}(\chi_{\rm CO})$ & $\mathcal{U}$($-20$, $-2$)$^{(a)}$ & $>-3.54$$^{(b)}$ & $>-3.79$ & $>-3.55$\\
$\log_{10}(\chi_{\rm H_2O})$ & $\mathcal{U}$($-20$, $-2$) & $-4.66^{+0.58}_{-0.60}$ & $-4.51^{+0.66}_{-0.79}$ & $-4.50^{+0.55}_{-0.66}$\\
$\log_{10}(\chi_{\rm Fe})$ & $\mathcal{U}$($-20$, $-2$) & $>-4.65$ & Unconstrained & $>-4.97$\\
$\rm [M/H]$ & - & $>-0.19$ (derived) & $>-0.64$ (derived) & $>-0.31$ (derived)\\
$\log_{10}(\rm C/O)$$^{(c)}$ & - & $-0.011^{+0.00}_{-0.01}$ (derived) & $-0.012^{+0.01}_{-0.01}$ (derived) & $-0.011^{+0.00}_{-0.01}$ (derived)\\
\hline
\end{tabular}
\tablefoot{$^{(a)}$ The choice of an upper prior limit on the VMRs of the chemical species is based on the assumption of a Jupiter-like atmosphere, where H$_2$ and He dominate. $^{(b)}$ Values with a `$>$' are 1-$\sigma$ lower limits and those with `-' were not included as free parameters for that particular retrieval. $^{(c)}$ The atmospheric metallicity and $\rm C/O$ are derived from the abundances and were not retrieved directly in this setup.}
\end{table*}

\begin{table*}[htbp]
\caption{Parameters recovered for the combined fits (CO+H$_2$O+Fe) of MASCARA-1b under the chemical equilibrium (CE) setup.}
\label{table:4}                     
\centering                                      
\renewcommand{\arraystretch}{1.5} 
\begin{tabular}{lcccc}
\hline \hline
Parameter [unit] & Prior & \multicolumn{3}{c}{Retrieved value (CE)} \\
\cline{3-5}
                 &       & Pre-eclipse & Post-eclipse & Pre+Post\\
\hline
$\alpha$ & - & fixed & fixed & fixed\\
$\beta$ & $\mathcal{U}$(0.1, 2) & 0.76$\pm$0.0003 & 0.82$\pm$0.0003 & -\\
$K_{\rm p}$ [$\rm km$ $\rm s^{-1}$] & $\mathcal{U}$(185, 215) & 194.6$\pm$2.90 & 197.04$\pm$2.4 & 197.8$\pm$0.4\\
$\Delta v_{\rm sys}$ [$\rm km$ $\rm s^{-1}$] & $\mathcal{U}$($-15$, 15) & 1.86$\pm$2.00 & $-0.80\pm$1.2 & $-0.34\pm$0.20\\
$W_{\rm conv}$ & $\mathcal{U}$(1, 50) & 1.54$^{+0.35}_{-0.29}$ & 1.53$\pm$0.30 & 1.52$^{+0.22}_{-0.24}$\\
$\log_{10}(\kappa_{\rm IR})$ [$\rm m^2$ $\rm kg^{-1}$] & $\mathcal{U}$($-5$, 0) & $-2.40^{+0.33}_{-0.47}$ & $-2.65^{+0.55}_{-0.66}$ & $-2.52^{+0.45}_{-0.58}$\\
$\log_{10}(\gamma)$ & $\mathcal{U}$($-2$, 2) & 0.81$^{+0.30}_{-0.17}$ & 1.10$^{+0.38}_{-0.27}$ & 0.91$^{+0.23}_{-0.17}$\\
$T_{\rm irr}$ [K] & $\mathcal{U}$(1000, 4100) & 2843$^{+273}_{-456}$ & 2392$^{+406}_{-491}$ & 2693$^{+281}_{-374}$\\
$T_{\rm int}$ [K] & - & fixed & fixed & fixed\\
$\log_{10}(P_{\rm cloud})$ [bar] & $\mathcal{U}$($-4$, 2) & 0.18$^{+1.21}_{-1.29}$ & 0.25$^{+1.20}_{-1.33}$ & 0.21$^{+1.20}_{-1.27}$\\
$\rm [M/H]$ & $\mathcal{U}$($-1$, 1) & $>0.07$$^{(a)}$ & $>-0.50$ & $>-0.33$\\
$\log_{10}(\rm C/O)$ & $\mathcal{U}$($-1$, 1) & $-0.17^{+0.08}_{-0.17}$ & $-0.13^{+0.06}_{-0.11}$ & $-0.13^{+0.06}_{-0.09}$\\
\hline
\end{tabular}
\tablefoot{$^{(a)}$ Values with a `$>$' are 1-$\sigma$ lower limits and those with `-' were not included as free parameters for that particular retrieval.}
\end{table*}

\section{Discussion}\label{sect:5}
\subsection{Chemistry of MASCARA-1b}\label{sect:5.1}
The composition (elemental abundances) of exoplanets, especially of highly-irradiated gas giants (e.g. UHJs), can provide clues about the planetary formation and migration histories within the protoplanetary disk \citep[][]{2011ApJ...743L..16O} via diagnostics like the bulk metallicity and the $\rm C/O$ ratio. Here, we infer the chemistry of MASCARA-1b in terms of the atmospheric C/O and metallicity obtained from our retrieval frameworks. These quantities can be derived from the inferred gas abundances or fitted directly from chemical equilibrium model fits. The former becomes invalid for UHJs due to significant altitude-dependent variations in chemical composition (e.g. Fig.~\ref{fig7}), highlighting the need to incorporate vertical chemical equilibrium when interpreting atmospheric properties.\\

Nonetheless, to assess potential biases introduced by the well-mixed assumption, we compute the ``free-retrieval'' based $\rm C/O$ and metallicity of MASCARA-1b derived from our retrieved gas abundance as follows:
\begin{equation}
    \mathrm{C/O} = \frac{n_{\rm C}}{n_{\rm O}} = \frac{n_{\rm CO}}{n_{\rm CO} + n_{\rm H_2O}} \label{eqn6}
\end{equation}
and
\begin{equation}
    \mathrm{[M/H]} = \log_{10}\left[\frac{2n_{\rm CO} + n_{\rm H_2O} + n_{\rm Fe}}{2n_{\rm H_2}[(n_{\rm O} + n_{\rm C} + n_{\rm Fe})/n_{\rm H}]_\odot}\right] \label{eqn7}
\end{equation}
We assume that CO and H$_2$O are the dominant carbon- and oxygen-bearing molecules in the atmosphere of MASCARA-1b and compute the planetary metallicity by normalising the elemental abundances relative to hydrogen (the numerator in Eqn.~\ref{eqn7}), relative to that in the Sun (the denominator in Eqn.~\ref{eqn7}). The solar elemental abundances are taken from \citet{2009ARA&A..47..481A} and the median values of these derived quantities for the datasets are listed in Table~\ref{table:3}. The derived metallicities for the pre-eclipse and combined datasets are consistent with the solar value within 1$\sigma$ and are broadly consistent with the stellar value within their uncertainties. For the post-eclipse data, despite the strong CO detection, the derived $\rm [M/H]$ appears lower than that of the pre-eclipse. This difference is likely driven by the absence of detectable iron in the post-eclipse data, combined with the slightly weaker H$_2$O detection/constraint. However, the pre- and post-eclipse metallicities remain statistically consistent.

Looking at the derived $\rm C/O$ ratios, we obtain very precise constraints with $\rm C/O = 0.98^{+0.01}_{-0.02}$, $0.97^{+0.01}_{-0.02}$ and $0.98^{+0.01}_{-0.01}$ for the pre-, post- and combined datasets. While there are no reported/measured values of the $\rm C/O$ ratio for the host star, MASCARA-1, the derived values are all markedly super-solar and are only constrained by our detection of CO and H$_2$O, assuming that every bit of C and O is within these molecules. However, this assumption could be incomplete. Qualitatively, if a significant portion of the oxygen is present in other species, such as OH or atomic oxygen, due to thermal dissociation of H$_2$O, it could influence the $\rm C/O$ ratio we are measuring. OH is expected to be present on the day side of MASCARA-1b, as it has been detected in colder planets such as WASP-121b and WASP-76b \citep{2024AJ....168..293S,2021A&A...656A.119L,2024MNRAS.530.2885G,2024AJ....168...14W}. However, since K-band CRIRES+ observations are not sensitive to OH lines and atomic oxygen is not observable in the near-infrared, our current retrievals may not fully account for these species. Therefore, we included potential C- and O-bearing species that exhibit sharp spectral lines in this wavelength setting but were not detected (e.g. OH, $^{13}$CO, HCN and CO$_2$; here $N_{\rm species} = 7$), in our free-chemistry retrieval and find that the inclusion of these species does not change the derived $\rm C/O$, which remains super-solar. Additionally, similar to \citet{2023AJ....165...91B}, if we account for missing oxygen by assuming atomic $\rm O$ with a VMR of $10^{-3}$, we indeed obtain a lower $\rm C/O{\approx}0.6$, although we are unable to provide a precise estimate of uncertainty for this value. Overall, our free-retrieval-based results strongly favoured a H$_2$O-depleted model over one with a similar abundance of CO and H$_2$O, thereby driving the $\rm C/O$ towards 1 resulting in precise constraints.

While the free-chemistry model offers flexibility by allowing the gas volume mixing ratios to be retrieved individually, the assumption of a well-mixed atmosphere does not accurately reflect the vertical variation in chemical composition, particularly for highly irradiated gas giants. This can lead to biases in the elemental abundance estimates \citep[e.g.][]{2023AJ....165...91B, 2023MNRAS.525.2985R}. Therefore, we fit directly for the $\rm C/O$ ratio and metallicity derived from chemical equilibrium fits (see Table~\ref{table:4} and Fig.~\ref{figD2}) using \textsc{FastChem}. This results in a $\rm C/O = \rm 0.68^{+0.12}_{-0.22}, \rm 0.75^{+0.11}_{-0.17}$ and $0.74^{+0.10}_{-0.14}$ for the pre-, post-, and pre+post eclipse datasets, respectively. The retrieved values are all consistent with the solar value within ${\approx}1.1$-$2\sigma$ and the combined dataset provides slightly tighter constraints on the retrieved values than the individual datasets. In addition, the post-eclipse seems to constrain the $\rm C/O$ ratio better than the pre-eclipse, though the difference is negligible. This could be because the uncertainties remain consistent across all datasets, as suggested by our injection-recovery tests (see Appendix~\ref{appendix:A}) where the injected and recovered values for both a constant-with-altitude abundance and equilibrium chemistry modelling setups are in good agreement.
Nevertheless, we speculate that the uncertainties in these parameters are likely dominated by degeneracies in the model and the wavelength range of the observations, as well as the limitations of our 1D models, which do not account for 3D effects such as spatial inhomogeneities in the planet's atmosphere. These factors may contribute to the consistency of the $\rm C/O$ constraints despite the lower data quality of the post-eclipse sequence. Moreover, although the metallicity prior in our chemical equilibrium retrieval is relatively narrow, once we broaden the metallicity constraints (from 0.001 to 1000$\rm x$ solar), the retrieved $\rm C/O$ ratio remains consistent for the post-eclipse data and the joint analysis. However, we find a strong correlation between $\rm C/O$ and metallicity for the pre-eclipse data (see Fig.~\ref{figD3}), leading to inconsistencies in the retrieved value. While this does not alter our overall conclusions, it highlights an important degeneracy we aim to explore further. Lastly, similar to our free-chemistry analysis, for completeness, we also included opacities for OH, CO$_2$ and HCN in our equilibrium chemistry retrievals and find that the inclusion of these species did not change the retrieved $\rm C/O$ value which remains consistent with the solar value within ${\approx}1\sigma$.
\begin{figure}[htbp]
   \centering
   \includegraphics[width=\hsize]{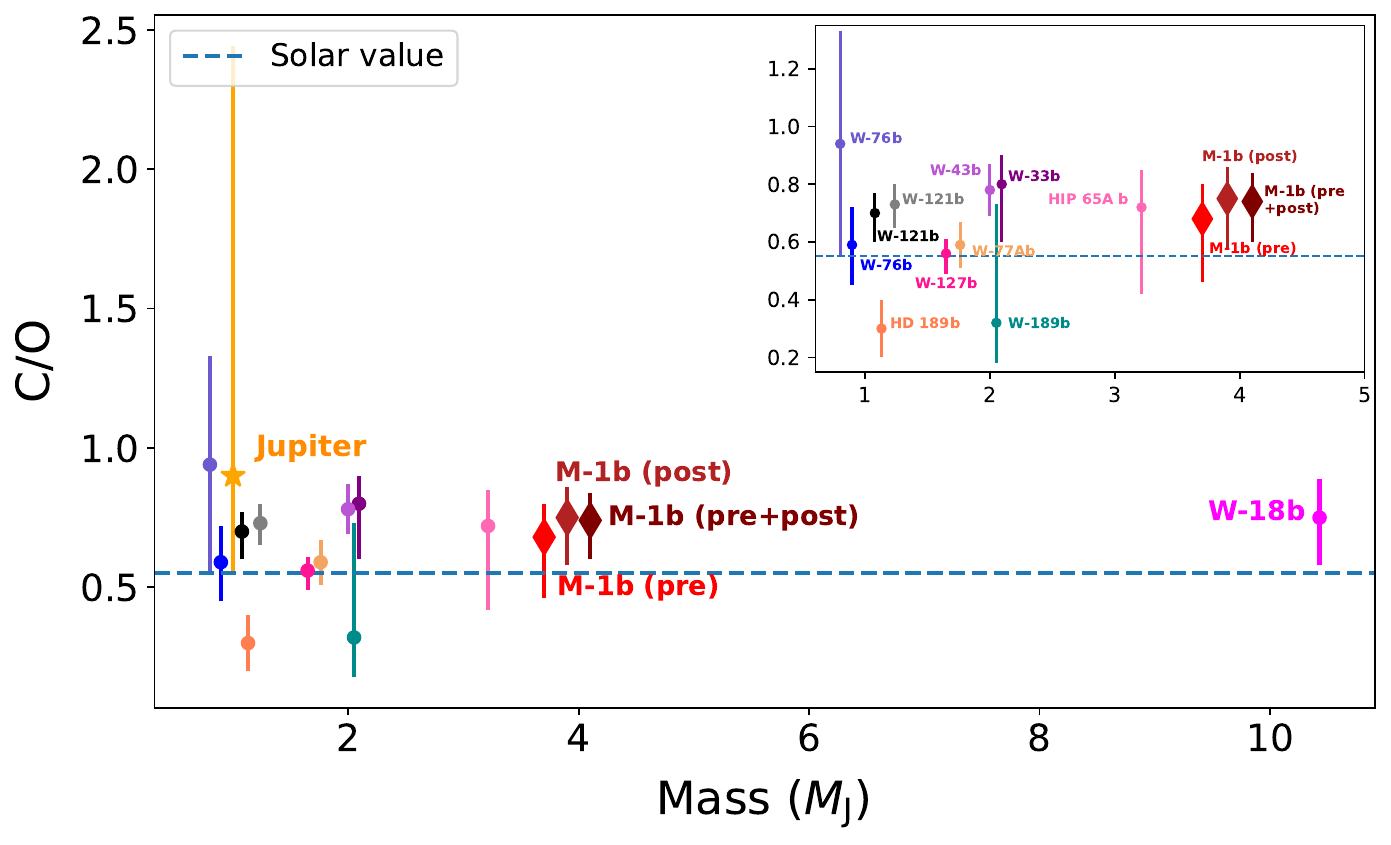}
        \caption{Planetary mass versus $\rm C/O$ ratio for hot/ultra-hot Jupiters with constrained abundances of C- and O-bearing species from ground-based high-resolution observations. Our retrieved $\rm C/O$ values for MASCARA-1b are shown as diamonds (labelled M-1b). Values for other planets are from previous studies using either free-chemistry retrievals or chemical equilibrium models (see Table~\ref{tableF1}). The dashed line indicates the solar $\rm C/O$ value, and Jupiter is included for reference \citep[from][]{2024AJ....168...14W}. To reduce clutter, the labels for planets excluding MASCARA-1b and WASP-18b are shown in the inset, which provides a zoomed-in view. For clarity, small mass offsets are applied to WASP-121b, WASP-76b, WASP-189b and MASCARA-1b.}
         \label{fig8}
   \end{figure}

Overall, our results indicate that the atmosphere of MASCARA-1b has a $\rm C/O$ ratio consistent with the solar value, with the chemical equilibrium retrieval providing more realistic and conservative constraints. While it would be interesting to frame the planet’s chemistry in terms of the chemistry of its parent star, the $\rm C/O$ ratio of MASCARA-1b alone is inconclusive in determining its formation as there are no reported or measured values of the $\rm C/O$ ratio for the host star, MASCARA-1. Figure~\ref{fig8} highlights the diversity of $\rm C/O$ ratios among hot and ultra-hot Jupiters using ground-based high-resolution observations. Most planets in the sample exhibit $\rm C/O$ ratios near or above the solar value, with the observed spread in the reported values reflecting the diversity of formation and migration pathways. MASCARA-1b adds to this population with ratios retrieved from pre-, post-, and combined datasets, all consistent with the solar value, making it a valuable addition to the sample of UHJs with well-constrained $\rm C/O$ ratios. While future observations could further refine these constraints, offering new insights into the planet's chemical inventory and its formation and migration history, it also underscores the challenges in interpreting these ratios in isolation since carbon and oxygen measurements alone are insufficient to determine the origin of a gas-giant planet due to the intrinsic degeneracies in the planet-building gas and solid compositions \citep[][]{2016ApJ...832...41M}.

Recent studies by \citet{2021ApJ...914...12L} and \citet{2023ApJ...943..112C} elucidated the diagnostic power of refractory elements such as Fe, Mg, Si, and Ca, and introduced the concept of the refractory-to-volatile ratio -- the abundance of refractory elements like iron relative to volatile species such as CO and H$_2$O -- as a valuable tool for tracing the formation and migration pathways of hot and ultra-hot Jupiters. This ratio provides a framework for predicting how a planet’s solid-to-gas ratio and accretion location within the protoplanetary disk influence its atmospheric composition and has very recently been applied to high-resolution observations of the ultra-hot Jupiter WASP-121b, enabling the exploration of potential formation and migration pathways for such planets \citep[][]{2025AJ....169...10P,2024AJ....168..293S}. In the case of MASCARA-1b, our pre-eclipse detection of iron allows us to estimate the refractory-to-volatile ratio within this framework. Assuming that the Fe signal is not spurious (see Sect.~\ref{sect:5.2}), we derive the refractory-to-volatile ratio to be $\rm [R/V] = 0.32^{+1.17}_{-1.49}$ from our free-chemistry retrievals.
However, it is important to note that attributing the entire refractory inventory to Fe alone or the volatile inventory solely to CO and H$_2$O likely introduces significant bias, as it excludes contributions from other important rock- and ice-forming elements. The non-detections of other refractory (e.g. Mg, Si) and potential volatile (e.g. OH and atomic O) species, partly due to the scarcity of strong spectral lines in the observed wavelength range, could lead to an incomplete representation of refractory and volatile content. Furthermore, our retrieved ratios are derived assuming a constant-with-altitude modelling regime. For these reasons, we choose not to over-interpret the derived refractory-to-volatile ratio and emphasise the need for future observations to better constrain the inventory of both refractory and volatile species in MASCARA-1b.

Nevertheless, UHJs present a tremendous opportunity to diagnose planet formation histories by revealing a wealth of volatile and refractory species in their atmospheres. Combining optical and near-infrared high-resolution emission spectroscopy datasets should provide interesting constraints on refractory-to-volatile elemental ratios for irradiated gas giants.

\subsection{Revisiting the iron detection}\label{sect:5.2}
\begin{figure}[htbp]
  \resizebox{\hsize}{!}{\includegraphics{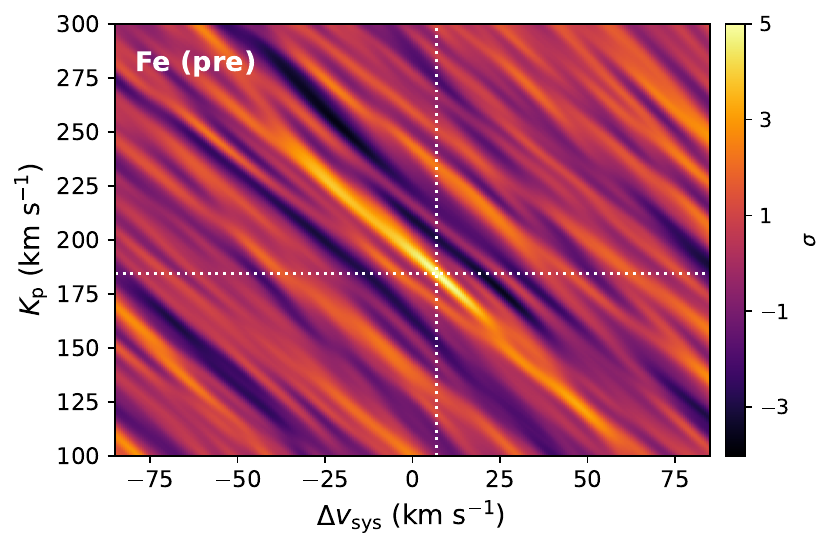}}
  \caption{$K_{\rm p}$-$\Delta v_{\rm sys}$ map of the pre-eclipse Fe signal, with the detection significance computed using a region where $K_{\rm p}{>}$210 $\rm km$ $\rm s^{-1}$ and $\Delta v_{\rm sys}{<}-40$ $\rm km$ $\rm s^{-1}$, avoiding the planetary signal near the expected $K_{\rm p}$ and $\Delta v_{\rm sys}$. While choosing a different region reduced the detection significance to ${\approx}5\sigma$, Fe remains significantly detected, with the alpha detection significance constrained to be non-zero at more than $4.2\sigma$, supporting the robustness of the Fe signal.}
  \label{fig9}
\end{figure}

The detection of iron in the pre-eclipse dataset \citep{2023MNRAS.525.2985R} presents an interesting, yet somewhat ambiguous, result. While a prominent peak is found around the consistent $K_{\rm p}$ and $\Delta v_{\rm sys}$ in the detection map (see top panel of Fig.~\ref{fig4}), albeit a little shifted, we cannot definitively conclude that it is a strong detection of Fe based on our CCF analysis alone. Specifically, the presence of significant negative peaks (e.g. the $-6\sigma$ feature at a $\Delta v_{\rm sys}$$\approx$23 $\rm km$ $\rm s^{-1}$ and $\approx$90 $\rm km$ $\rm s^{-1}$) in the CCF map raises reasons to question the robustness of this detection. We also note that the reported detection significance does not have to be $8\sigma$ because the choice of the region used to calculate this is arbitrary and can result in varied values (e.g. Fig.~\ref{fig9}). Furthermore, when we perform a free retrieval analysis with a model containing Fe alone, we cannot constrain it, likely because the approach does not account for other species (e.g. CO, H$_2$O) that can mask certain spectral lines or alter the effective line strengths, which could potentially weaken the Fe signal. Nonetheless, the presence of a positive CCF peak in our $K_{\rm p}$-$\Delta v_{\rm sys}$ map combined with constraints in our retrieval analysis with all the species seems to indicate that Fe signal in the pre-eclipse data is genuine. In addition, the detection significance derived from the likelihood (alpha value) is constrained to be non-zero at more than $4.2\sigma$ (see penultimate panel of Fig.~\ref{fig4}), which further boosts our confidence in its detection.

Assuming the pre-eclipse detection is real, the absence of iron in the post-eclipse data becomes particularly intriguing. To explore this, we performed injection-recovery tests with the best-fitting Fe model, injected at a negative $K_{\rm p}$. For the pre-eclipse data, the injected and recovered values are in agreement (see Appendix~\ref{appendix:G}), with the detection significance computed from the $K_{\rm p}$-$\Delta v_{\rm sys}$ map and the alpha detection significance being ${\approx}10.92\sigma$ and ${\approx}8.31\sigma$, respectively. However, we note that the `actual' signal--corresponding to the Fe detection originally observed in the CCF analysis--is not visible in the recovered map. This absence may result from the injected signal adding noise, which can weaken the effective signal, particularly pronounced for weaker detections, such as iron, which has an alpha detection significance of only ${\approx}4.4\sigma$.
In contrast, we do not detect iron in the post-eclipse dataset and performing an injection/recovery test yields detection significance and alpha detection significance values of $\approx$7.73$\sigma$ and $\approx$7.53$\sigma$, respectively. These values are not significantly different from those of the pre-eclipse dataset; however, they are slightly lower, indicating that the data quality is somewhat reduced, and the sensitivity to iron is consequently diminished. This suggests that the non-detection in the post-eclipse phase could result from the weaker Fe signal falling below the detection threshold due to the slightly poorer signal-to-noise ratio. The non-detection could, therefore, be attributed to the data quality rather than an inherent absence of iron in this phase sequence.
\begin{figure*}
\begin{minipage}[htbp]{0.333\textwidth}
  \includegraphics[width=\linewidth]{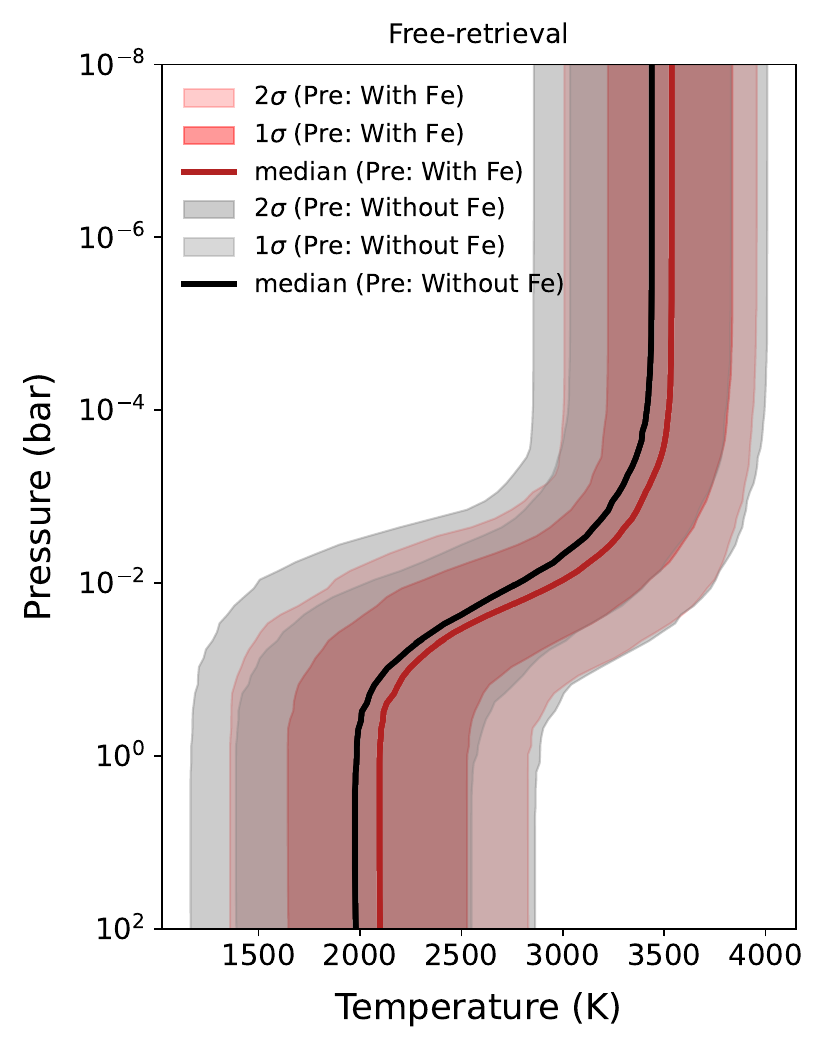}
\end{minipage}%
\hfill 
\begin{minipage}[htbp]{0.333\textwidth}
  \includegraphics[width=\linewidth]{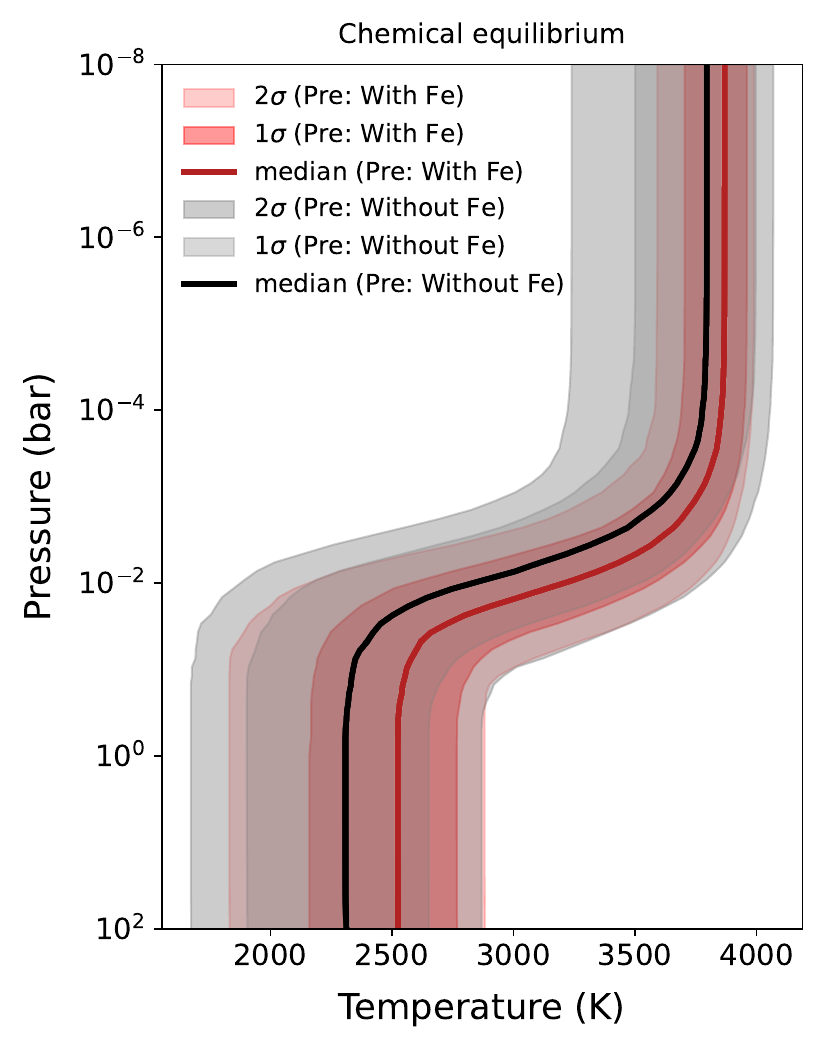}
\end{minipage}%
\hfill 
\begin{minipage}[htbp]{0.333\textwidth}
  \includegraphics[width=\linewidth]{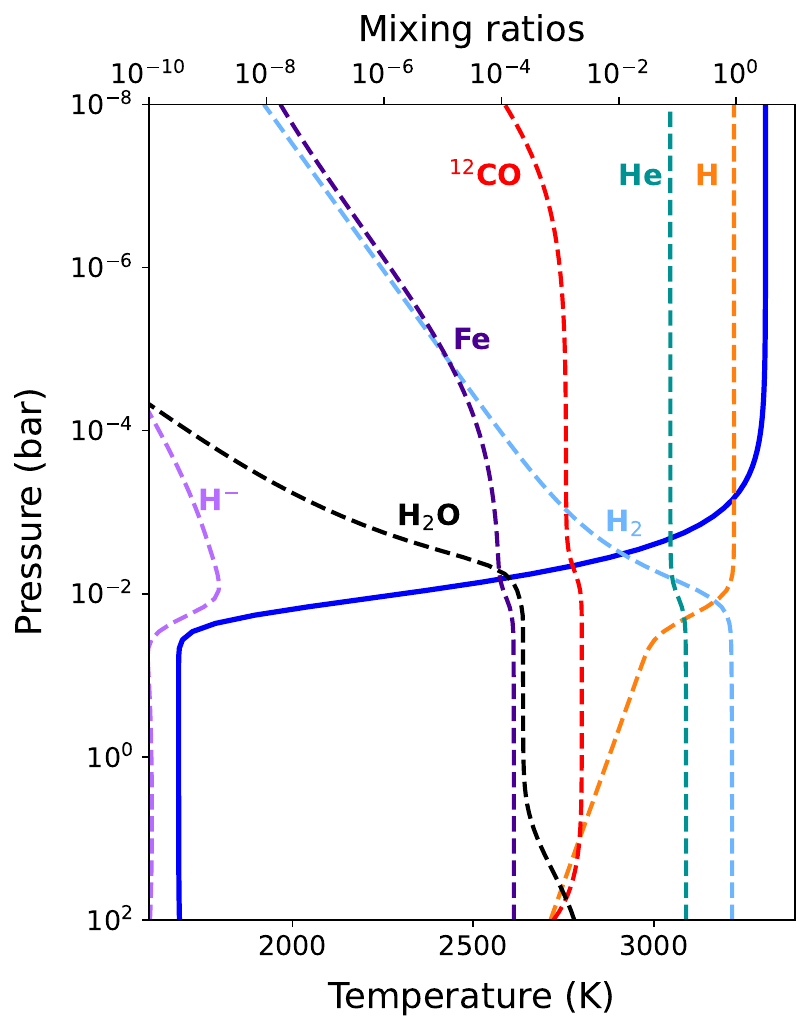}
\end{minipage}%
\caption{\emph{Left and middle panel:} retrieved $T$-$P$ profiles \citep{2010A&A...520A..27G} for the pre-eclipse data with the inclusion of iron (i.e. CO+H$_2$O+Fe; red) and without iron (i.e. CO+H$_2$O; black) under the two retrieval setups outlined in Sect.~\ref{sect:4.1}. The red and grey shading marks the 1$\sigma$ and 2$\sigma$ recovered distribution computed from 10,000 samples from the MCMC. \emph{Right panel:} the atmospheric structure for the pre-eclipse dataset from the best-fitting CO+H$_2$O only model. The volume mixing ratio profiles for the continuum and detected species are shown as dashed lines (calculated using \textsc{FastChem}), with the parametric $T$-$P$ profile shown as a solid blue line.}
\label{fig10}
\end{figure*}

Lastly, to assess whether the inclusion of Fe impacts our inference of other species, we performed retrievals identical to those described in Sect.~\ref{sect:4.2}, but excluding Fe from the model (i.e. assuming CO and H$_2$O as the only opacity sources). The results are shown in Appendix~\ref{appendix:H}. The retrieved temperature profiles with and without Fe are consistent (Fig.~\ref{fig10}), suggesting that including iron does not significantly alter the atmospheric structure of MASCARA-1b. This consistency indicates that the model is not artificially forcing iron into the atmosphere by substantially affecting the $T$-$P$ profile, which might occur if the iron detection were spurious. Furthermore, the absolute abundances of CO and H$_2$O remain unchanged across both models. Our chemical model also predicts iron to exist at these temperatures (right panel of Fig.~\ref{fig10}), and the corresponding retrieved $T$-$P$ profiles remain consistent. Additionally, the abundance retrieved from our free-chemistry retrieval analysis agrees with the predictions from our chemical model. Thus, even if we were to ignore iron, we would still expect to see it in the atmosphere at these temperatures.

\begin{figure*}
\begin{minipage}[htbp!]{0.333\textwidth}
  \includegraphics[width=\linewidth]{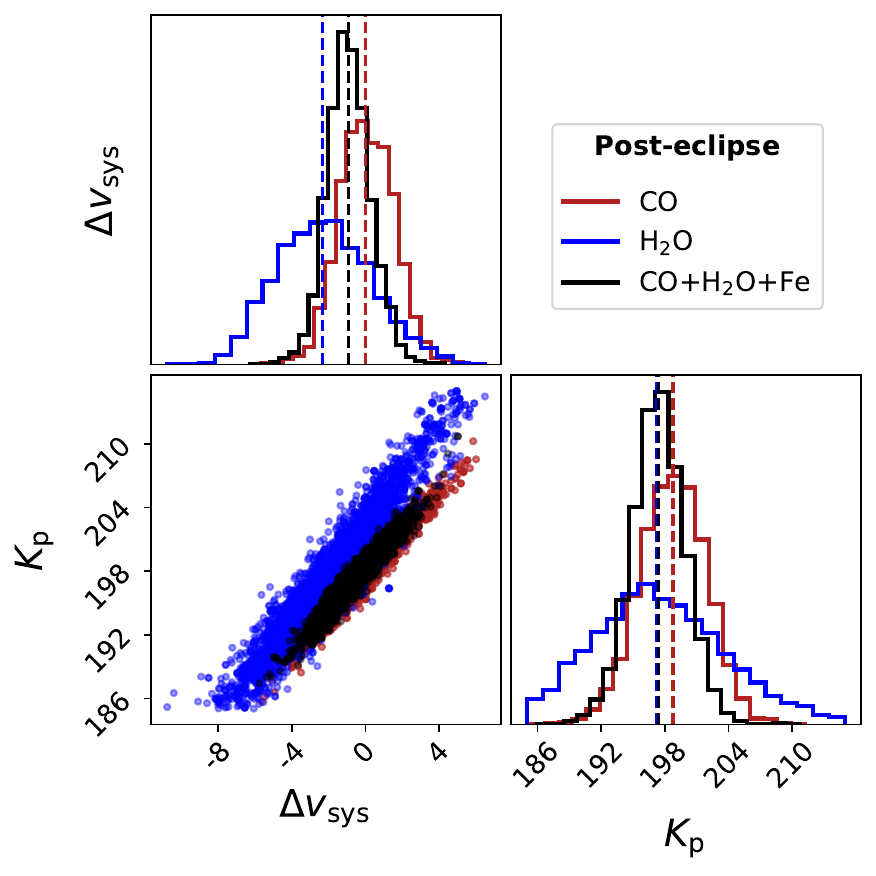}
\end{minipage}%
\hfill
\begin{minipage}[htbp!]{0.333\textwidth}
  \includegraphics[width=\linewidth]{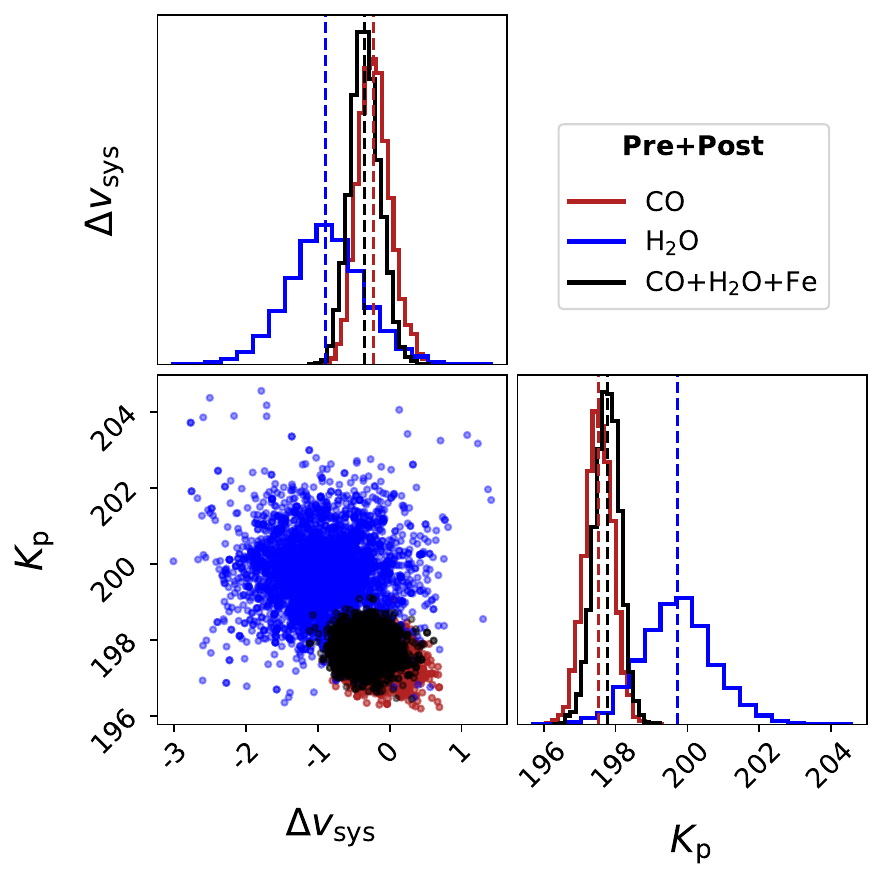}
\end{minipage}%
\hfill
\begin{minipage}[htbp!]{0.333\textwidth}
  \includegraphics[width=\linewidth]{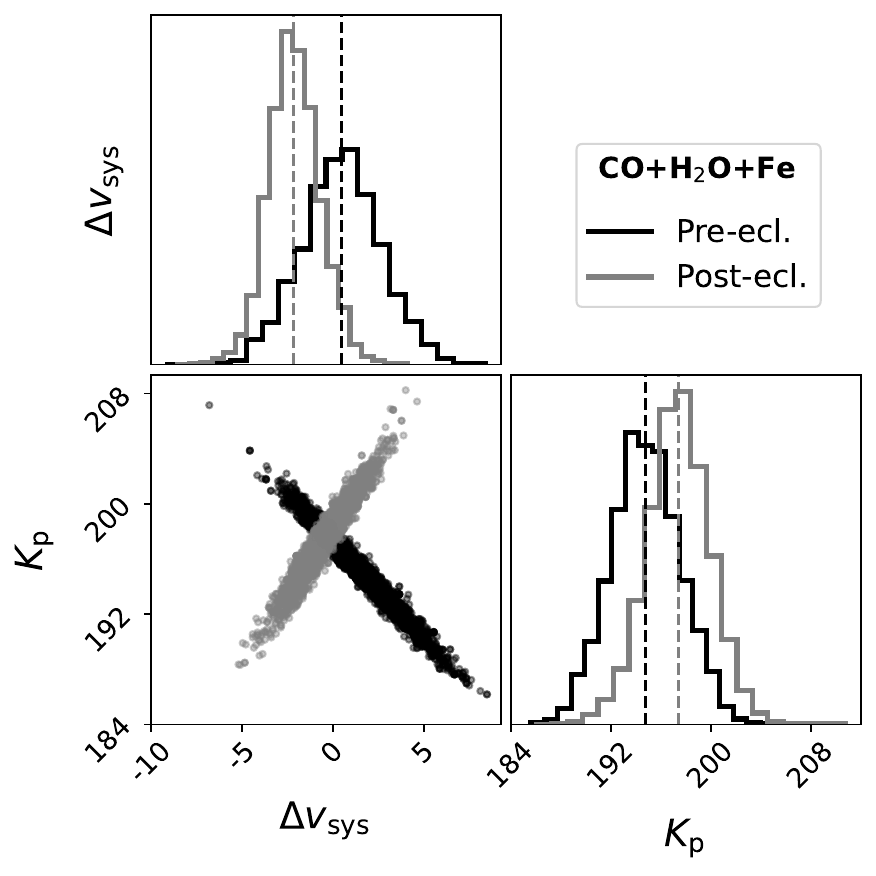}
\end{minipage}%
\caption{Searching for signatures of atmospheric dynamics. \emph{Left and middle panel:} Posterior distributions for the planetary velocities of CO, H$_2$O and the combined spectrum (i.e. CO+H$_2$O+Fe) for the post-eclipse and pre+post datasets. \emph{Right panel:} Marginalised posterior distributions of the combined atmosphere for the pre- and post-eclipse datasets.}
\label{fig11}
\end{figure*}
\subsection{Signatures of atmospheric dynamics}\label{sect:5.3}
\begin{figure}[h!]
  \resizebox{\hsize}{!}{\includegraphics{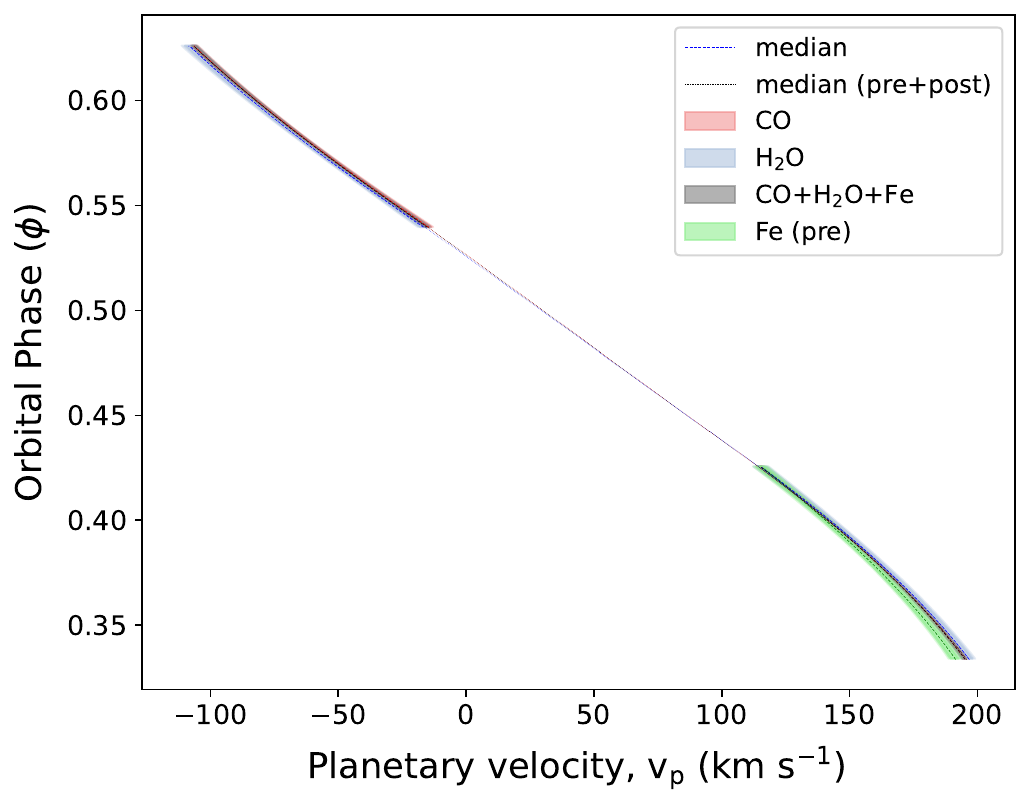}}
   \caption{The retrieved planetary orbital velocity ($v_{\rm p}$) for the detected species from our observations, computed from 10,000 random samples of the MCMC. Dashed lines mark the median, with shaded regions showing $1\sigma$ and $2\sigma$ contours. For clarity, only the median values (dotted lines) are plotted for the pre+post data.}
    \label{fig12}
\end{figure}

We look for shifts in the retrieved $K_{\rm p}$ and $\Delta v_{\rm sys}$ as proxies for possible atmospheric dynamics, searching for any velocity offsets between the detected species \citep[e.g.][]{2023AJ....165...91B, 2021A&A...651A..33C} or between the pre- and post-eclipse datasets \citep[e.g.][]{2022A&A...668A.176P}. Figure~\ref{fig11} illustrates the 2D posteriors of the detected species for the pre-, post- and pre+post-eclipse datasets.
We first search for shifts between the detected species in the individual datasets. For the post-eclipse and pre+post data, CO and the combined species (CO+H$_2$O+Fe) are approximately centred around the expected rest-frame velocity ($\Delta v_{\rm sys}$). However, H$_2$O appears slightly shifted in the 2D parameter space but when we examine the 1D marginalised distributions for H$_2$O across the datasets, the results remain consistent with no significant shift. We note that correlations between $K_{\rm p}$ and $\Delta v_{\rm sys}$ in the 2D posteriors can be obscured when marginalised, making species-specific shifts challenging to interpret. While H$_2$O seems to be shifted in the 2D space, we have not performed statistical tests to quantify its significance. Therefore, we remain cautious in over-interpreting these shifts based purely on posterior distributions.
However, in our analysis of the pre-eclipse data \citep{2023MNRAS.525.2985R}, tentative evidence for shifts is seen for the Fe feature in both $K_{\rm p}$ and $\Delta v_{\rm sys}$. Specifically, while the cross-correlation analysis shows that the peak/mean of the map is red-shifted in $\Delta v_{\rm sys}$ for Fe, the 2D posteriors from the retrieved samples suggest that Fe in general is blue-shifted relative to CO and H$_2$O. If the iron signal is real, this blue shift may indicate that iron is tracing a different dynamical region of the atmosphere, which may also be linked to its non-detection post-eclipse. In addition, to investigate potential differences in dynamics between species, we examine the planetary velocity ($v_{\rm p}$) as a function of the orbital phase for the detected species in the pre-, post-, and combined datasets (Fig.~\ref{fig12}).
This plot shows that while the CO and H$_2$O signals exhibit consistent trails across the observed phases, the iron signal appears slightly offset during the earlier phases of the pre-eclipse dataset (see also Fig.~\ref{figI2} for a zoomed-in version), with the trace overlapping towards the end (i.e. close to the secondary eclipse), where it appears to be more strongly detected.

To further explore these trends and potential longitudinal variations in the atmospheric parameters, we split the pre- and post-eclipse datasets into two subsets based on the orbital phase: phases near quadrature and phases close to the secondary eclipse (see Appendix~\ref{appendix:I}). A joint retrieval performed on these subsets indicates that iron is better constrained during the phases close to the secondary eclipse, as corroborated by the $K_{\rm p}$-$\Delta v_{\rm sys}$ maps (Appendix~\ref{appendix:I}), and aligns with the trace plot. This raises two possibilities: (i) the stronger detection of iron near the eclipse may be due to localised atmospheric phenomena on the planet, or (ii) when combined with the already weak detection (based on the alpha-detection significance), pre-processing steps like \textsc{SysRem} is likely getting rid of the signal more efficiently at the lower/shorter phases when the planet is more static, which could be weakening the iron signal at this phase subset. Both scenarios merit further investigation to disentangle their contributions. Nonetheless, it is intriguing that Fig.~\ref{fig12} combined with the phase-resolved retrieval shows a slightly different trail for iron, which might imply that there are some 3D effects at play which may lead to the non-detection or could in principle lead to a lower signal post-eclipse. That said, we exercise caution in interpreting these features without detailed statistical tests and, therefore, only speculate—underscoring the need for follow-up studies using high-resolution spectroscopy and 3D atmospheric models to explore these phenomena fully.

\begin{figure}[h!]
   \centering
   \includegraphics[width=\hsize]{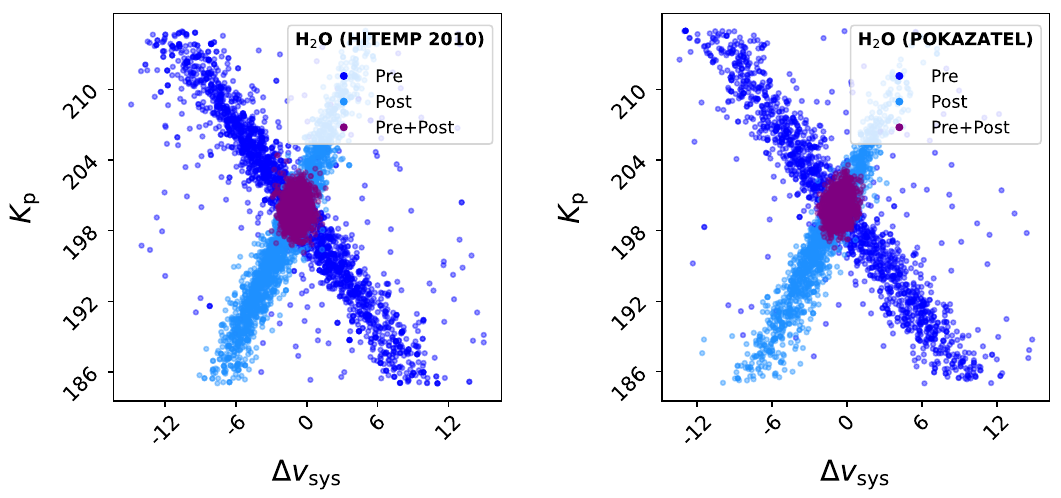}
      \caption{The 2D posteriors of H$_2$O for the two datasets using the HITEMP and POKAZATEL line lists (see text).}
         \label{fig13}
   \end{figure}

Following this, we also checked for offsets between the pre- and post-eclipse sequences for each species and found no signatures of atmospheric dynamics (e.g. right panel of Fig.~\ref{fig11}).
While it is known that different line lists can influence our ability to detect planetary signals, discrepancies in line strengths and positions might affect the S/N \citep[e.g.][]{2020MNRAS.494..108W, 2021ApJ...910L...9N, 2024AJ....168..106R} or retrieved atmospheric properties. Nevertheless, previous analyses, such as that of \citet{2020MNRAS.495..224G}, confirm that the line lists for CO and H$_2$O are appropriate for high-resolution studies up to spectral resolutions of $R{=}100{,}000$. To assess the robustness of our detections and to determine whether the choice of line list impacts our retrieval results, we performed cross-correlation analysis using the POKAZATEL line list from Exomol for H$_2$O \citep[][]{2018MNRAS.480.2597P}. We found no significant differences between the retrieved parameters or the detection significance computed from the conditional likelihood distribution (see Appendix~\ref{appendix:J}) when comparing the POKAZATEL line list with HITEMP 2010 \citep{2010JQSRT.111.2139R}, which we use for the results presented in Section~\ref{sect:3.3}. Additionally, we look for shifts in the retrieved $K_{\rm p}$ and $\Delta v_{\rm sys}$ for H$_2$O alone using two line lists and did not observe any significant shifts in velocity, with all the datasets remaining centred around the expected systemic velocity (see Fig.~\ref{fig13}).

\subsection{Phase/time-dependence}\label{sect:5.4}
Exoplanet atmospheres are inherently 3D structures characterised by dynamic processes, including circulation, chemical mixing, and temperature variations. This is especially true for tidally locked, highly irradiated planets, such as ultra-hot Jupiters, where significant day-to-night temperature gradients can occur. However, our retrieval frameworks currently utilise 1D forward models that assume a globally uniform atmosphere where the atmospheric signal is constant over time. Therefore, it is necessary to exercise caution when interpreting the atmospheric properties retrieved from 1D retrievals, as they may not accurately represent the global chemistry of the planet, particularly in cases where phase-dependent variations are expected. For example, recent emission spectroscopy analyses of the UHJ, WASP-33b, detected a phase dependence found via the model scaling parameter, $\alpha$ \citep[e.g.][]{2022AJ....163..248H, 2023MNRAS.522.2145V}, and reported that a larger scaling is required to best model the observations after the secondary eclipse. Similarly, \citet{2022A&A...668A.176P} explored atmospheric dynamics in KELT-9b by tracking the planet’s velocity as a function of time using phase-resolved CCFs. Such analyses have demonstrated the power of high-resolution spectroscopy phase curves in probing atmospheric dynamics, temperature gradients, and even potential ``3D-ness''.

Our analysis of MASCARA-1b suggests possible variations in the strength of the cross-correlation signal over time, particularly in the pre-eclipse data (see Fig.~\ref{fig6}). Notably, the strength of the CCF trail changes at the beginning and end of the observing sequences, with the lower half of the plot (i.e. the start of the observations) showing a diminished trail. This is most likely attributed to the \textsc{SysRem} pre-processing since de-trending algorithms can remove significant parts of the planet signal. While our filtering technique \citep{2022MNRAS.512.4618G} accounts for these alterations by applying the same pre-processing to the forward model as we do to the data, we will always reduce the signal a little in a phase-dependent way with \textsc{SysRem}/PCA. Nonetheless, the filtering approach works well within our retrieval frameworks, allowing us to recover consistent posteriors with reduced sensitivity to how aggressively the data is filtered until we reach a higher number of \textsc{SysRem} passes \citep[e.g.][Fig.10]{2023MNRAS.525.2985R}.
Additionally, there appears to be a slight dependence on the signal-to-noise, which may also contribute to the variation seen in the CCF. For instance, the pre-eclipse signal disappears towards the end of the sequence, coinciding with a significant increase in airmass and a decrease in S/N. This drop in signal at phases furthest from the eclipse is also expected due to the reduced visibility of the hot day side from the field of view. However, it does not drop off rapidly (see right panel of Fig.~\ref{fig1}).

\begin{figure}[h!]
   \centering
   \includegraphics[width=\hsize]{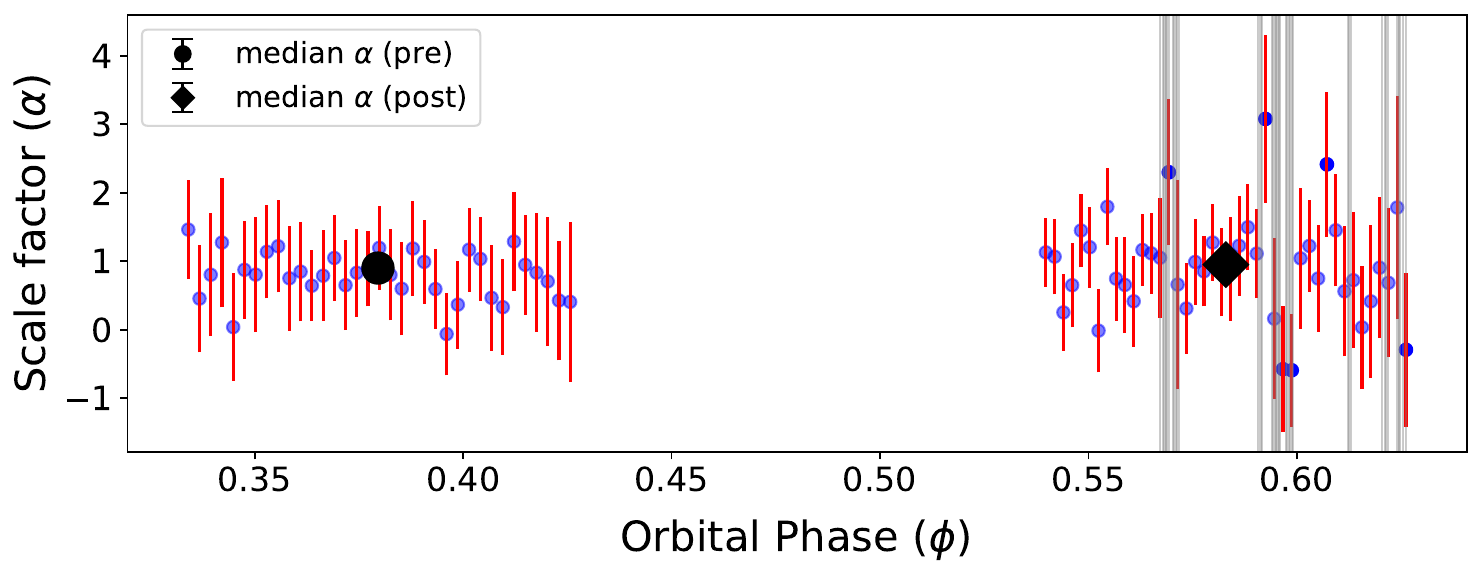}
        \caption{A plot of the model scaling factor ($\alpha$) versus the orbital phase for MASCARA-1b. For visualisation purposes, frames from the individual dataset have been binned together. The $\alpha$ values (blue dots) and their uncertainties for each frame are derived from the conditional likelihood distribution (see text). The black markers show the retrieved median $\alpha$ for the individual datasets. The dark blue points denote outliers corresponding to cloud-affected frames, located within the regions marked by vertical grey bands.}
         \label{fig14}
   \end{figure}

To determine whether the variation in the strength of the CCF trail is solely due to fluctuations in signal-to-noise or if it indicates a physical change in the planet's signal, we chose not to fix the model scaling parameter, $\alpha$, at 1. Instead, we treated it as a free parameter in our combined (pre+post) retrieval fits. This approach allowed us to examine potential time/phase dependence before considering a phase-dependent alpha implementation within the likelihood equation. Upon doing so, we find the retrieved values for $\alpha$ before and after the eclipse to be consistent, with $\alpha_{\rm pre} = 0.90^{+0.08}_{-0.08}$ and $\alpha_{\rm post} = 0.95^{+0.09}_{-0.09}$. Building on this, we further investigated potential phase-dependent trends by computing a likelihood map as a function of $\alpha$ using a combination of Equations~\ref{eqn4} and~\ref{eqn5}. We begin with our cross-correlation map ($\phi$ vs $\Delta v_{\rm sys}$) and calculate it as a function of $\alpha$ (expansion of Eqn.~\ref{eqn5}). Next, we Doppler-shift it to the planet rest frame using Eqn.~\ref{eqn2} (conditioned on the optimal values of all atmospheric parameters). Instead of summing over time, we calculate it for each phase point. This was then used to compute a log-likelihood map as a function of alpha for every frame, which was converted into a likelihood to obtain a conditional likelihood distribution of $\alpha$, similar to the method outlined in Sects.~\ref{sect:3.3} and~\ref{sect:4.1} (see also Fig.~\ref{fig4}), but for every frame. Lastly, we determined $\alpha$ and its corresponding uncertainties for each phase by calculating the mean and standard deviation of the distribution. A plot of $\alpha$ versus orbital phase for the pre- and post-eclipse datasets is shown in Figure~\ref{fig14}. The $\alpha$ values are centred around 1, with no significant trend seen between the two datasets, suggesting that the line contrast is constant over time. However, given the degeneracy between the temperature gradient and chemical abundances (+ continuum), it does not necessarily mean the atmosphere is constant over time. Nonetheless, more sophisticated methods and high-resolution observations may be required to better constrain and probe the phase dependence of the model scaling factor.

\subsection{Comparison with phase curve measurements}
Recent analyses of Spitzer’s 4.5 $\mu$$\rm m$ phase curve of MASCARA-1b \citep{2021MNRAS.504.3316B,2025AJ....169...32D} have measured day-side and night-side temperatures of $3000{\pm}100$ K and $1300{\pm}300$ K, respectively, showing a fairly inefficient recirculation, which is consistent with expectations for ultra-hot Jupiters. Building on these findings, we calculated equilibrium temperatures for different heat redistribution scenarios given by the heat redistribution factor, $f$, as follows:
\begin{equation}
    T_{\rm p,eq} = T_*\left(\frac{R_*}{a} \right)^\frac{1}{2}\left[\frac{f}{4}(1-A_{\rm B}) \right]^\frac{1}{4}
\end{equation}
where $T_*$, $R_*$, $a$ and $A_{\rm B}$ denote the stellar effective temperature, stellar radius, semi-major axis and the bond albedo, respectively. The heat redistribution factor, $f$, accounts for the fact that the absorbed energy may not all be re-radiated uniformly over the planet’s surface, with $f{=}1$ corresponding to uniform distribution (e.g. full heat redistribution across the planet), $f{=}2$ for re-radiation over half the surface of the planet (e.g. day-side only re-radiation) and $f{=}\frac{8}{3}$ corresponding to instant re-emission (Lambertian sphere). For an $A_{\rm B} {=} 0.32^{+0.09}_{-0.11}$ \citep{2025AJ....169...32D}, we calculate the equilibrium temperature to be {$\approx$} 2376$\pm$35K, 2826$\pm$41K and 3036$\pm$44K for $f{=}1$, $f{=}2$ and $f{=}\frac{8}{3}$, respectively. On comparing these with the irradiation temperature from our retrieval analysis, we find that while the results are broadly consistent with $f{=}1$, particularly for the free-chemistry setup, given the uncertainties, our results also overlap with $f{=}2$, allowing for consistency with any of the heat redistribution scenarios within ${\approx}1\sigma$. However, our uncertainties on the $T$-$P$ profile remain consistent with the Spitzer phase curve analysis. Additionally, in the context of high-resolution spectroscopy, our analysis of the phase-dependent scaling factor $\alpha$--which represents variations in line contrast as a function of the orbital phase--reveals no significant trends. This further suggests that we do not observe any substantial changes in the atmosphere of MASCARA-1b as it rotates, likely explaining the consistency in our retrieved parameters between the pre- and post-eclipse datasets. It is also worth noting that the near-symmetric phase coverage of the pre- and post-eclipse datasets may further contribute to the observed consistency. For instance, if the planet's hotspot exhibits minimal longitudinal offset or its spatial variations are symmetric relative to the substellar point, retrievals from both phases could appear similar even with strong spatial variation. This consistency between pre- and post-eclipse also aligns with the Spitzer phase curve observations, which do not show any significant phase offset \citep{2025AJ....169...32D}.

\section{Conclusions} \label{sect:6}
In this work, we have presented high-resolution emission spectroscopy observations of the UHJ MASCARA-1b with CRIRES+ in the K-band, covering the post-eclipse phases of the planet's orbit. We applied the standard cross-correlation technique and a Bayesian retrieval framework that includes both pre-eclipse and post-eclipse datasets and learned the following about the thermal and chemical properties of the planet:
\begin{itemize}
    \item In addition to the CRIRES+ pre-eclipse phase-curve observations presented in \citet{2023MNRAS.525.2985R}, we present here an additional night taken two years apart, capturing the planet's day-side emission in the post-eclipse phases. Using the traditional cross-correlation method, we detected the presence of CO and H$_2$O in the post-eclipse data, with the features also identified at high significance in the joint (pre+post) analysis (Sect.~\ref{sect:3.3}).
    
    \item The molecular species are seen as emission signatures through which we also confirmed the presence of a thermal inversion layer in the day-side atmosphere of MASCARA-1b. No significant signature of the chemical species CO$_2$, $^{13}$CO, OH, Mg, HCN and CH$_4$ could be detected in our pre-, post-, and pre+post-eclipse data (Sect.~\ref{sect:3.3}).
    
    \item The likelihood framework introduced in \citet{2020MNRAS.493.2215G,2022MNRAS.512.4618G} was applied to obtain quantitative constraints on the atmospheric properties, such as the parametric $T$-$P$ profile, planetary velocities ($K_{\rm p}$, $\Delta v_{\rm sys}$), abundances of the species as well as the atmospheric $\rm C/O$ ratio while simultaneously marginalising over the noise properties of the dataset (Sect.~\ref{sect:4}).

    \item We implemented two different approximations in our retrieval framework to describe the chemical composition of MASCARA-1b's atmosphere (Sect.~\ref{sect:4.1}). While our free-chemistry retrievals, which assume a well-mixed atmosphere, indicate a super-solar $\rm C/O$ ratio, incorporating a chemical equilibrium model results in a $\rm C/O$ of 0.75$^{+0.11}_{-0.17}$ for the post-eclipse data and 0.74$^{+0.10}_{-0.14}$ for the joint analysis, both consistent with solar values within $\approx$1.1-2$\sigma$ (Sect.~\ref{sect:4.2}). The chemical equilibrium model, which accounts for altitude-dependent variations in composition, is the preferred model for ultra-hot Jupiters, given their complex atmospheric dynamics.

    \item Despite the difference in signal-to-noise between the pre-and post-eclipse datasets, the retrieved model parameters, under both free-chemistry and chemical equilibrium setups, are consistent with the pre-eclipse values, though less precise. Combining the two nights of the CRIRES+ data enabled slightly tighter constraints on the retrieved parameters than the individual datasets. Overall, the retrieved values for the $T$-$P$ profile parameters between the two retrieval frameworks remain consistent across the datasets, falling within $\approx$1.1$\sigma$ (Sect.~\ref{sect:4.2}).

     \item We revisit the detection of the Fe signal in our pre-eclipse data (Sect.~\ref{sect:5.2}) and found no signatures of atmospheric dynamics in the post-eclipse and combined datasets (Sect.~\ref{sect:5.3}).
    
    \item A phase-dependent analysis of the scaling factor $\alpha$ showed no significant variation between the pre- and post-eclipse phases, with the retrieved median $\alpha$ values centred around $\approx$1 (Sect.~\ref{sect:5.4}). 

    \item We find no strong evidence for 3D effects in the atmosphere of MASCARA-1b, and given the large uncertainties on the $T$-$P$ profile, our findings can be explained by any of the $f$ values and are not in conflict with the Spitzer phase curve (Sect.~\ref{sect:5.4}). Nonetheless, we note that our observations cover only a subset of the planet’s rotation, which may limit our sensitivity to detecting longitudinal temperature variations or other 3D effects.
    
\end{itemize}

While the consistency between pre- and post-eclipse retrievals may not be particularly intriguing from a 3D perspective, it remains interesting that this agreement persists even though the observations were taken two years apart, with the post-eclipse data probing the atmosphere of MASCARA-1b after the planet had rotated by approximately $\approx$106 degrees compared to the pre-eclipse observations. Future high-resolution observations with broader phase coverage and/or complementary wavelengths could further enhance our understanding of longitudinal and temporal variability in ultra-hot Jupiter atmospheres. Such observations, in synergy with space-based facilities such as the JWST, could help provide critical insights into the chemical and thermal dynamics shaping these extreme planetary environments.

\begin{acknowledgements}
      We thank the anonymous referee for careful reading of the manuscript and providing helpful comments. This work relied on the observations collected at the European Organisation for Astronomical Research in the Southern Hemisphere under ESO programs 107.22TQ.001 (PI: Gibson) and 112.260X.001 (PI: Gibson, dPI: Ramkumar). We are extremely grateful to the CRIRES+ instrument teams and observatory staff who made these observations possible. S.R. gratefully acknowledges support from the Provost's PhD Award from Trinity College Dublin. N.P.G is supported by Science Foundation Ireland and the Royal Society in the form of a University Research Fellowship and S.K.N is supported by JSPS KAKENHI grant No. 22K14092. We are grateful to the developers of the NumPy, SciPy, Matplotlib, iPython, corner, \texttt{petitRADTRANS}, \textsc{FastChem}, and Astropy packages, which were used extensively in this work \citep{2020Natur.585..357H, 2020NatMe..17..261V, 2007CSE.....9...90H, 2007CSE.....9c..21P, 2016JOSS....1...24F, 2019A&A...627A..67M, 2018MNRAS.479..865S, 2022MNRAS.517.4070S, 2022ApJ...935..167A}.
\end{acknowledgements}

%
\bibliographystyle{aa} 


\begin{appendix}




\onecolumn
\section{Injection-recovery tests}\label{appendix:A}
\begin{figure*}[h!]
    \centering
     \resizebox{17cm}{21cm}
    {\includegraphics {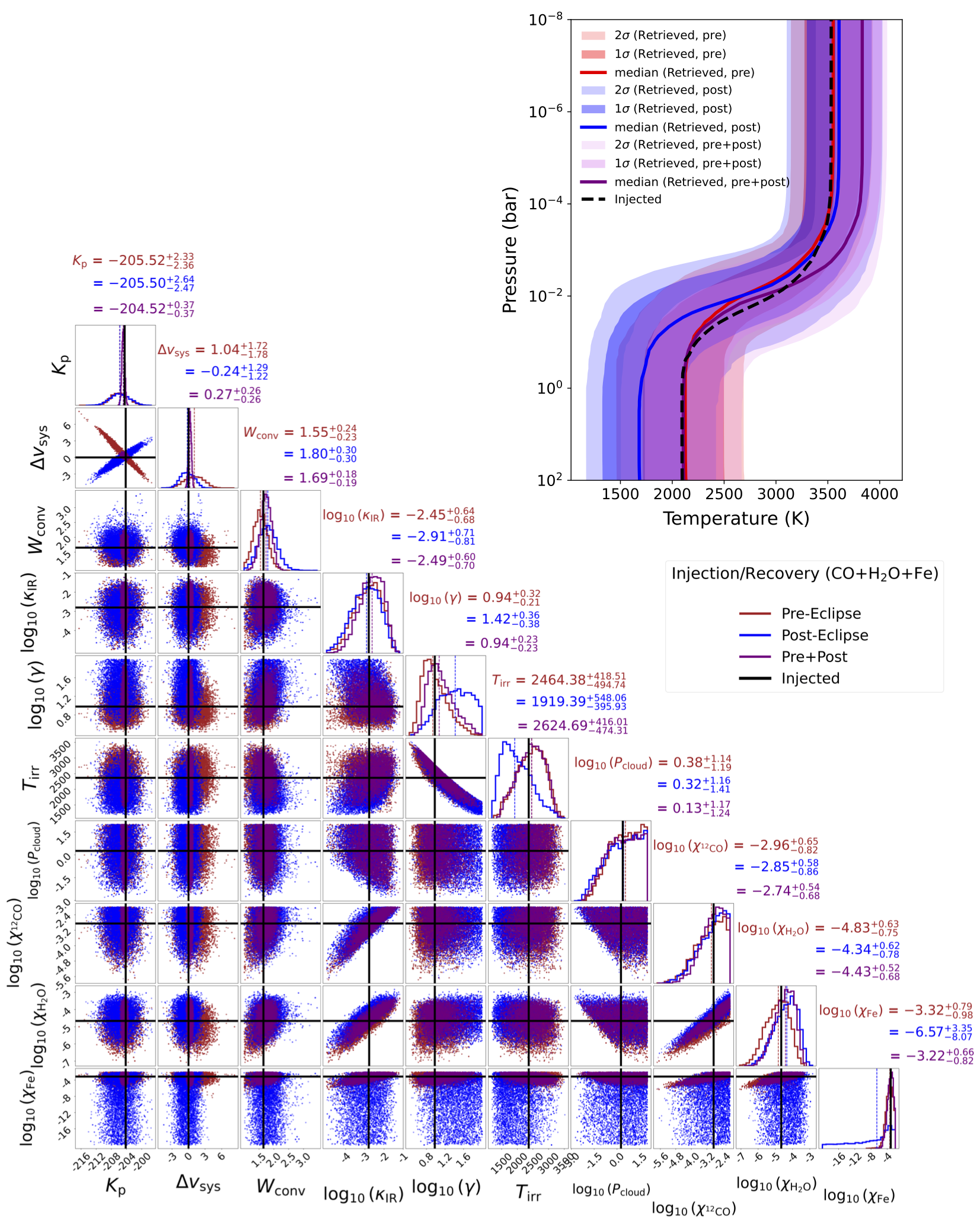}}
     \caption{Summary of our injection tests for pre-, post- and pre+post-eclipse datasets outlined in Sect.~\ref{sect:3.2}, with the 1D and 2D marginalised posterior distributions of each of our model parameters displayed. The red, blue and black posterior distributions represent the datasets, with the injected values highlighted by horizontal and vertical solid purple lines. The dotted lines show the retrieved median values. \emph{Upper right:} The retrieved $T$-$P$ profiles are computed from 10,000 random samples of the MCMC; the solid curves show the median profiles and the shaded regions show the 1$\sigma$ and 2$\sigma$ contours. The dashed black curve shows the injected $T$-$P$ profile.}
      \label{figA1}
\end{figure*}

\FloatBarrier 

\onecolumn
\begin{figure*}[h!]
    \centering
     \resizebox{17cm}{22cm}
    {\includegraphics {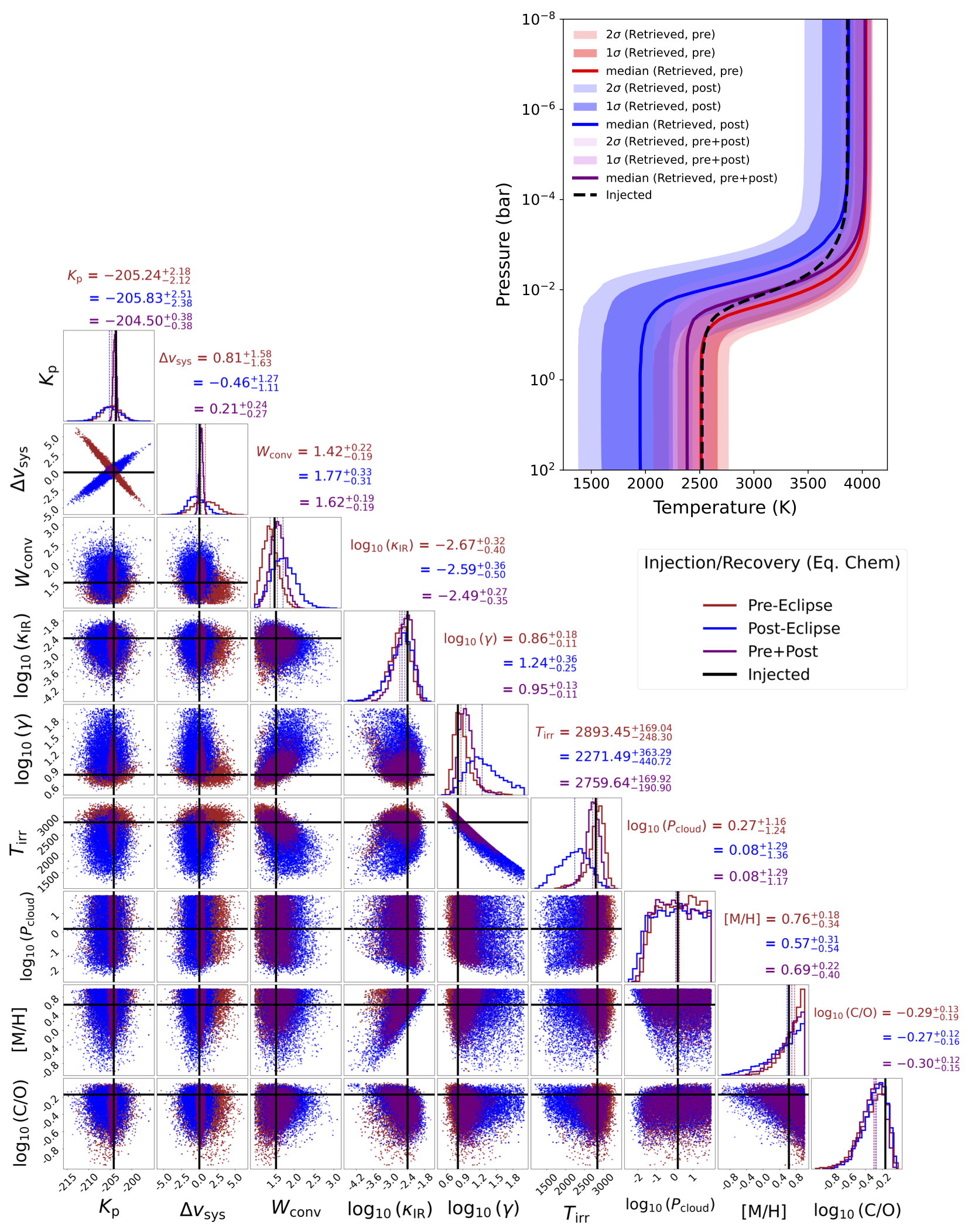}}
     \caption{Same as Fig.~\ref{figA1} but for a chemical model.}
      \label{figA2}
\end{figure*}

\FloatBarrier

\onecolumn
\section{Additional retrieval analyses including the cloud-affected frames in the post-eclipse data}\label{appendix:B}

\begin{figure*}[h!]
    \centering
     \resizebox{9.5cm}{10.3cm}
    {\includegraphics {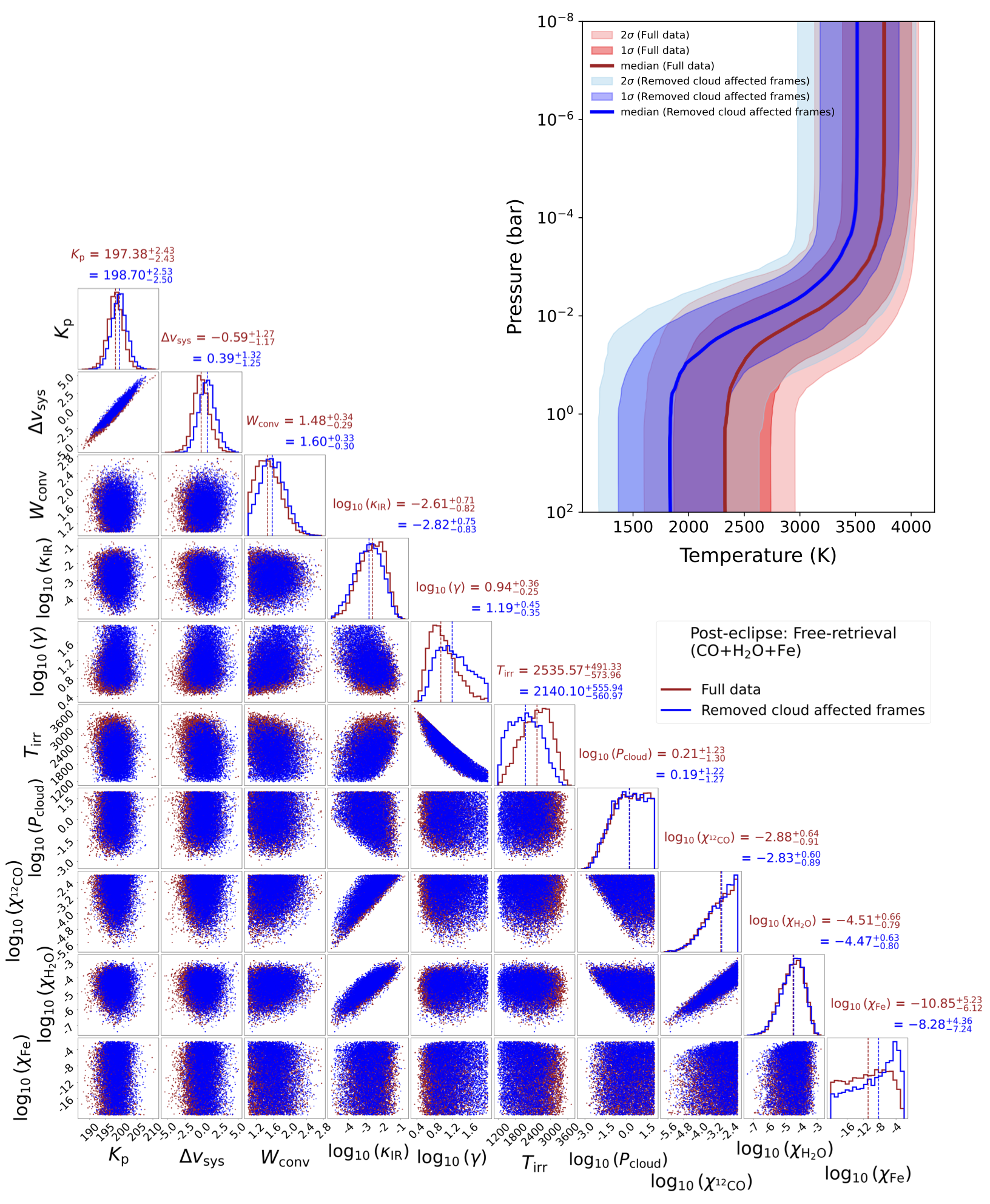}}
     \caption{Free-retrieval results for the post-eclipse data \emph{with} (red) and \emph{without} (blue) the inclusion of cloud-affected frames (Sect.~\ref{sect:3.2}). While excluding these frames slightly improved the Fe constraint, it is still a long-tailed distribution and did not impact the overall conclusions (Fe remains a non-detection).}
      \label{figB1}
\end{figure*}
\begin{figure*}[h!]
    \centering
     \resizebox{9.5cm}{10.3cm}
    {\includegraphics {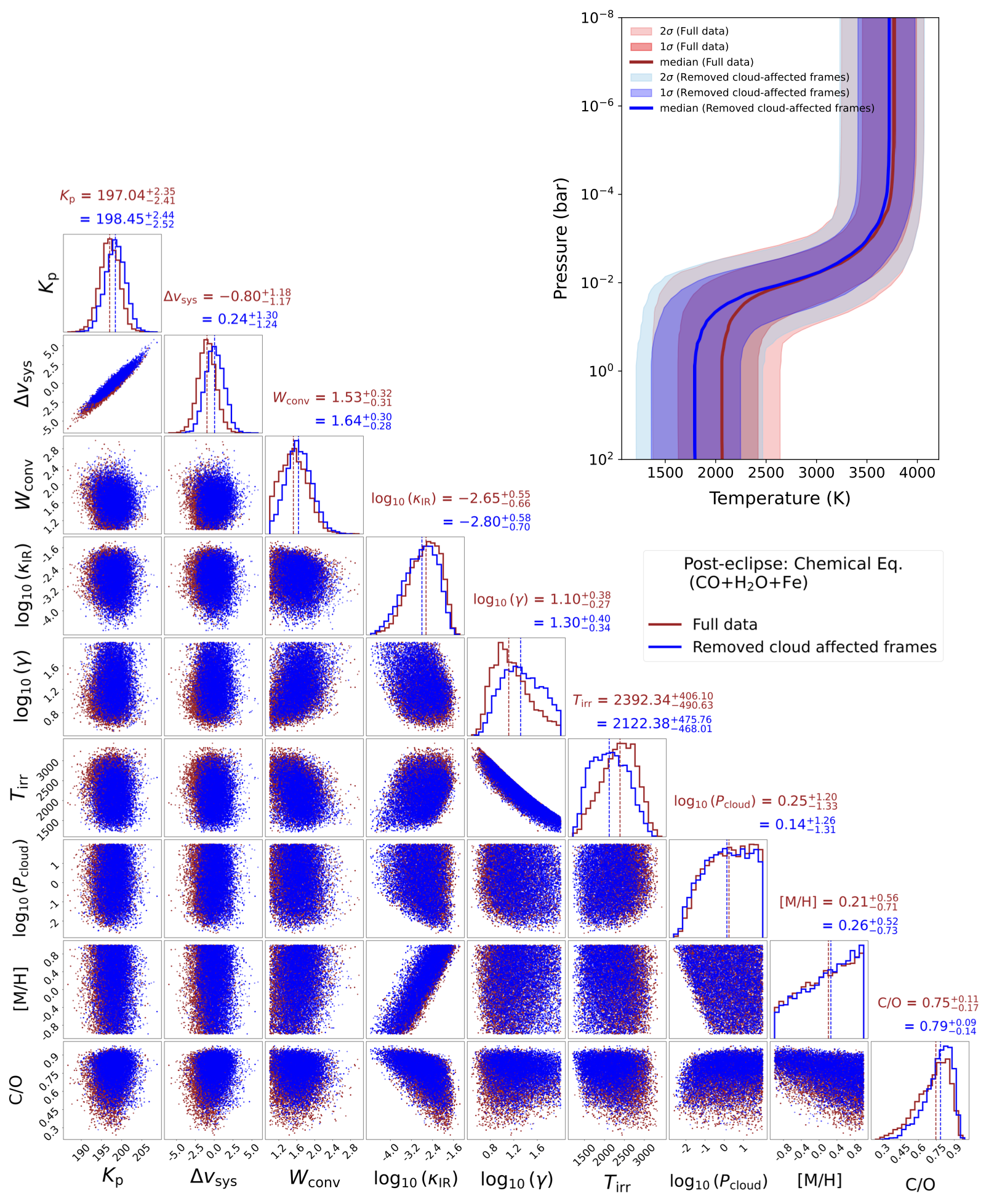}}
     \caption{Same as Fig.~\ref{figB1} but assuming chemical equilibrium.}
      \label{figB2}
\end{figure*}

\FloatBarrier
\onecolumn
\section{Additional $K_{\rm p}$-$\Delta v_{\rm sys}$ maps}\label{app:c}
\begin{figure*}[h!]
    \centering
    \resizebox{4.5cm}{3cm}
    {\includegraphics {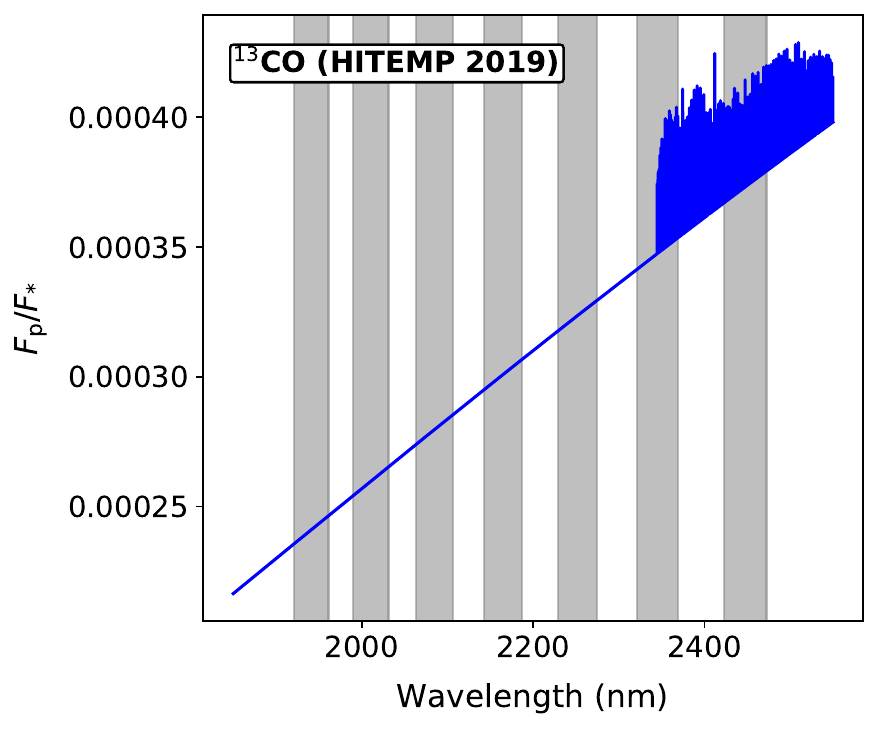}}
    \resizebox{4.5cm}{3cm}
    {\includegraphics {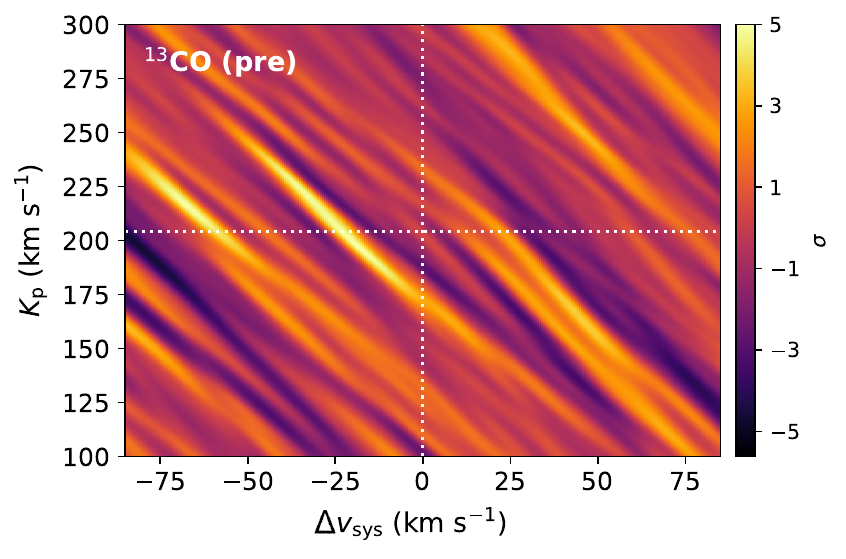}}
    \resizebox{4.5cm}{3cm}
    {\includegraphics {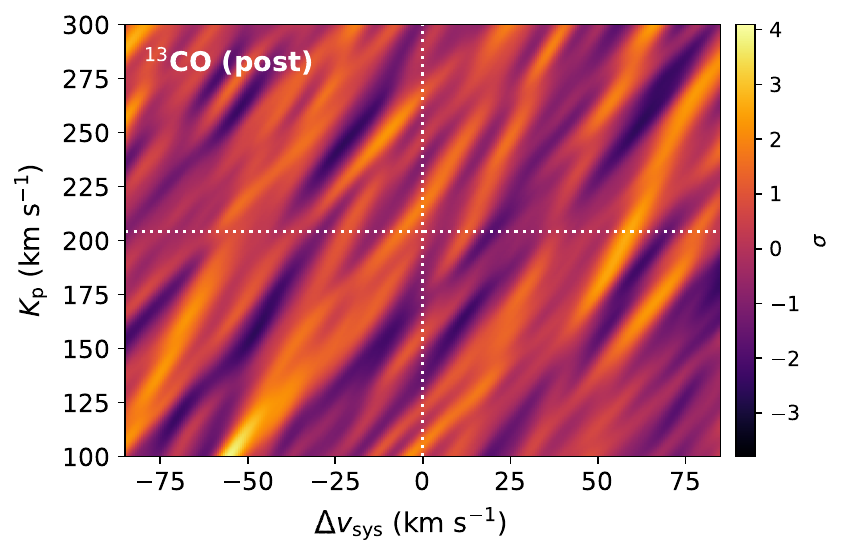}}
    \resizebox{4.5cm}{3cm}
    {\includegraphics {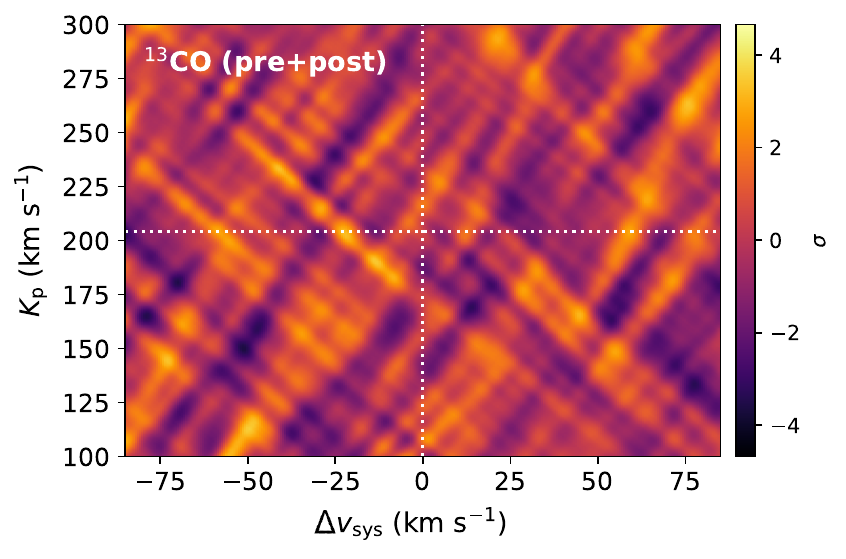}}
    
    \vspace*{0.02cm}
    
     \resizebox{4.5cm}{3cm}
    {\includegraphics {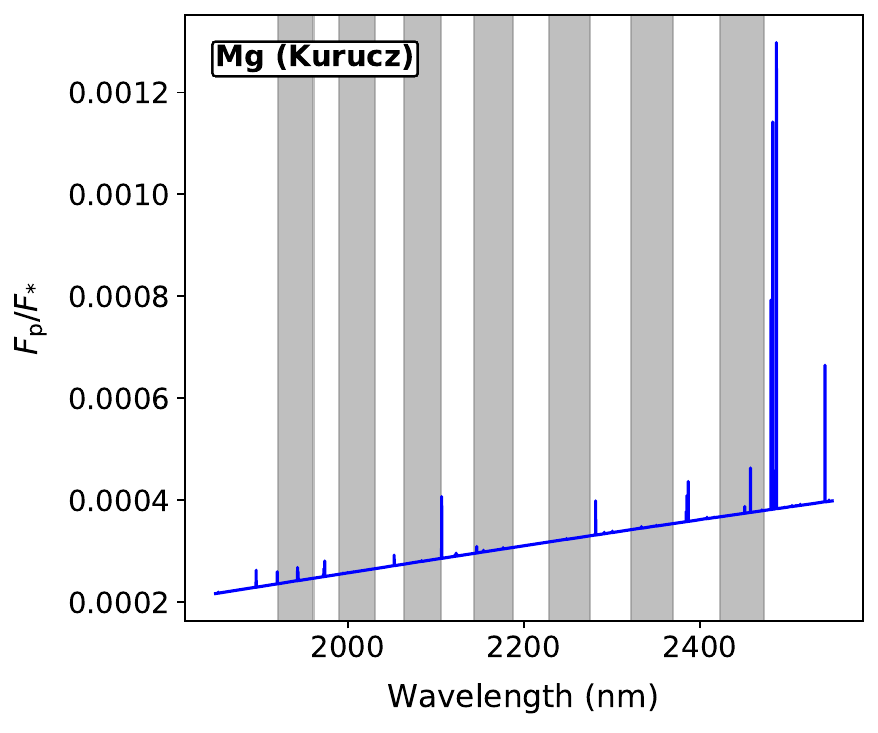}}
    \resizebox{4.5cm}{3cm}
    {\includegraphics {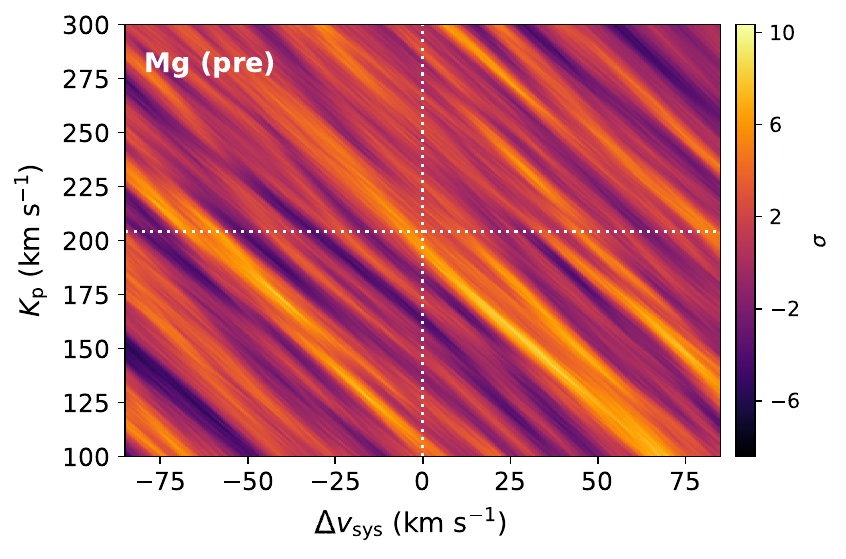}}
    \resizebox{4.5cm}{3cm}
    {\includegraphics {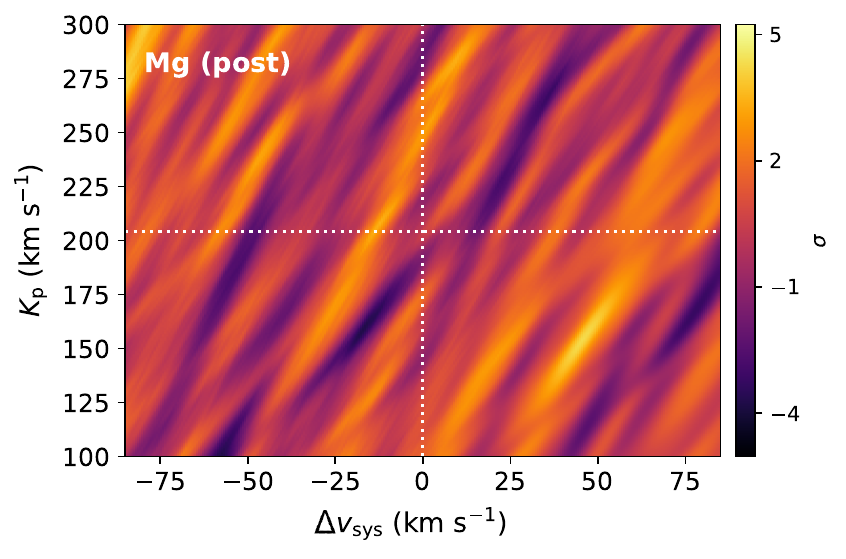}}
    \resizebox{4.5cm}{3cm}
    {\includegraphics {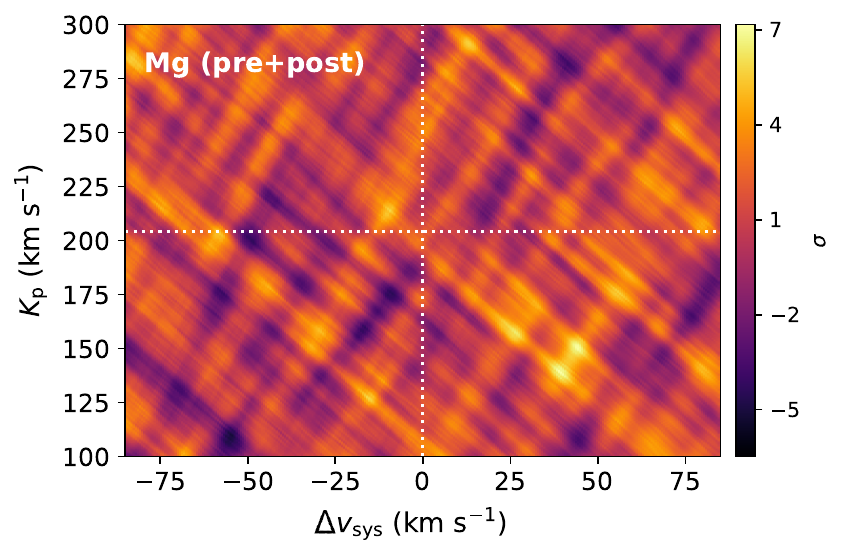}}

    \vspace*{0.02cm}

    \resizebox{4.5cm}{3cm}
    {\includegraphics {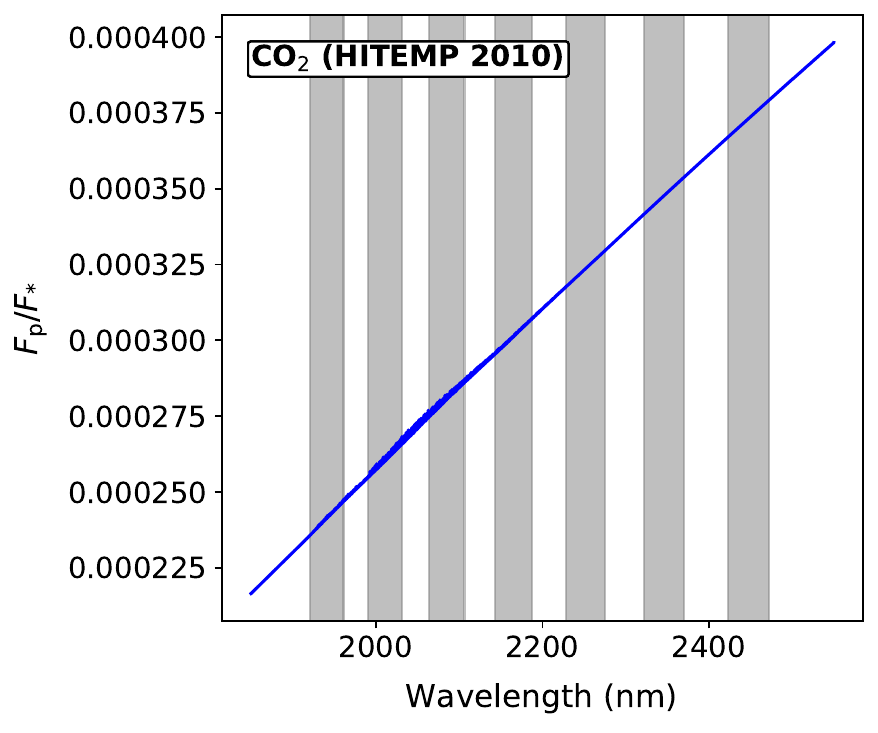}}
    \resizebox{4.5cm}{3cm}
    {\includegraphics {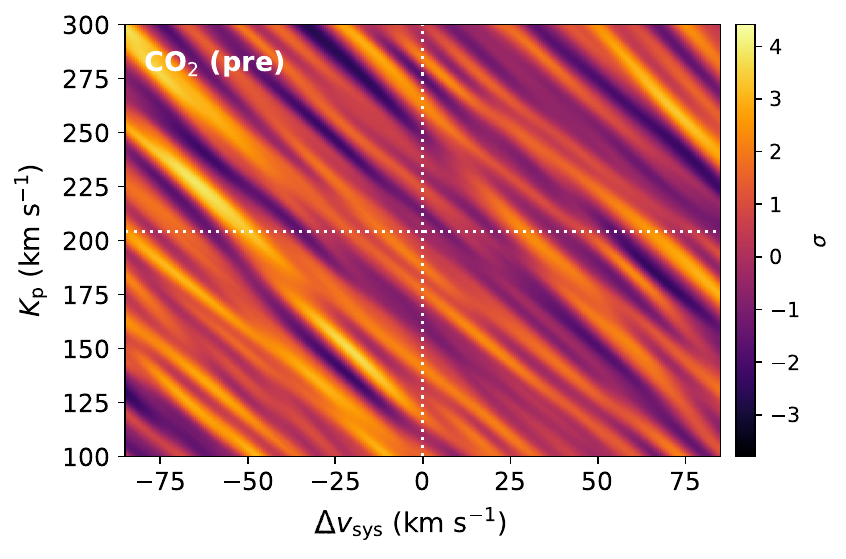}}
    \resizebox{4.5cm}{3cm}
    {\includegraphics {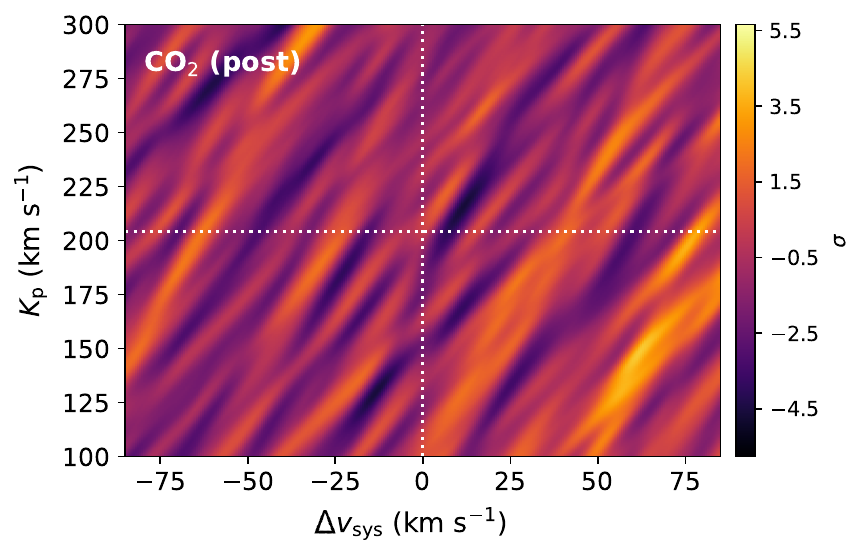}}
    \resizebox{4.5cm}{3cm}
    {\includegraphics {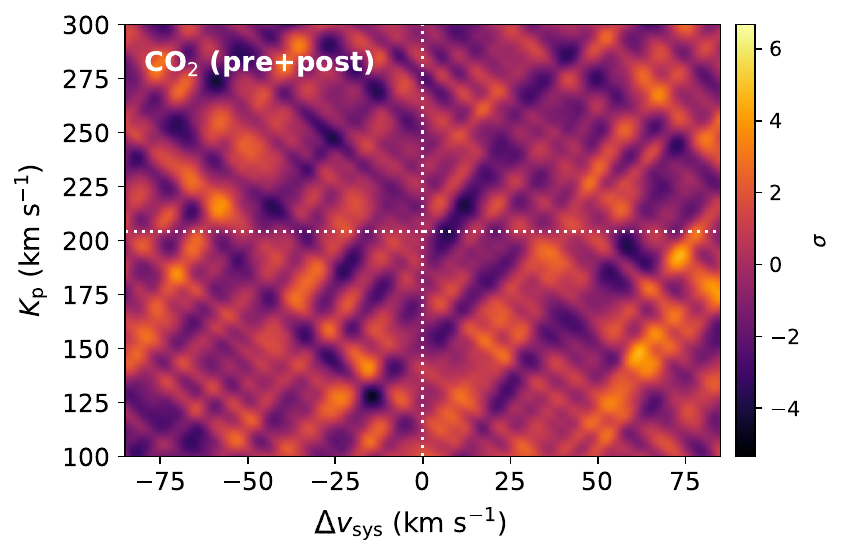}}

    \vspace*{0.02cm}

    \resizebox{4.5cm}{3cm}
    {\includegraphics {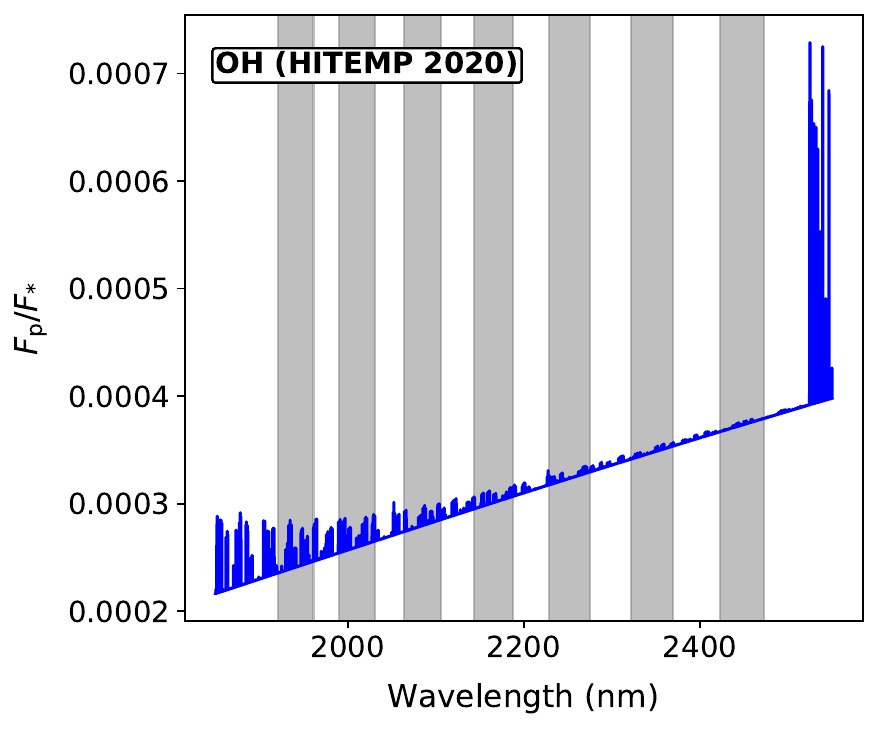}}
    \resizebox{4.5cm}{3cm}
    {\includegraphics {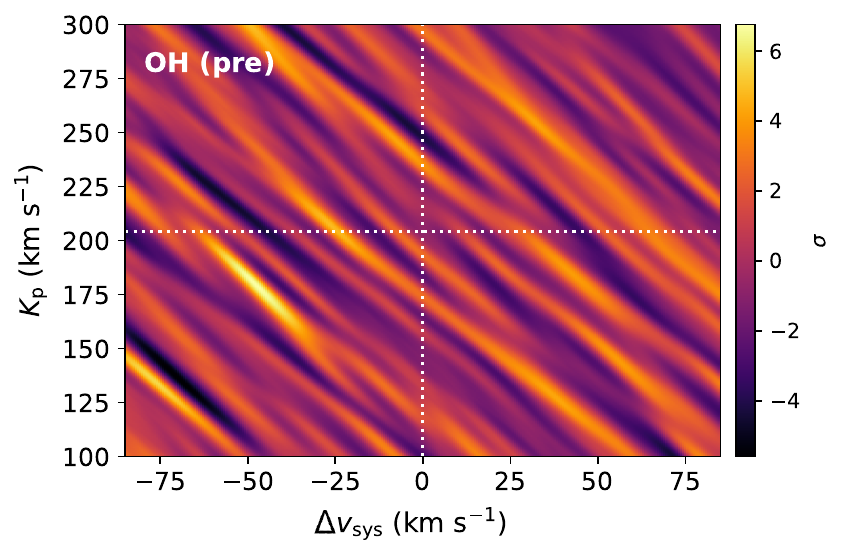}}
    \resizebox{4.5cm}{3cm}
    {\includegraphics {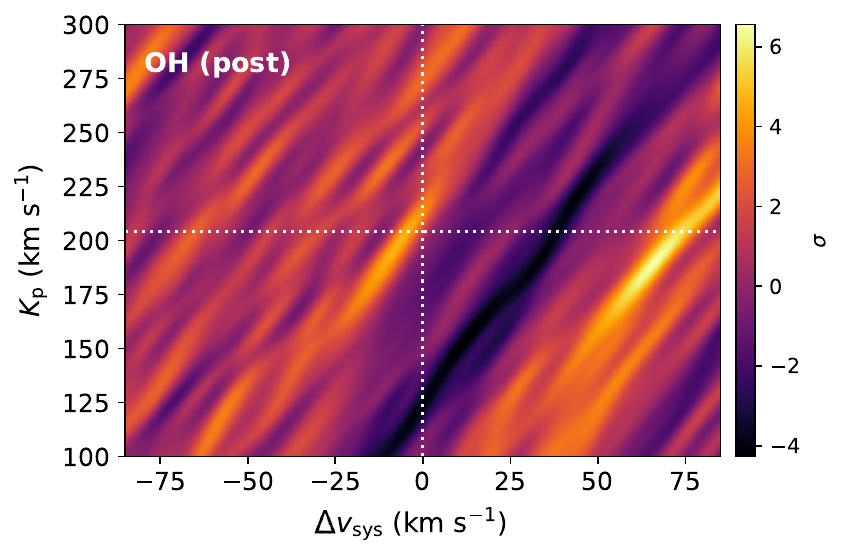}}
    \resizebox{4.5cm}{3cm}
    {\includegraphics {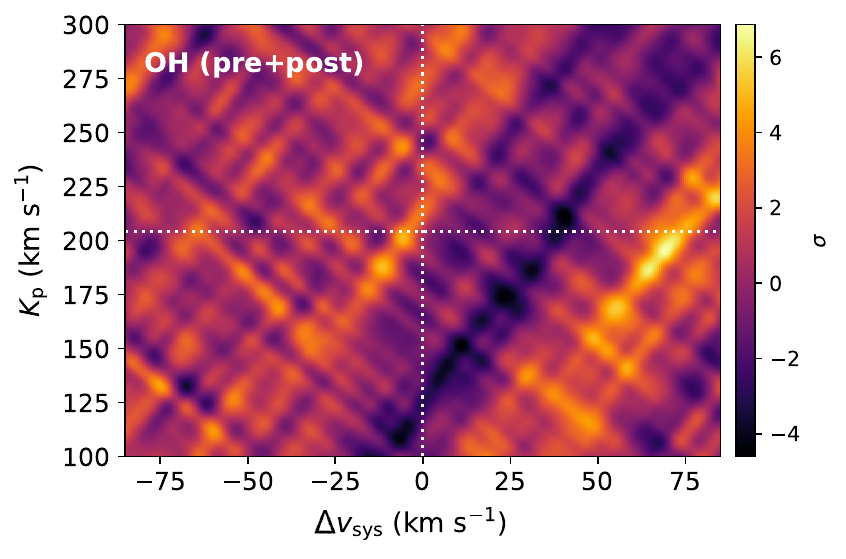}}

    \vspace*{0.02cm}

    \resizebox{4.5cm}{3cm}
    {\includegraphics {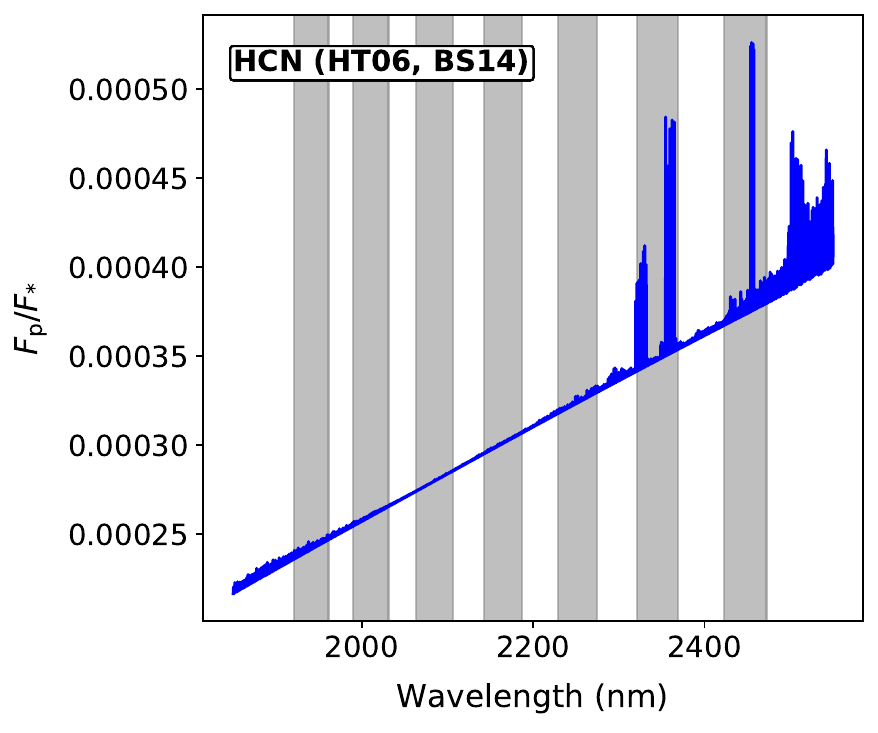}}
    \resizebox{4.5cm}{3cm}
    {\includegraphics {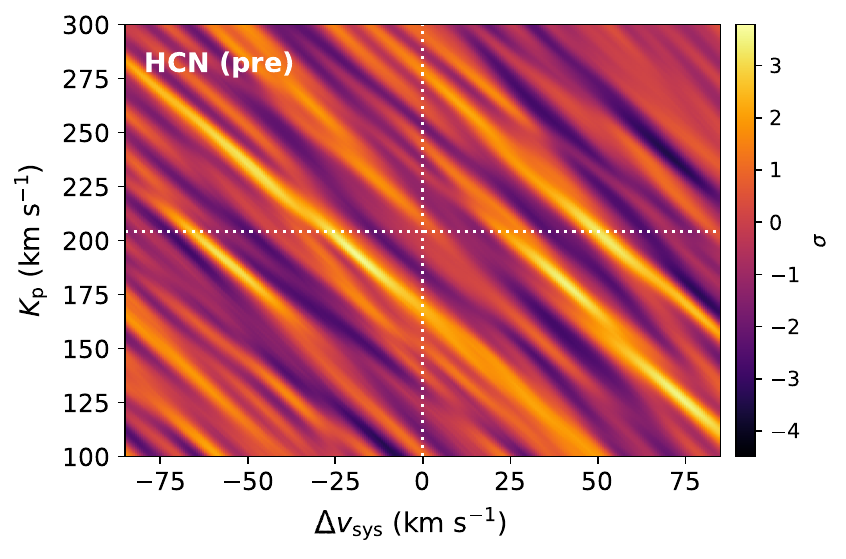}}
    \resizebox{4.5cm}{3cm}
    {\includegraphics {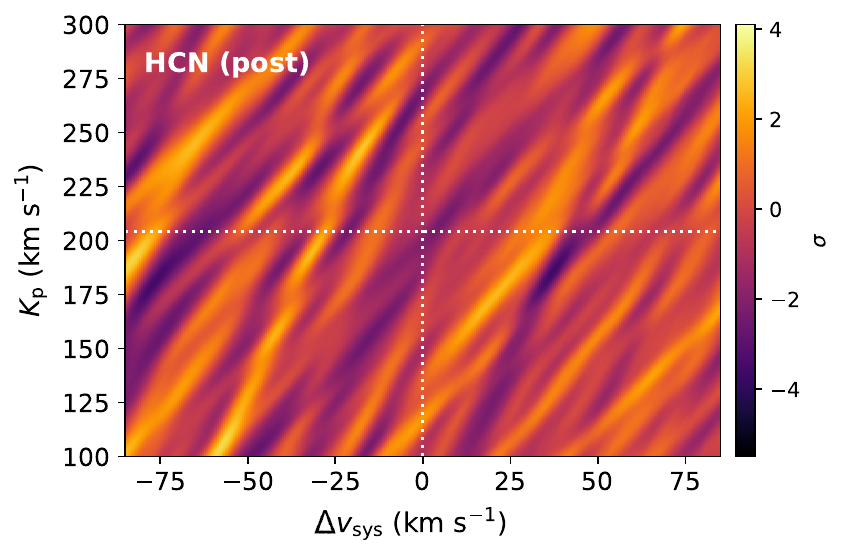}}
    \resizebox{4.5cm}{3cm}
    {\includegraphics {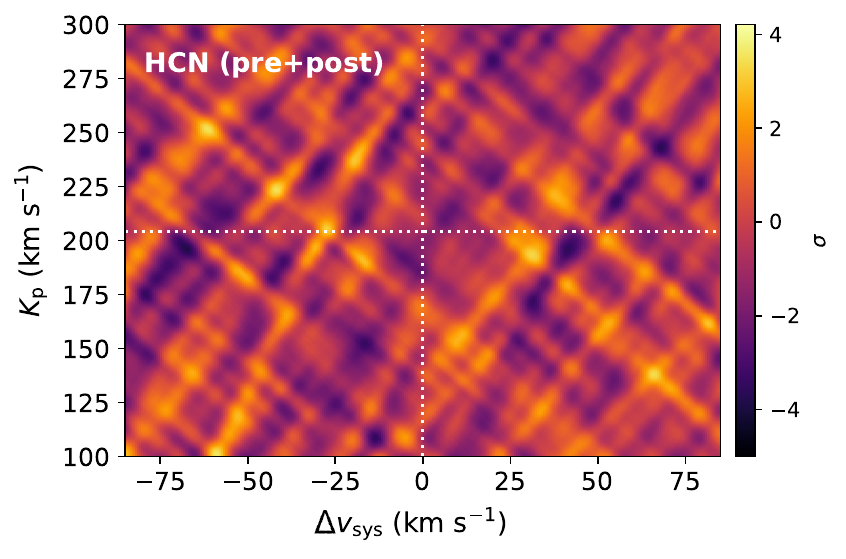}}

    \vspace*{0.02cm}

    \resizebox{4.5cm}{3cm}
    {\includegraphics {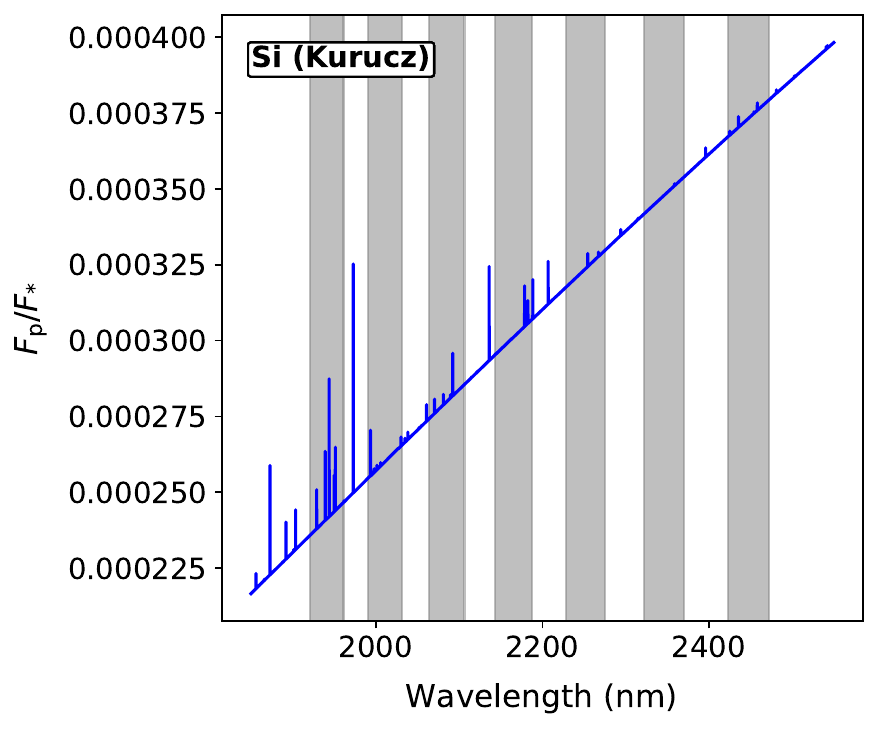}}
    \resizebox{4.5cm}{3cm}
    {\includegraphics {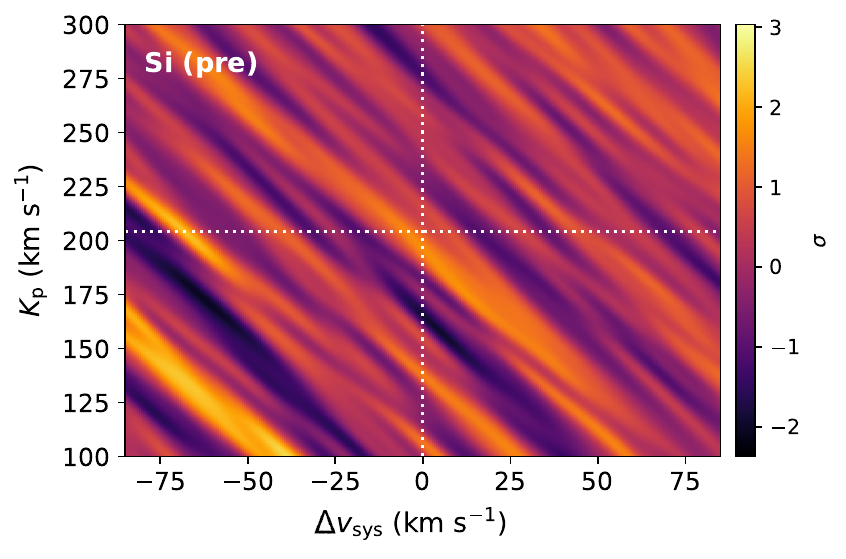}}
    \resizebox{4.5cm}{3cm}
    {\includegraphics {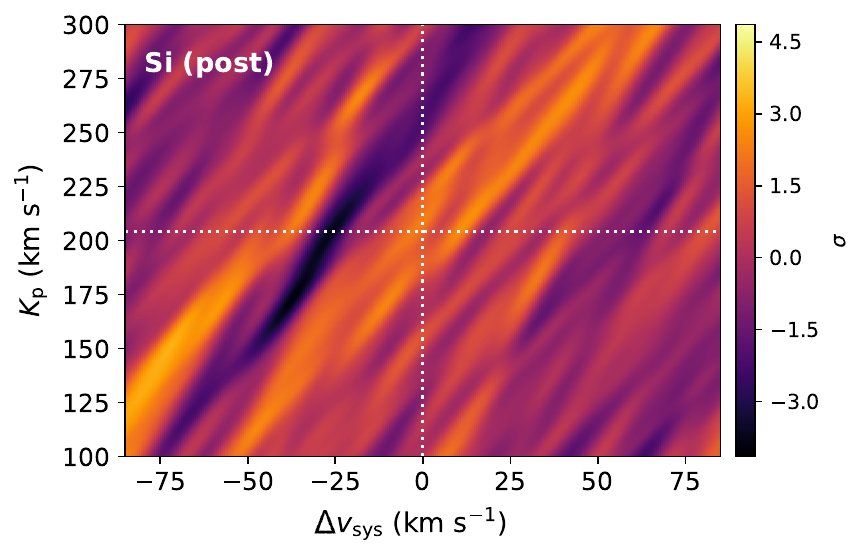}}
    \resizebox{4.5cm}{3cm}
    {\includegraphics {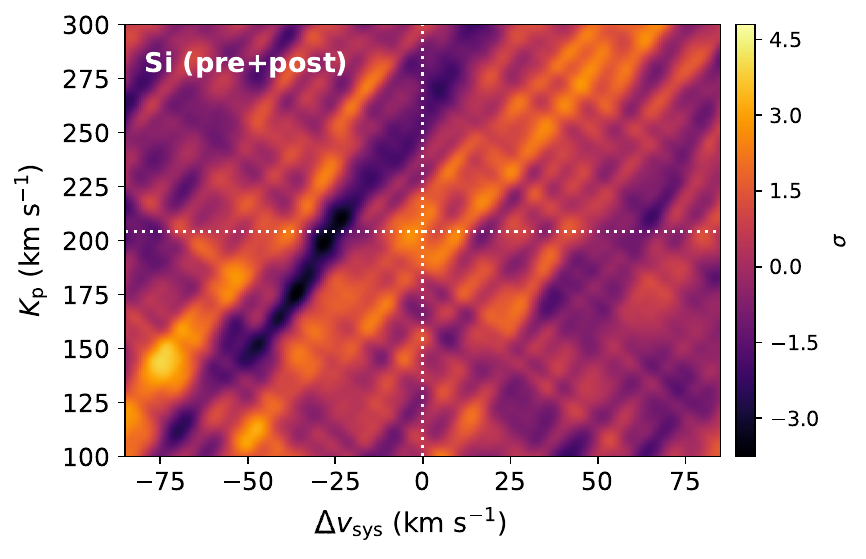}}

    \vspace*{0.02cm}

    \resizebox{4.5cm}{3cm}
    {\includegraphics {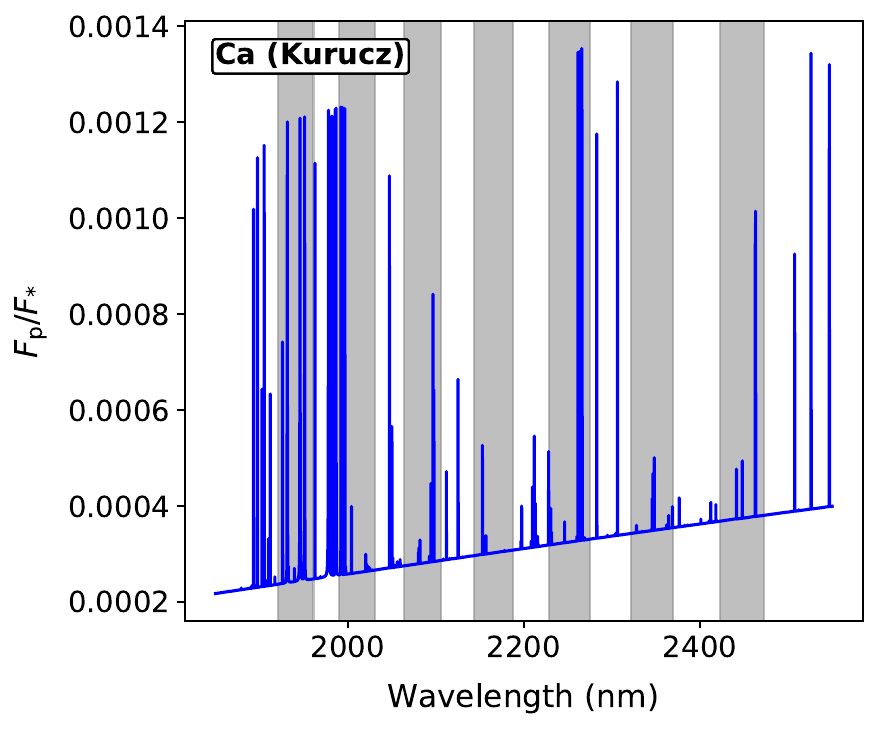}}
    \resizebox{4.5cm}{3cm}
    {\includegraphics {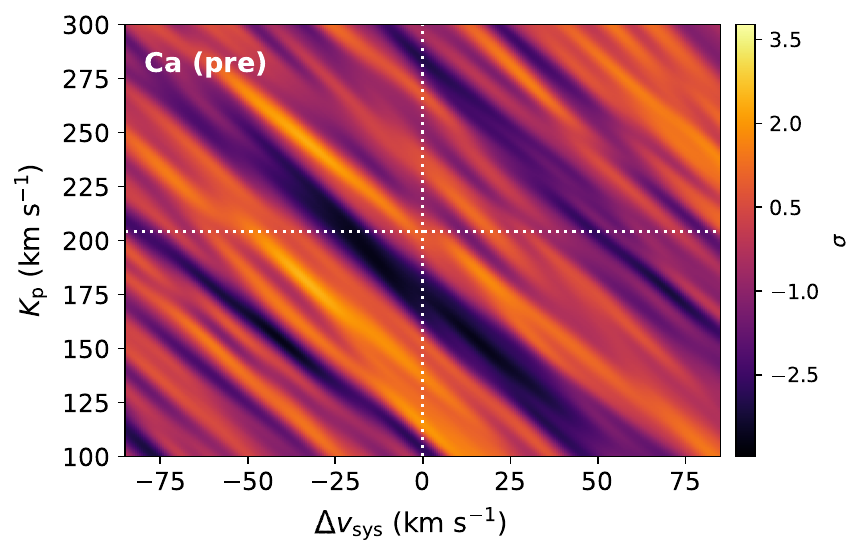}}
    \resizebox{4.5cm}{3cm}
    {\includegraphics {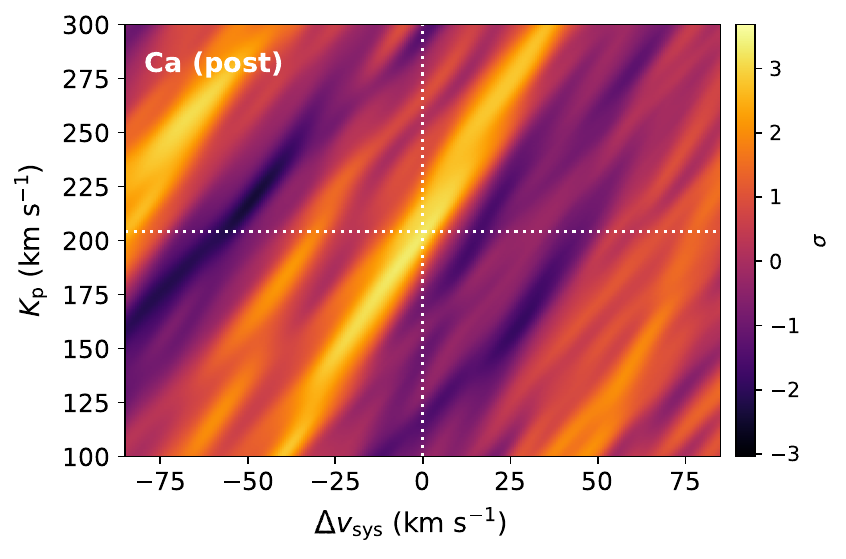}}
    \resizebox{4.5cm}{3cm}
    {\includegraphics {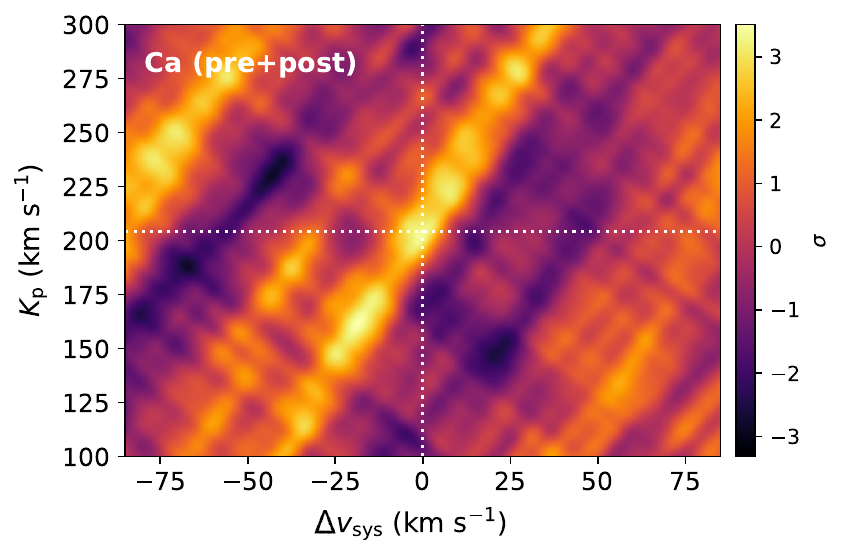}}
    
     \caption{$K_{\rm p}$-$\Delta v_{\rm sys}$ maps for the non-detected species. \emph{Left panels:} emission model of each species with the wavelengths covered by CRIRES+ K2166 setting used in this work shown as vertical grey bands. \emph{Middle and right panels:} the detection maps for the pre-ecl., post-ecl., and combined nights of the CRIRES+ data (Sect.~\ref{sect:3.4}). The high-resolution opacities were obtained from the following sources: $^{13}$CO, CO$_2$ and OH from the HITEMP line list \citep{2010JQSRT.111.2139R}, metal opacities from \citep{2018ASPC..515...47K} and HCN opacities from ExoMol database \citep{2006MNRAS.367..400H, 2014MNRAS.437.1828B}.}
      \label{figC1}
\end{figure*}

\FloatBarrier
\onecolumn
\section{Atmospheric retrievals of MASCARA-1b}\label{appendix:D}
\begin{figure*}[h!]
    \centering
     \resizebox{16cm}{16cm}
    {\includegraphics {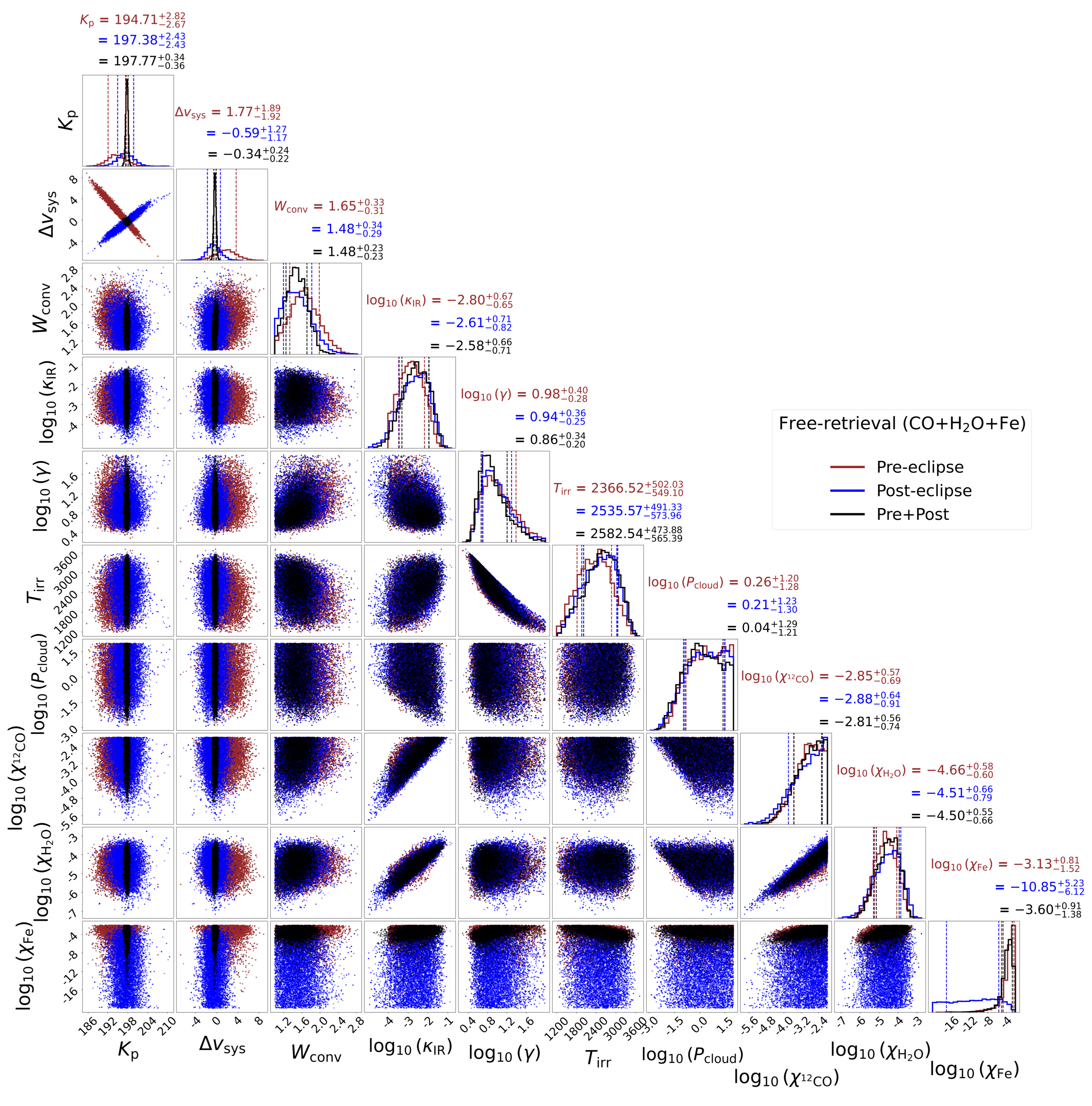}}
     \caption{A summary of our `free-retrieval' results for the combined atmosphere with the 1D and 2D marginalised posterior distributions of each model parameter displayed within a corner plot. The red, blue and black distributions represent the pre-, post- and pre+post eclipse datasets (Sect.~\ref{sect:4.2}).}
      \label{figD1}
\end{figure*}

\FloatBarrier
\onecolumn
\begin{figure*}[h!]
    \centering
     \resizebox{11cm}{11cm}
    {\includegraphics {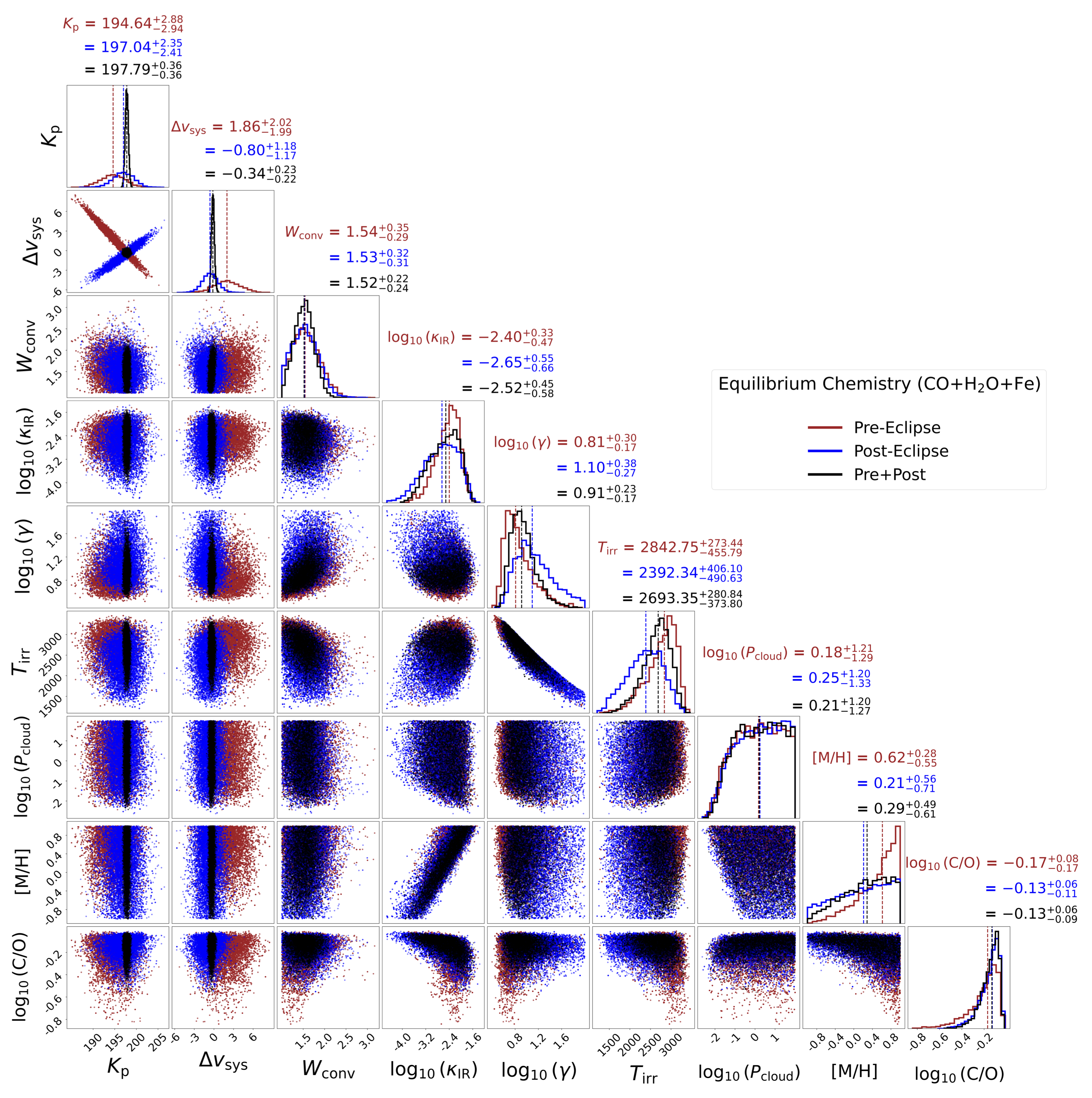}}
     \caption{Same as Figure~\ref{figD1}, but for the chemical retrieval described in Section~\ref{sect:4.2.2}.}
      \label{figD2}
\end{figure*}

\begin{figure*}[h!]
    \centering
     \resizebox{11cm}{11cm}
    {\includegraphics {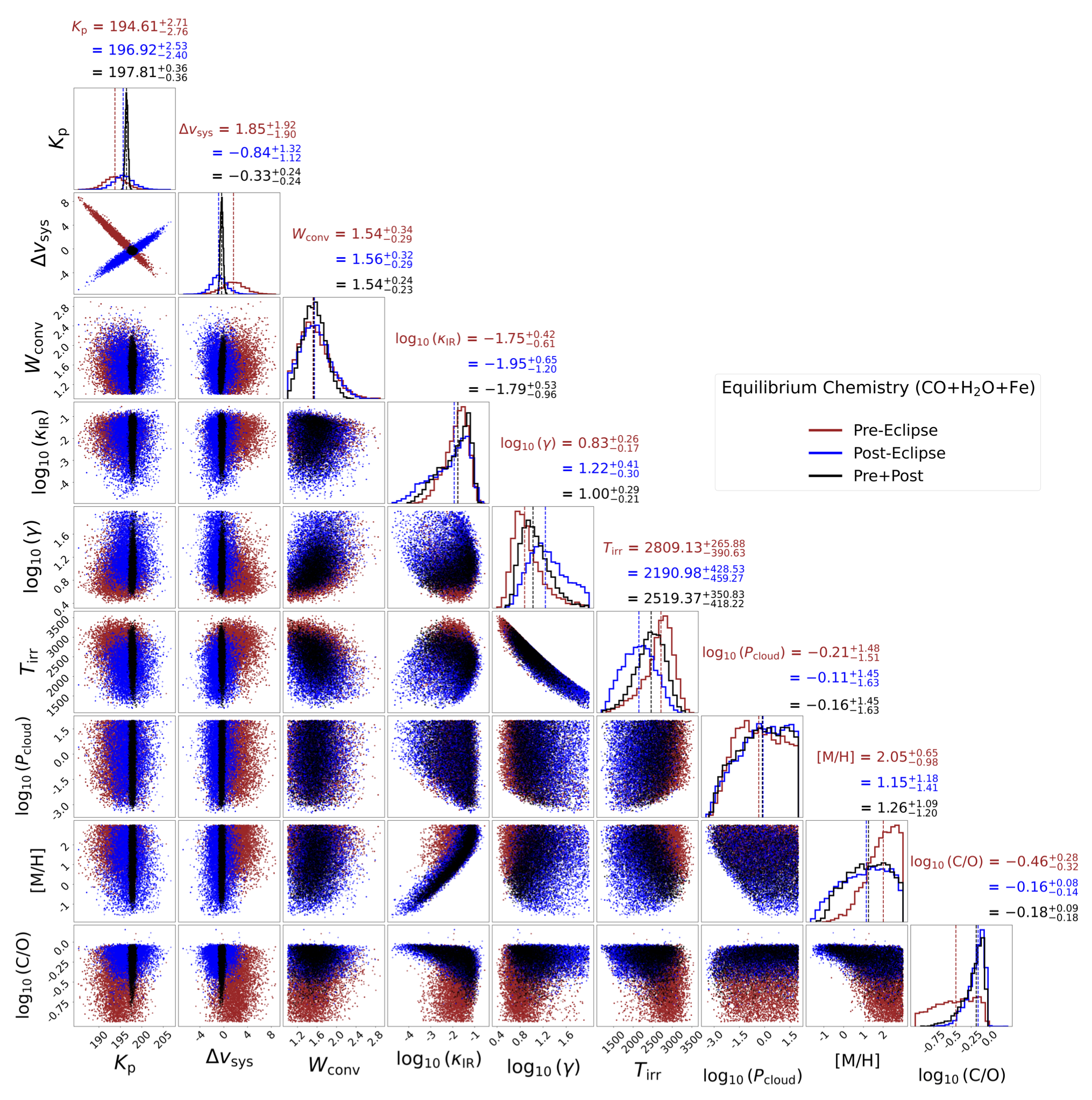}}
     \caption{Same as Figure~\ref{figD2}, but with a broader metallicity prior (from 0.001 to 1000$\rm x$ solar).}
      \label{figD3}
\end{figure*}

\FloatBarrier
\onecolumn
\section{Consistency between \textsc{SysRem} iterations on the post-eclipse and pre+post datasets}\label{appendix:E}

\begin{figure*}[h!]
    \centering
     \resizebox{10.4cm}{10.4cm}
    {\includegraphics {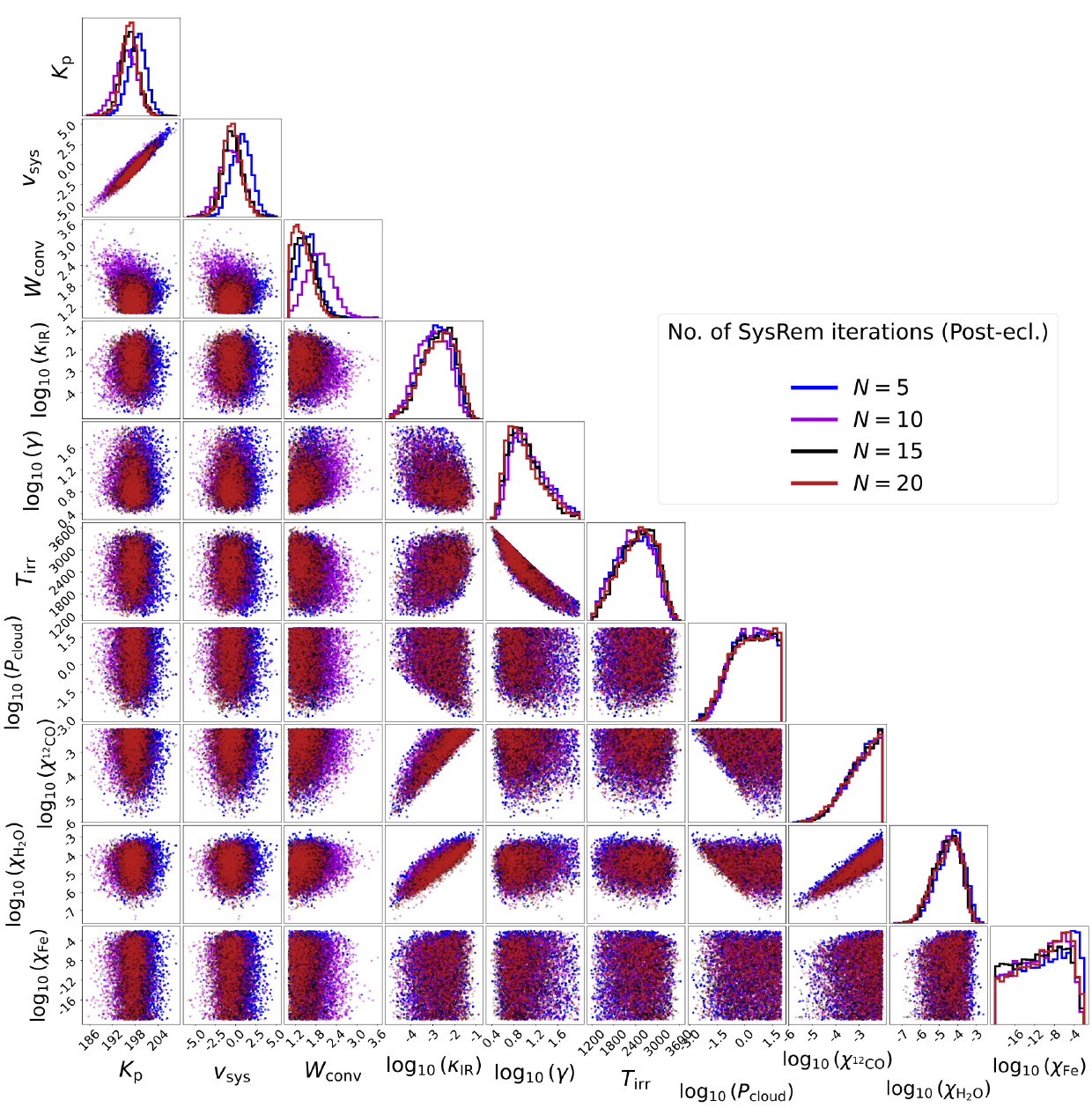}}
     \caption{A summary of our free-retrieval results with the 1D and 2D marginalised posterior distributions of the parameters from the MCMC fit. The different colours show the samples obtained using four values for the \textsc{SysRem} iterations (Sect.~\ref{sect:4.2.1}).}
      \label{figE1}
\end{figure*}

\begin{figure*}[h!]
    \centering
     \resizebox{10.4cm}{10.4cm}
    {\includegraphics {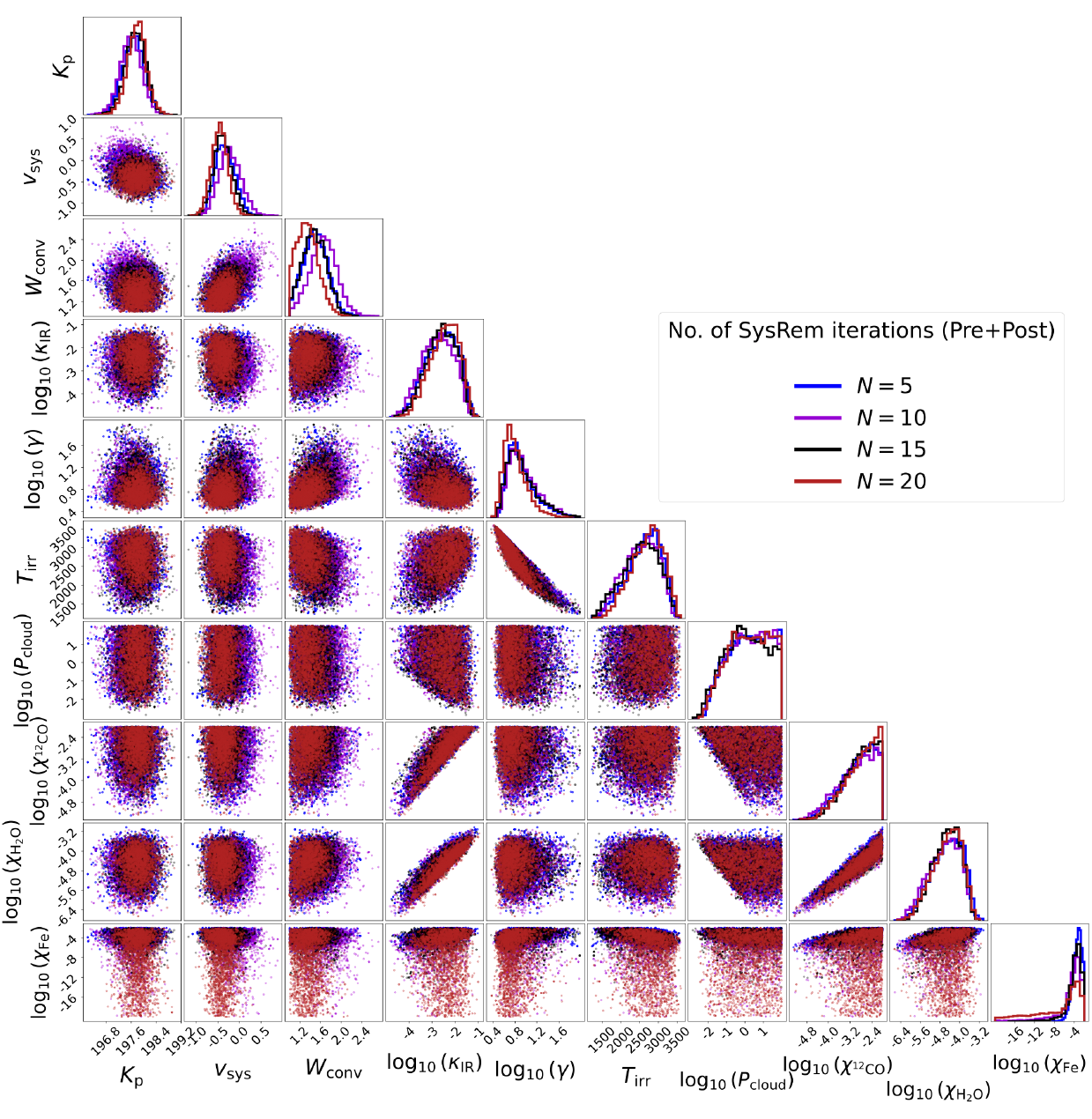}}
     \caption{Same as Fig.~\ref{figE1} but for the pre+post dataset.}
      \label{figE2}
\end{figure*}

\FloatBarrier
\onecolumn
\section{Ground-based high-resolution measurements of well-constrained $\rm C/O$ ratios}\label{appendix:F}

\begin{table*}[htbp]
\caption{$\rm C/O$ ratios for hot/ultra-hot Jupiters included in Fig.~\ref{fig10} with constrained abundances of C- and O-bearing species from ground-based high-resolution measurements.}
\label{tableF1}                     
\centering                                      
\renewcommand{\arraystretch}{2} 
\begin{tabular}{lcccc}
\hline \hline
Planet & Planet Mass [$M_{\rm J}$] & $\rm C/O$ ratio & Instrument [retrieval framework] & Reference\\
\hline
HD 189733b & 1.13$\pm$0.08 & 0.3$\pm$0.1 & KPIC [free-retrieval] & \citet{2024AJ....167...43F}\\
HIP 65A b & 3.213$\pm$0.078 & 0.72$^{+0.13}_{-0.30}$ & IGRINS [chemical eq.] & \citet{2024AJ....167..206B} \\
MASCARA-1b & 3.7$\pm$0.9 & 0.68$^{+0.12}_{-0.22}$ (pre) & CRIRES+ [chemical eq.] & \citet{2023MNRAS.525.2985R}\\
& & 0.75$^{+0.11}_{-0.17}$ (post) & CRIRES+ [chemical eq.] & This work\\
& & 0.74$^{+0.10}_{-0.14}$ (pre+post) & CRIRES+ [chemical eq.] & This work\\
WASP-18b$^{(a)}$ & 10.4296$\pm$0.69 & 0.75$^{+0.14}_{-0.17}$ & IGRINS [free-retrieval] & \citet{2023AJ....165...91B} \\
WASP-33b & 2.093$\pm$0.139 & 0.8$^{+0.1}_{-0.2}$ & KPIC [free-retrieval] & \citet{2023AJ....166...31F}\\
WASP-43b & 1.998$\pm$0.079 & 0.78$\pm$0.09 & CRIRES+ [chem eq.] & \citet{refId1}\\
WASP-76b & 0.894$^{+0.014}_{-0.013}$ & 0.59$^{+0.13}_{-0.14}$ & IGRINS [grid retrieval] & \citet{2024AJ....168...14W}\\
& & 0.94$\pm$0.39 & SPIRou [free-retrieval] & \citet{refId2}\\
WASP-77Ab & 1.76$\pm$0.06 & 0.59$\pm$0.08 & IGRINS [free-retrieval] & \citet{2021Natur.598..580L}\\
WASP-121b & 1.157$\pm$0.070 & 0.73$^{+0.07}_{-0.08}$ & CRIRES+ \& ESPRESSO [chem eq.] & \citet{2025AJ....169...10P}\\
& & 0.70$^{+0.07}_{-0.10}$ & IGRINS [chemical eq.] & \citet{2024AJ....168..293S} \\
WASP-127b$^{(b)}$ & 1.647$^{+0.0214}_{-0.0172}$ & 0.56$^{+0.05}_{-0.07}$ & CRIRES+ [chemical eq.] & \citet{refId11}\\
& & $0.34^{+0.08}_{-0.09}$ $^{(c)}$ & IGRINS [grid retrieval] & \citep{2024AJ....168..201K}\\
WASP-189b & $1.99^{+0.16}_{-0.14}$ & $0.32^{+0.41}_{-0.14}$ & CRIRES+ [chemical eq.] & \citet{refId0}\\
\hline
\end{tabular}
\tablefoot{$^{(a)}$ We include the free-retrieval-based derived value since the authors only reported an upper limit on the $\rm C/O$ ratio retrieved assuming RCTE grid retrievals. $^{(b)}$ We note that there is a C/O measurement for WASP-127b from RCPE grid retrievals using IGRINS \citep[e.g.][]{2024AJ....168..201K}, but it is not included here as the authors reported only an upper limit. $^{(c)}$ Although the authors report a bounded value using RCTE grid retrievals, they emphasise in their study that their main conclusions are drawn from the RCPE grid retrievals, which provided an upper limit on the $\rm C/O$ ratio ($< 0.68$). As such, we include the bounded value for completeness.}
\end{table*}

\FloatBarrier
\onecolumn
\section{Injection/recovery test for Fe}\label{appendix:G}
\begin{figure*}[h!]
    \centering
     \resizebox{14cm}{14cm}
    {\includegraphics {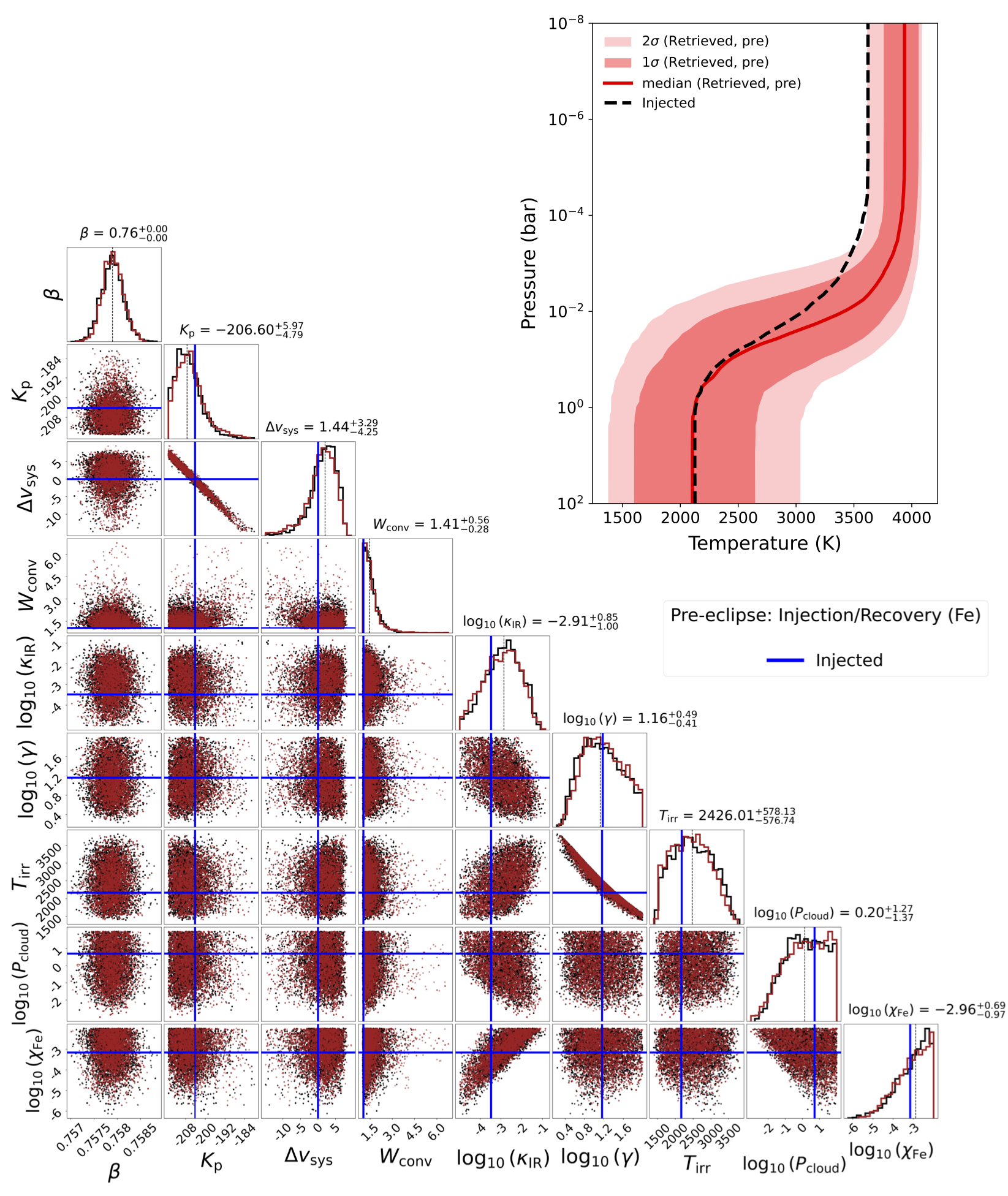}}
     \caption{Results of the injection and recovery test for the pre-eclipse Fe (Sect.~\ref{sect:5.2}). The red and black posterior distributions represent independent sub-chains of the same MCMC chain, both converging to similar distributions. \emph{Upper right:} the retrieved $T$-$P$ profile computed from 10,000 random samples of the MCMC; the solid red curve shows the median profile and the shaded regions show the 1$\sigma$ and 2$\sigma$ contours. The dashed black curve shows the injected $T$-$P$ profile.}
      \label{figG1}
\end{figure*}
\begin{figure*}[h!]
    \centering
     \resizebox{8cm}{5.5cm}
    {\includegraphics {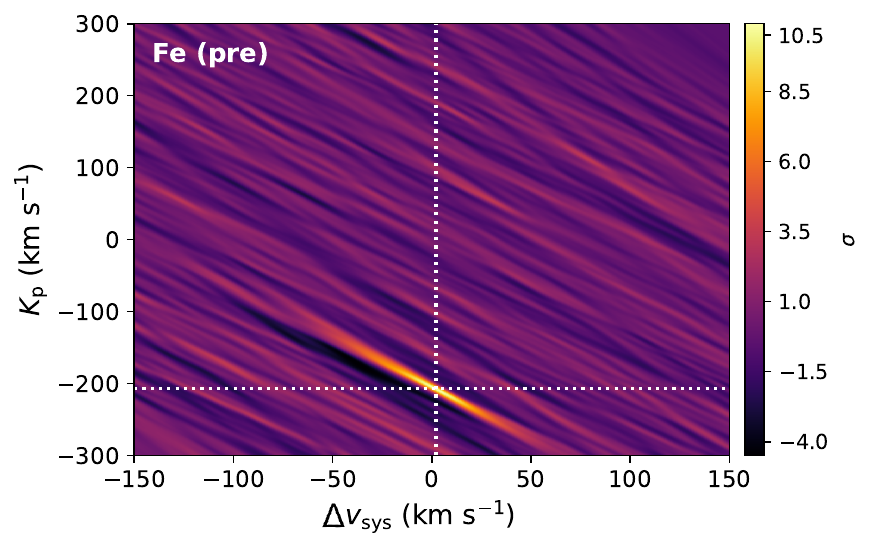}}
    \resizebox{8cm}{5.5cm}
    {\includegraphics {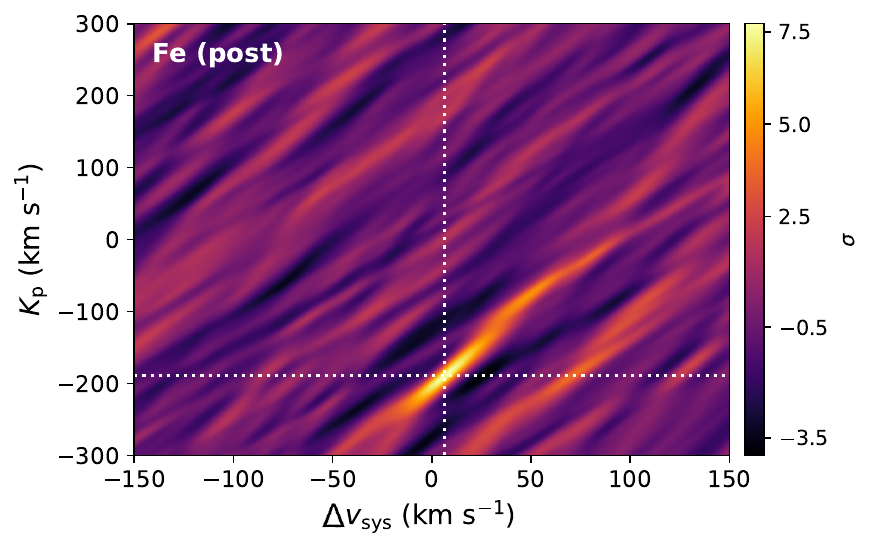}}
     \caption{Recovered $K_{\rm p}$-$\Delta v_{\rm sys}$ maps of iron for the pre- and post-eclipse datasets. The white dotted lines mark the peak of detection.}
      \label{figG23}
\end{figure*}

\FloatBarrier
\onecolumn
\section{Revisiting the pre-eclipse iron detection}\label{appendix:H}
\begin{figure*}[h!]
    \centering
     \resizebox{10.4cm}{10.4cm}
    {\includegraphics {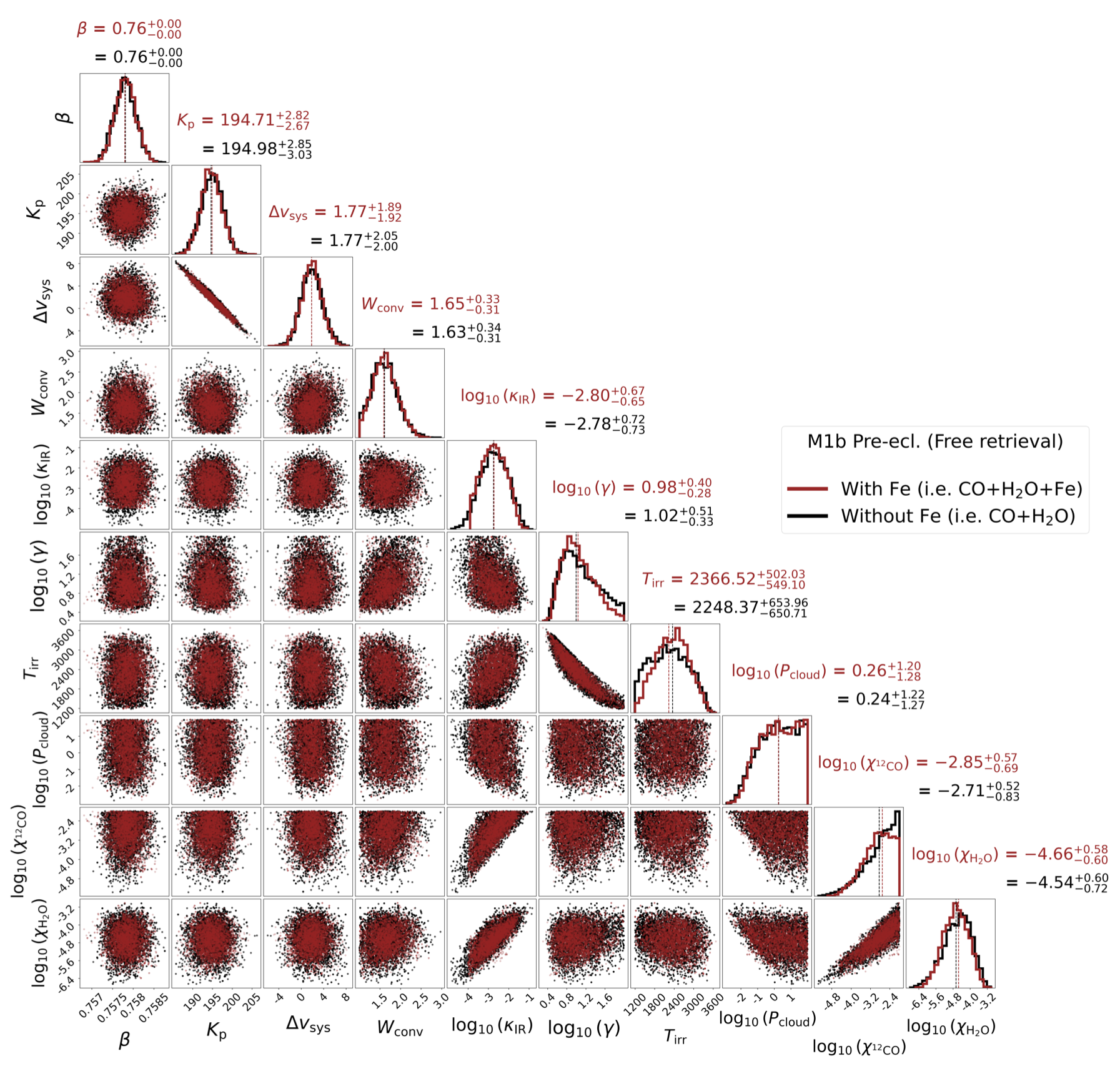}}
     \caption{Results of the free-retrieval for the pre-eclipse dataset with (red) and without (black) the inclusion of Fe as described in Sect.~\ref{sect:5.2}.}
      \label{figH1}
\end{figure*}
\begin{figure*}[h!]
    \centering
     \resizebox{10.4cm}{10.4cm}
    {\includegraphics {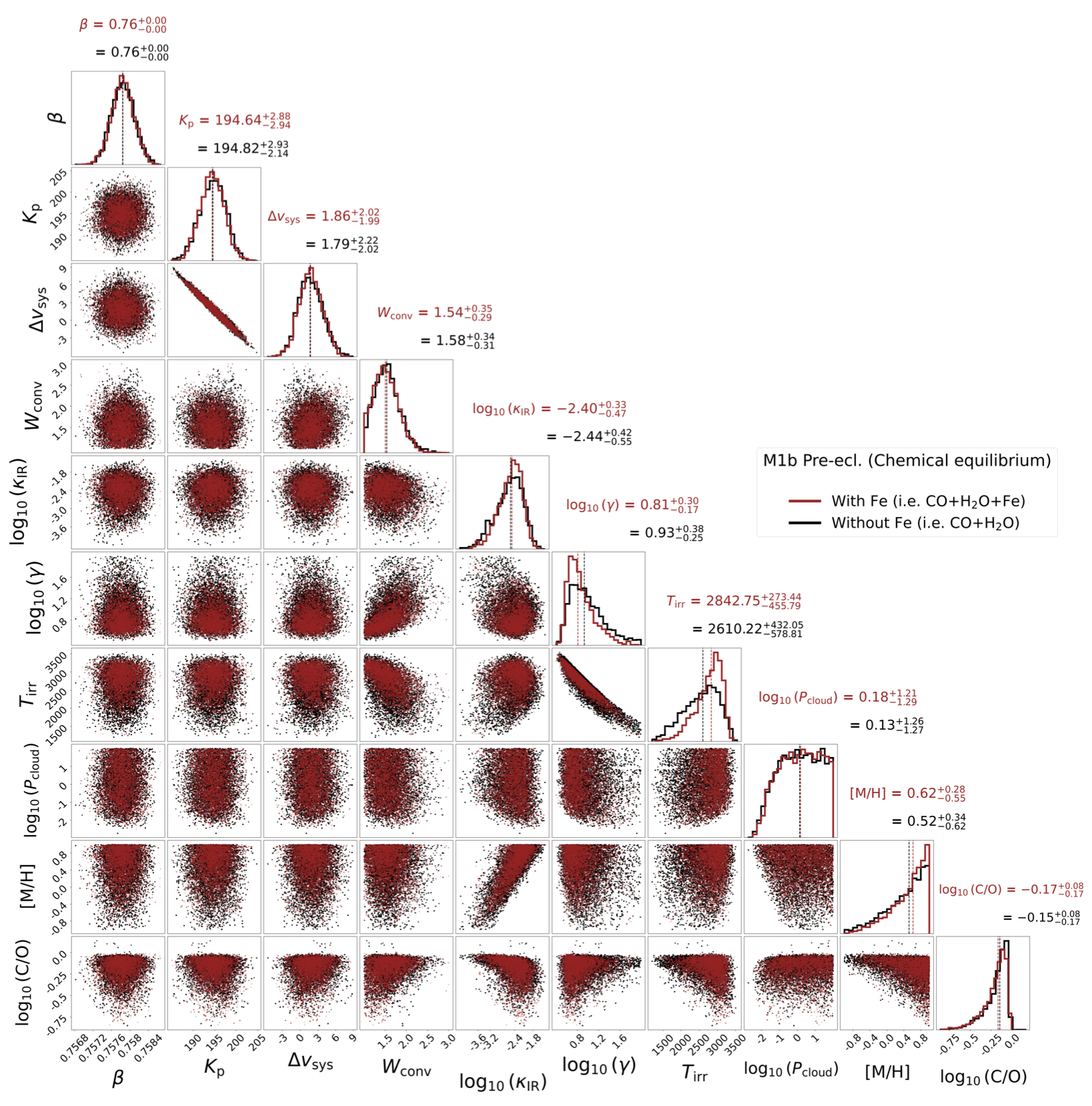}}
     \caption{Same as Fig.~\ref{figH1} but for a chemical model.}
      \label{figH2}
\end{figure*}

\FloatBarrier
\onecolumn
\section{Phase-dependent retrieval}\label{appendix:I}
\begin{figure*}[h!]
    \centering
     \resizebox{13cm}{13cm}
    {\includegraphics {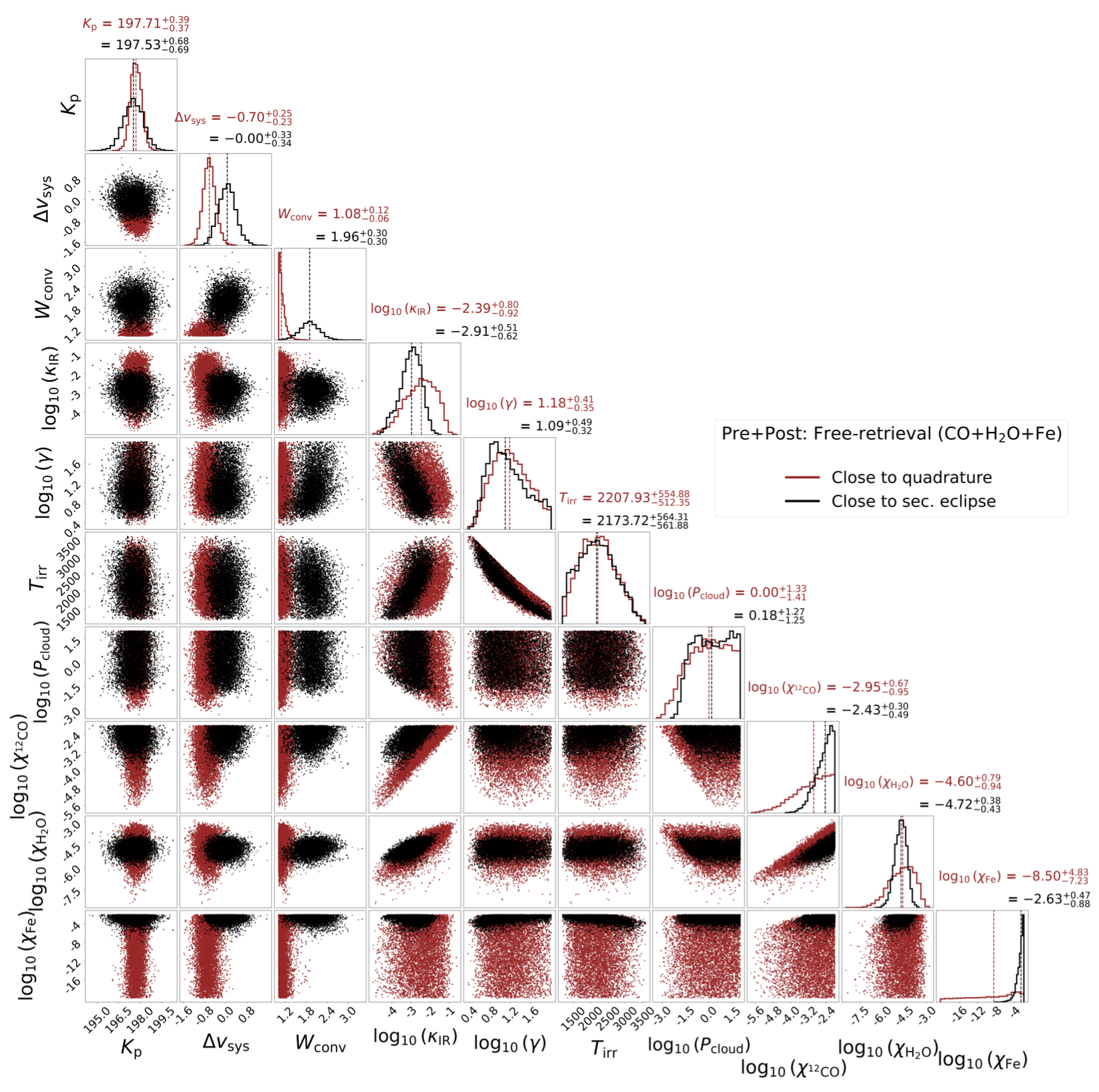}}
     \caption{Results of the joint (pre+post) phase-dependent free-retrieval as described in Sect.~\ref{sect:5.2}. The subsets of phases close to the quadrature are shown in red and phases close to the eclipse are shown in black. \emph{Top right:} retrieved $T$-$P$ profile.}
      \label{figI1}
\end{figure*}
\begin{figure*}[h!]
    \centering
     \resizebox{10cm}{7cm}
    {\includegraphics {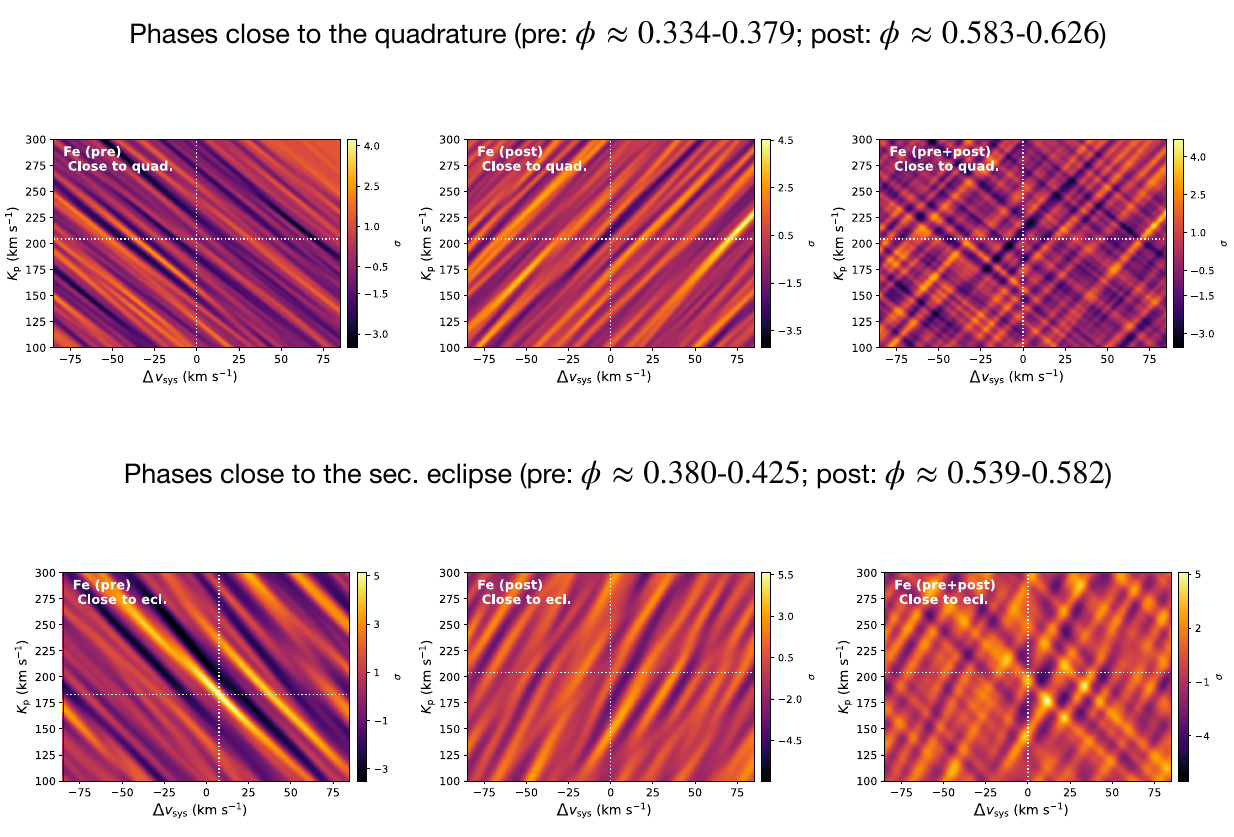}}
    \hspace{0.4cm}
    \resizebox{7cm}{6cm}
    {\includegraphics {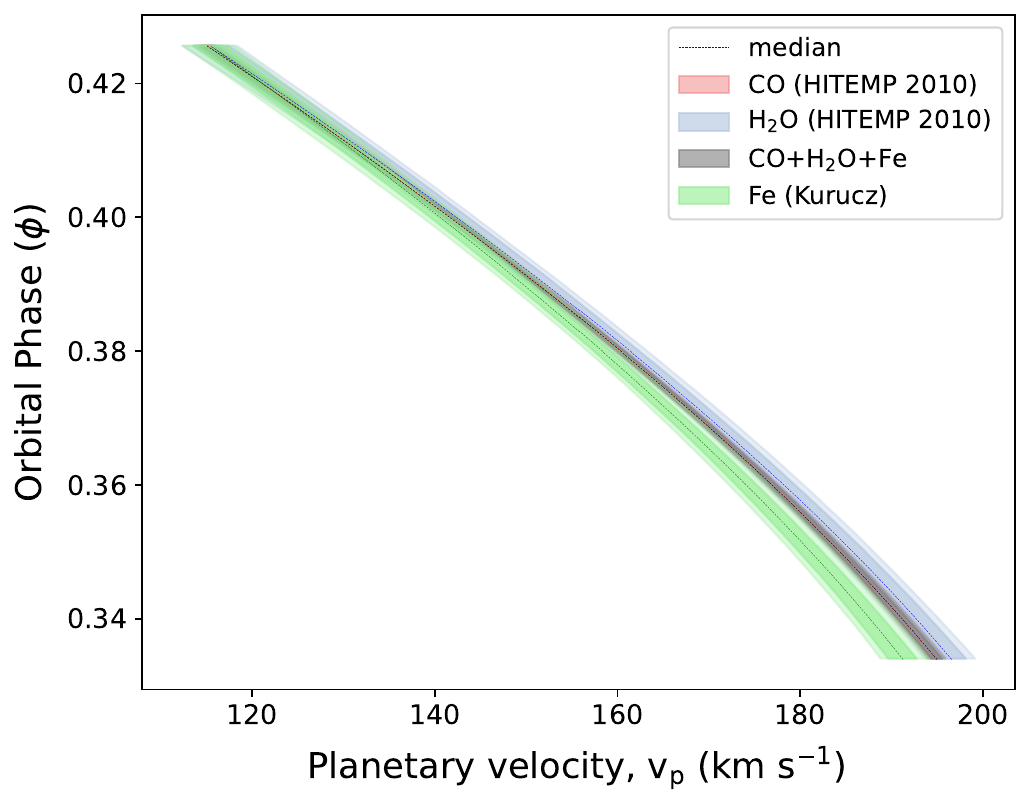}}
     \caption{\emph{Left:} $K_{\rm p}$-$\Delta v_{\rm sys}$ maps for the iron signal using the phase-dependent retrieval setup. \emph{Right:} The retrieved planetary orbital velocity ($v_{\rm p}$) of the detected species for the pre-eclipse data alone, computed from 10,000 random samples of the MCMC. The dashed lines show the median value, and the shaded regions show the $1\sigma$ and $2\sigma$ contours (see Sect.~\ref{sect:5.3}).}
      \label{figI2}
\end{figure*}

\FloatBarrier
\onecolumn
\section{Plot of `alpha' detection significance using different line lists for the pre-eclipse, post-eclipse and pre+post datasets}\label{appendix:J}
\begin{figure*}[h!]
    \centering
     \resizebox{7cm}{5.5cm}
    {\includegraphics {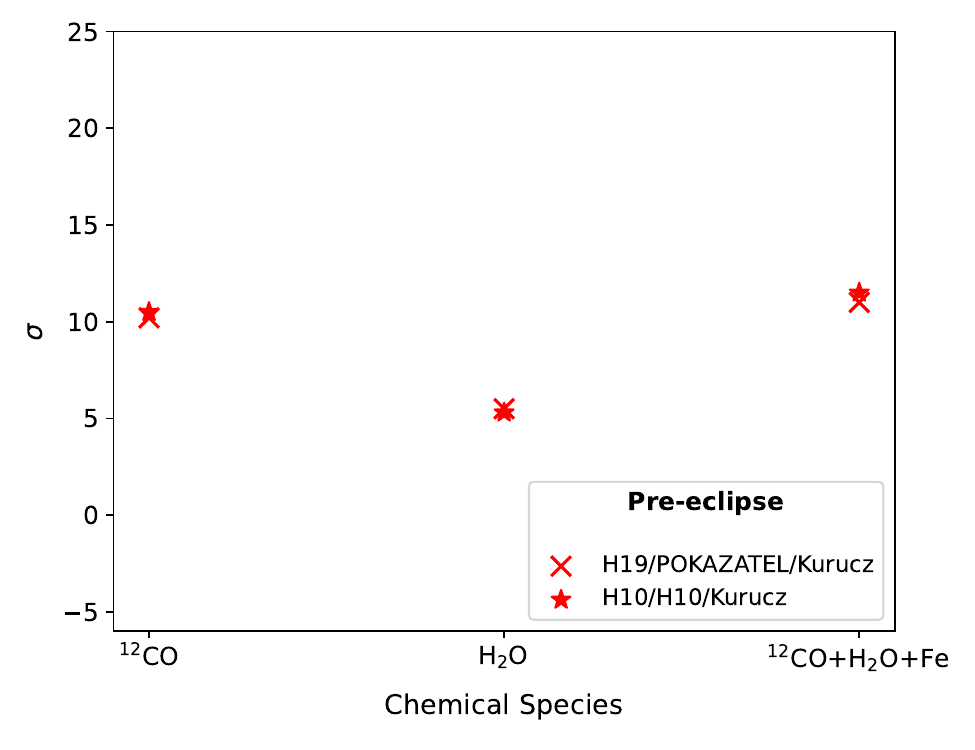}}
    \resizebox{7cm}{5.5cm}
    {\includegraphics {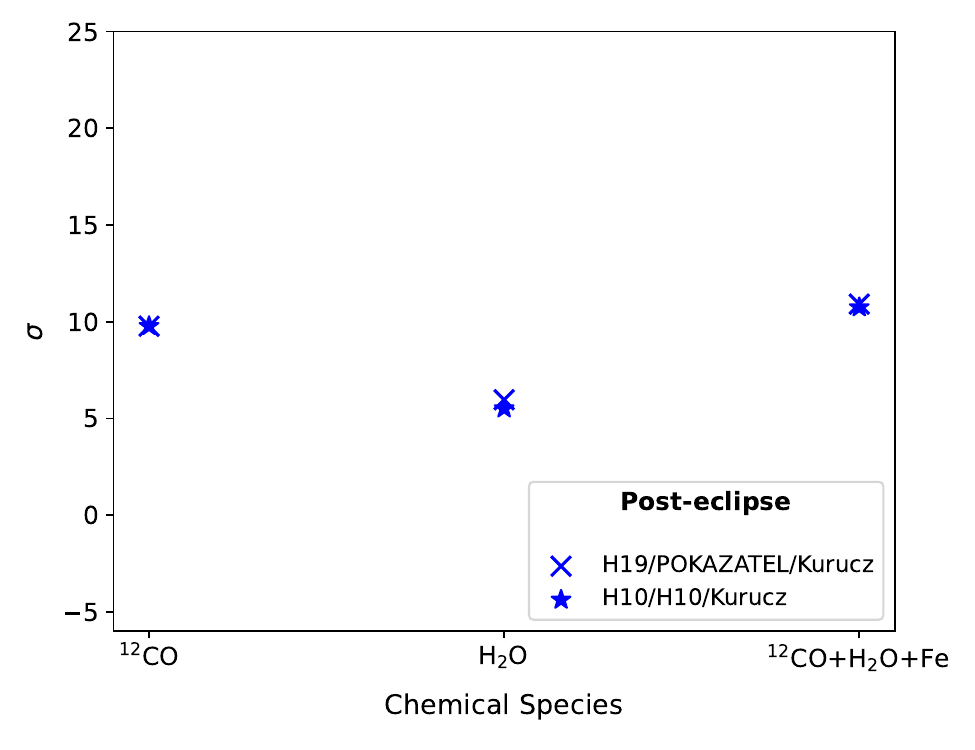}}
    \resizebox{7cm}{5.5cm}
    {\includegraphics {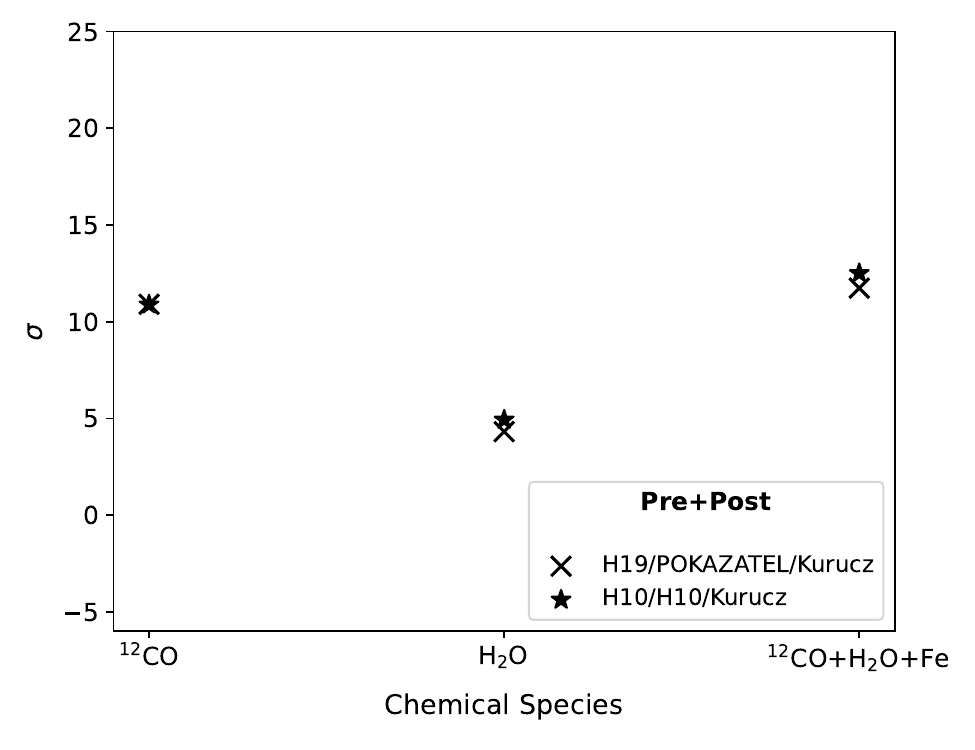}}
     \caption{Plot of ``alpha'' detection significance computed from the conditional likelihood distribution (detailed in Sects.~\ref{sect:3.3} and ~\ref{sect:4}) using different line lists (see Sect.~\ref{sect:5.3}) shown for the pre-eclipse, post-eclipse and pre+post datasets. The labels `H10' and `H19' correspond to the HITEMP 2010 and HITEMP 2019 line lists, respectively.}
      \label{figJ1}
\end{figure*}

\FloatBarrier 
\clearpage

\end{appendix}
\end{document}